\documentclass{aa}
\newif\iffigure
\figurefalse
\figuretrue

\newcommand{\add}[1]{\textcolor{black}{#1}}
\newcommand{\addsec}[1]{\textcolor{black}{#1}}
\newcommand{\addthi}[1]{\textcolor{black}{#1}}
\newcommand{\addfor}[1]{\textcolor{black}{#1}}
\newcommand{\addfif}[1]{\textcolor{black}{#1}}
\newcommand{\rev}[1]{\textcolor{black}{#1}}
\newcommand{\revsec}[1]{\textcolor{black}{#1}}
\usepackage{graphicx,natbib,url,twoopt}
\usepackage[varg]{txfonts}           %% A&A font choice
\usepackage{hyperref}              %% for pdflatex
\usepackage{pdfcomment}              %% for popup acronym meanings
\usepackage{acronym}                 %% for popup acronym meanings
\usepackage{comment}
\usepackage{cases}
\usepackage{empheq}                  %Figs. 1,2 
\usepackage{here}
\usepackage{bm}
\usepackage[version=3]{mhchem}
\hypersetup{% hyperref option
 colorlinks=true,%
 linkcolor=blue,
 citecolor=blue,
}

\usepackage{framed}
\usepackage{mdframed}

\defcitealias{kuwahara2022dust}{Paper {\rm I}}
\defcitealias{KK23}{KK23}
\usepackage{lineno}
\newcommand*\patchAmsMathEnvironmentForLineno[1]{
  \expandafter\let\csname old#1\expandafter\endcsname\csname #1\endcsname
  \expandafter\let\csname oldend#1\expandafter\endcsname\csname end#1\endcsname
  \renewenvironment{#1}
     {\linenomath\csname old#1\endcsname}
     {\csname oldend#1\endcsname\endlinenomath}}
\newcommand*\patchBothAmsMathEnvironmentsForLineno[1]{
  \patchAmsMathEnvironmentForLineno{#1}
  \patchAmsMathEnvironmentForLineno{#1*}}
\AtBeginDocument{
\patchBothAmsMathEnvironmentsForLineno{equation}
\patchBothAmsMathEnvironmentsForLineno{align}
\patchBothAmsMathEnvironmentsForLineno{flalign}
\patchBothAmsMathEnvironmentsForLineno{alignat}
\patchBothAmsMathEnvironmentsForLineno{gather}
\patchBothAmsMathEnvironmentsForLineno{multline}
}
\bibpunct{(}{)}{;}{a}{}{,}    %% natbib cite format used by A&A and ApJ

\makeatletter
\newcommand{\bibnote}[2]{\global\@namedef{#1note}{#2}}
\newcommand{\biblink}[2]{\global\@namedef{#1link}{#2}}
\newcommand{\Tabref}[1]{Table~\ref{#1}}
\newcommand{\Equref}[1]{Eq.~(\ref{#1})}
\newcommand{\Figref}[1]{Fig.~\ref{#1}}
\makeatother

\makeatletter
 \newcommandtwoopt{\citeads}[3][][]{%
   \nonstopmode%              %% fix to not stop at error message in latex
   \href{http://adsabs.harvard.edu/abs/#3}%
        {\def\hyper@linkstart##1##2{}%
         \let\hyper@linkend\@empty\citealp[#1][#2]{#3}}%   %% Rutten, 2000
   \biblink{#3}{\href{http://adsabs.harvard.edu/abs/#3}{ADS}}%
   \errorstopmode}            %% fix to resume stopping at error messages 
 \newcommandtwoopt{\citepads}[3][][]{%
   \nonstopmode%              %% fix to not stop at error message in latex
   \href{http://adsabs.harvard.edu/abs/#3}%
        {\def\hyper@linkstart##1##2{}%
         \let\hyper@linkend\@empty\citep[#1][#2]{#3}}%     %% (Rutten 2000)
   \biblink{#3}{\href{http://adsabs.harvard.edu/abs/#3}{ADS}}%
   \errorstopmode}            %% fix to resume stopping at error messages
 \newcommandtwoopt{\citetads}[3][][]{%
   \nonstopmode%              %% fix to not stop at error message in latex
   \href{http://adsabs.harvard.edu/abs/#3}%
        {\def\hyper@linkstart##1##2{}%
         \let\hyper@linkend\@empty\citet[#1][#2]{#3}}%     %% Rutten (2000)
   \biblink{#3}{\href{http://adsabs.harvard.edu/abs/#3}{ADS}}%
   \errorstopmode}            %% fix to resume stopping at error messages 
 \newcommandtwoopt{\citeyearads}[3][][]{%
   \nonstopmode%              %% fix to not stop at error message in latex
   \href{http://adsabs.harvard.edu/abs/#3}%
        {\def\hyper@linkstart##1##2{}%
         \let\hyper@linkend\@empty\citeyear[#1][#2]{#3}}%  %% 2000
   \biblink{#3}{\href{http://adsabs.harvard.edu/abs/#3}{ADS}}%
   \errorstopmode}            %% fix to resume stopping at error messages 
\makeatother

\newacro{ADS}{Astrophysics Data System}
\newacro{NLTE}{non-local thermodynamic equilibrium}
\newacro{NASA}{National Aeronautics and Space Administration}

\defcitealias{Kuwahara:2020}{Paper {\rm I}}

\begin{document}
%\linenumbers
\authorrunning{A. Kuwahara et al.}
\titlerunning{Dust ring and gap formation by gas flow}
   \title{Dust ring and gap formation by gas flow induced by low-mass planets embedded in protoplanetary disks}
   \subtitle{$\rm II$. \add{Time-dependent model}}
      \author{Ayumu Kuwahara\inst{1,2,3} 
          \and \addsec{Michiel Lambrechts}\inst{1}
          \and Hiroyuki Kurokawa\inst{4,\addsec{5}}
          \and Satoshi Okuzumi\inst{2}
          \and Takayuki Tanigawa\inst{\addsec{6}}}

   \institute{Center for Star and Planet Formation, GLOBE Institute, University of Copenhagen, Øster Voldgade 5-7, 1350 Copenhagen, Denmark
        \and
              Department of Earth and Planetary Sciences, Tokyo Institute of Technology, 2-12-1 Ookayama, Meguro-ku, Tokyo 152-8551, Japan
         \and
             Earth-Life Science Institute, Tokyo Institute of Technology, 2-12-1 Ookayama, Meguro-ku, Tokyo 152-8550, Japan
         \and Department of Earth Science and Astronomy, Graduate School of Arts and Sciences, The University of Tokyo, 3-8-1 Komaba, Meguro-ku, Tokyo 153-8902, Japan
         \and \add{Department of Earth and Planetary Science, Graduate School of Science, The University of Tokyo, 3-8-1 Komaba, Meguro-ku, Tokyo 153-8902, Japan}
         \and National Institute of Technology, Ichinoseki College, Takanashi, Hagisho, Ichinoseki-shi, Iwate 021-8511, Japan}

   \date{Received September XXX; accepted YYY}

% 5 {} token are mandatory
 
  \abstract{The observed dust rings and gaps in protoplanetary disks \addthi{could be} \add{imprints} of forming planets. \addfor{Even low-mass planets in the one-to-ten Earth-mass regime, that do not yet carve deep gas gaps, can generate such dust rings and gaps by driving a radially-outwards gas flow, as shown in previous work. However, understanding the creation and evolution of these dust structures is challenging due to dust drift and diffusion, requiring an approach beyond previous steady state models.} Here we investigate the time evolution of the dust surface density influenced by the planet-induced gas flow, \addthi{based on post-processing three-dimensional hydrodynamical simulations.} We find that planets \addthi{larger than a} dimensionless thermal mass \addthi{of $m=0.05$}, corresponding to $0.3$ Earth mass at 1 au \addthi{or 1.7 Earth masses at 10 au}, generate dust rings and gaps, \addthi{provided that solids have small Stokes numbers} (${\rm St}\lesssim10^{-2}$) \addthi{and that the disk midplane is} weakly turbulent ($\alpha_{\rm diff}\lesssim10^{-4}$). \addfor{As dust particles pile up outside the orbit of the planet, the interior gap expands with time, when the advective flux dominates over diffusion. Dust gap depths range from a factor a few, to several orders of magnitude, depending on planet mass and the level of midplane particle diffusion.} \addthi{We construct \addfor{a semi-analytic model describing the width} of the dust ring and gap, and then compare \addfor{it} with the observational data. We find that up to \rev{65\%} of the observed wide-orbit gaps could be explained as resulting from the presence of a low-mass planet, assuming \addfor{$\alpha_{\rm diff}=10^{-5}$} and \addfor{${\rm St}=10^{-3}$}. \addfif{However, it is more challenging to explain the observed wide rings, which in our model would require the presence of a population of small particles (${\rm St=10^{-4}}$).} Further work is \rev{needed} to explore the role of pebble fragmentation, planet migration, and the effect of multiple planets.}} 

   \maketitle
%---------------------------------------------------------
%---------------------------------------------------------
%---------------------------------------------------------

\section{Introduction}\label{sec:Introduction}
%---------------------------------------------------------
Observations of protoplanetary disks have revealed substructures in dust profiles at \addthi{distances outside of 10} au \citep[e.g.,][]{ALMA:2015,andrews2018-DSHARP1}. \addfif{An unbiased protoplanetary disk survey in the Taurus star-forming region, where approximately $75\%$ of solar-mass stars have disks \citep{luhman2009disk}, exhibits a substructure occurrence rate as high as  $40$\% \citep{long2018gaps}.} \addthi{The most common type of the dust substructures are annular depletions and enhancements in the continuum emissions, which are referred to as dust rings and gaps.}

Several mechanisms have been proposed to explain the dust rings and gaps, such as various types of instabilities, processing of dust at the snow lines, magnetohydrodynamic effects, \add{and pressure maxima in a radial pressure profile} \citep[][and references therein]{bae2023structured}. %Pressure maxima in a radial pressure profile are one of them, though the properties of the pressure maxima such as the origin and the lifetime are still under debate. The dust particles drift from the outer to inner disk regions with a velocity proportional to the pressure gradient of the disk gas \citep[the so-called radial drift of dust;][]{Whipple:1972,Weidenschilling:1977}. The drifting dust is trapped at the pressure maxima, which leads to the formation of the dust rings and gaps.
\add{In addition to the preceding mechanisms,} \addthi{a widely accepted formation channel is by planets carving gas gaps with masses typically $\gtrsim15\,M_\oplus$ (Earth masses). We will hereafter refer to this mechanism as the gas-gap mechanism \citep[e.g.,][]{paardekooper2006dust}.}
%Planets with typically $\gtrsim15\,M_\oplus$ (Earth masses) can open a gas gap in a disk, generating a pressure maximum at the gas gap edge (e.g., Parrdekooper+ 2006, Lambrechts & Johansen 2014). 
\addthi{If} an observed dust gap at $\gtrsim10$ au is caused by an unseen planet, the planet mass can be estimated from the results of the disk-planet interaction simulations \citep[][]{zhang2018-DSHARP7,lodato2019newborn,wang2021architecture}. The inferred masses of putative planets are distributed in a range of a few Earth-masses to $\sim$10 Jupiter-masses, \add{$\sim70\%$ of which} have $>0.1$ Jupiter-mass \citep{bae2023structured}. \addfif{Considering the fraction of disks with substructures, these estimates suggest that the occurrence fraction of planets with masses exceeding $0.1$ Jupiter-mass in wide orbits is approximately 20$\%$. This value appears to be in tension with the low occurrence rate of cold gas giants as suggested by the current observed period-mass diagram of exoplanets \citep[$<10\%$;][]{Fernandes:2019,fulton2021california} and predicted occurrence rates in population synthesis models \citep[][]{mordasini2018handbook,emsenhuber2021new}. The current period and mass distribution of exoplanets could be reproduced if these putative planets undergo the large-scale inward migration \citep{lodato2019newborn,mulders2021mass,van2021stellar}. However, the feasibility of this scenario could be low due to inefficient type-$\rm II$ migration in low-viscosity disks \citep{ndugu2019observed,muller2022emerging,tzouvanou2023all}.}

\rev{Dust substructures can be created by low-mass, no-gas-gap-opening planets ($\lesssim10\,M_\oplus$) in disks. A dust gap forms due to the gravitational interaction between the planet and the dust \citep{Muto:2009,dipierro2016two,dipierro2017opening}. In our previous work, \cite{kuwahara2022dust} (hereafter \citetalias{kuwahara2022dust}), we showed that the gas flows driven by low-mass planets can create dust substructures in disks with low turbulent viscosity at the disk midplane (hereafter referred to as the gas-flow mechanism).}
%In our previous work, \cite{kuwahara2022dust} (hereafter \citetalias{kuwahara2022dust}), we \addthi{showed that dust substructures can be created} by low-mass, non-gap-opening planets ($\lesssim10\,M_\oplus$) in disks with low turbulent viscosity at the disk midplane \addsec{(hereafter referred to as the gas-flow mechanism)}. 
\add{A} low-mass protoplanet (typically $\sim0.1\text{--}10\,M_\oplus$) embedded in a disk induces a three-dimensional (3D) gas flow \addsec{\citep[e.g.,][]{Ormel:2015b,Fung:2015,Kuwahara:2019}}. %In particular, a midplane outflow of the gas plays an important role in the dust motion. %Small dust grains in disks are sensitive to the gas flow. 
\addsec{If} the disk midplane is weakly turbulent, as suggested by recent studies \citep{villenave2022highly,jiang2024grain}, the radially-outward outflow of the gas generates a congestion of dust outside the planetary orbit, \addthi{because} the radially-inward outflow blows dust away from the planetary orbit. \addthi{This} leads to the formation of \addthi{a dust ring outside the planetary orbit, and a gap interior to it. This mechanism thus differs from the dust substructures generated by carving gas gaps, as done around higher mass planets}. \add{The dust ring and gap formation by low-mass, \rev{no}-gas-gap-opening planets could \addthi{therefore reconcile the frequently observed dust gaps seen in disks that have no corresponding gas gaps  \citep{zhang2021molecules,jiang2022no}. Moreover, it may \rev{be} a fresh perspective on the proposed large fraction of} low-mass planets ($\lesssim10\,M_\oplus$) at wide orbits ($\gtrsim10^3$ days) inferred from a population synthesis model \citep{drazkowska2023planet}.}

%The radially-outward (inward) outflow inhibits (enhances) the inward drift of small dust ($\lesssim1$ cm), leading to the formation of the dust ring and gap with their radial extent of $\sim1\text{--}10$ times gas scale height. 

\citetalias{kuwahara2022dust} assumed a steady state for simplicity and computed the dust surface density perturbed by the planet-induced gas flow. %\erase{Because the 3D structure of the gas flow has the complex dependence on the parameters such as the planetary mass and the pressure gradient of the disk gas \citep{Ormel:2015b,Kurokawa:2018}, the simplification was carried out as an initial step to investigate the response of the dust surface density to the gas flow.} 
However, the validity of the steady-state assumption is nontrivial. \addthi{The 3D structure of the gas flow has a complex dependence on different parameters such as the planetary mass and the pressure gradient of the disk gas \citep{Ormel:2015b,Kurokawa:2018}, which in turn regulates how dust piles-up outside the orbit of the planet and gets depleted interior to it.} %The time evolution of the dust surface density should depend on the drift and diffusion timescales of dust, which depends on the dust size and the turbulence strength in a disk. 
It is \addthi{therefore} still unclear how the profiles of dust ring and gap vary with time and \addsec{compare with observed disks whose ages are typically a few Myrs} \citep{Haisch2001ApJ}.

In this study, we investigate the time evolution of the dust surface density perturbed by the planet-induced gas flow (Sect. \ref{sec:Numerical results}). We extend the parameter space and conduct a more comprehensive investigation than in \citetalias{kuwahara2022dust}. In addition, we introduce semi-analytic models describing the properties of the dust rings and gaps such as the widths and the depths based on the results of numerical simulations (Sect. \ref{sec:Semi-analytic models of dust rings and gaps}). These approaches allow us to efficiently explore the disk parameter space where low-mass planets can create dust gaps and rings that are comparable in magnitude to those observed in young disks. In Sect. \ref{sec:Discussion}, we discuss the implications for planet formation and observations of protoplanetary disks, \add{showing that \addthi{up to \rev{$\sim65\%$ ($\sim15\%$)} of} the observed dust gaps \addthi{(rings)} could be caused by the gas flow induced by low-mass planets at wide-orbits}. We conclude in Sect. \ref{sec:Conclusions}. \rev{For readers who want to quickly go through our key results, Eqs. (\ref{eq:W gap SA}), (\ref{eq:delta SA}), and (\ref{eq: W ring SA}) are the semi-analytic models describing the widths of the dust ring and gap and the depth of the dust gap. We compared these models with the observational data in Figs. \ref{fig:gap_width_vs_obs} and \ref{fig:ring_width_vs_obs} in Sect. \ref{sec:Discussion}.}

%---------------------------------------------------------
\section{Numerical method\add{s}}\label{sec:Numerical method}

%\section{Overview of dust ring and gap formation by the gas flow}
%In this study, we focus on the influence of the gas flow around an embedded planet on the spatial distribution of dust. Planets embedded in disks perturb the surrounding disk gas. We focus in particular on the midplane outflow of the gas, which occurs in the radial direction to the disk. Small dust grains in disks are sensitive to the gas flow and thus their radial drift velocity is perturbed by the outflow of the gas. As a result, the gas flow induced by planets generates a congestion of dust outside the planetary orbit and (or) blow dust away from the planetary orbit, leading to the formation of a dust substructure.

In \citetalias{kuwahara2022dust}, we investigated the dust substructure formation by \addsec{the gas-flow mechanism} in three steps: (1) we first performed 3D hydrodynamical simulations of the gas flow around an embedded planet and obtained the gas flow velocity field. (2) \addthi{By} the post-processing \addthi{these} simulations, we calculated the radial drift velocity of dust perturbed by the gas flow, where the obtained gas velocity field was used to compute the dust motion. (3) Finally, assuming a steady state, we computed the dust surface density by incorporating the obtained perturbed radial drift velocity of dust into a one-dimensional (1D) advection-diffusion equation \citepalias[see also Fig. 1 of][]{kuwahara2022dust}.

In this study, we followed the same procedures described above, but we investigated the time-dependent dust surface density in the step 3 described above. The following sections summarize our numerical approach.  %The steps 1 and 2 are described in detail in \citetalias{kuwahara2022dust}. We summarize the hydrodynamic simulations and the calculation of the radial drift velocity of dust below.

\subsection{Non-dimensionalization}\label{sec:Non-dimensionalization}
\add{As} in \citetalias{kuwahara2022dust}, in our simulations, the length, times, velocities, and densities are normalized by the disk gas scale height, $H$, the reciprocal of the orbital frequency, $\Omega^{-1}$, isothermal sound speed, $c_{\rm s}$, and the unperturbed
gas density at the location of the planet, $\rho_\infty$.

In this dimensionless unit system, we defined the dimensionless thermal mass of the planet: 
\begin{align}
    m\equiv\frac{R_{\rm Bondi}}{H}=\frac{M_{\rm p}}{M_{\rm th}},\label{eq:m}
\end{align}
where $R_{\rm Bondi}=GM_{\rm p}/c_{\rm s}^2$ is the Bondi radius, $G$ is the gravitational constant, $M_{\rm p}$ is the mass of the planet, $M_{\rm th}=M_\ast h^3$ is the thermal mass, $M_\ast$ is the stellar mass, and $h$ is the aspect ratio of the disk. The Hill radius is given by $R_{\rm Hill}=(m/3)^{1/3}\,H$. \add{In \citetalias{kuwahara2022dust}, we considered \addthi{three} planetary masses: $m=0.03,\,0.1$, and 0.3. In this study, we considered \addthi{eight} planetary masses ranging from $m=0.03$ to 0.5 (\Tabref{tab:hydro simulations})}, which corresponds to planets with \addthi{$M_{\rm p}\simeq0.2$--$3.3\,M_\oplus$} orbiting a solar-mass star at 1 au (\addthi{$M_{\rm p}\simeq3.9$--$66\,M_\oplus$} at 50 au; Eq. \ref{eq:Mpl}). \addsec{Throughout the paper, when we convert the dimensionless quantities into dimensional ones, we considered the typical steady accretion disk model with a constant turbulence strength \citep{Shakura:1973}, including viscous heating due to the accretion of the gas and irradiation heating from the central star \citep[Appendix \ref{sec:Conversion to dimensional quantities};][]{Ida:2016}.}

Our planet revolves with the Keplerian speed, $v_{\rm K}$, on a fixed circular orbit. Because the disk gas rotates with the sub-Keplerian speed due to the global pressure gradient, planet experiences the headwind of the gas. We defined the Mach number of the headwind as:\addthi{
\begin{align}
    \mathcal{M}_{\rm hw}\equiv-\frac{h}{2}\Bigg(\frac{\mathrm{d}\ln p}{\mathrm{d}\ln r}\Bigg)%\frac{v_{\rm hw}}{c_{\rm s}},\label{eq:Mhw}
\end{align}
where $p$ is the pressure.}  %$v_{\rm hw}=\eta v_{\rm K}$ is the headwind speed and $\eta=-1/2(c_{\rm s}/v_{\rm K})^2(\mathrm{d}\ln p/\mathrm{d}\ln r)$ is a dimensionless quantity characterizing the global pressure gradient. 
We considered \add{\addthi{three} Mach numbers}: $\mathcal{M}_{\rm hw}=0.01,\,0.03$, and $0.1$ \addthi{(\Figref{fig:params})}. 

The global pressure gradient of the disk gas causes the radial drift of dust. The unperturbed drift velocity is given by \citep{Weidenschilling:1977,Nakagawa:1986}:
\begin{align}
    v_{\rm drift}=-\frac{2{\rm St}}{1+{\rm St}^2}\mathcal{M}_{\rm hw},\label{eq:vdrift}
\end{align}
where  
\begin{align}
    {\rm St}=t_{\rm stop}\Omega,\label{eq:Stokes number}
\end{align}
is the Stokes number of dust and $t_{\rm stop}$ is the stopping time of dust. \addthi{Because \citetalias{kuwahara2022dust} found that the apparent dust ring and gap form when ${\rm St}\lesssim10^{-2}$,} we considered ${\rm St}=10^{-4}$--$10^{-2}$, which corresponds to $\sim0.37$--$37$ mm-sized dust grains at 1 au ($\sim7.6\times10^{-3}$--$0.76$ mm at 50 au; Appendix \ref{sec:Conversion to dimensional quantities}).

\subsection{Hydrodynamical simulations}\label{sec:Hydrodynamical simulations}
Assuming a compressible, inviscid, non-self-gravitating sub-Keplerian gas disk with the vertical stratification due to the stellar gravity, we performed 3D nonisothermal hydrodynamical simulations using Athena++ code\footnote{https://github.com/PrincetonUniversity/athena} \citep{stone2020athena++}. \add{Our methods of hydrodynamical simulations are the same as described in \citetalias{kuwahara2022dust}, but this study handles \addthi{a} broader and more detailed parameter space \addthi{compared to} \citetalias{kuwahara2022dust} in terms of the planetary mass (\Tabref{tab:hydro simulations}).} Our hydrodynamical simulations were performed in the local frame co-rotating with the planet \citepalias[see also Fig. 1 of][]{kuwahara2022dust}. Radiative cooling was implemented by using the so-called $\beta$ cooling model, where the radiative cooling occurs on a finite timescale, $\beta$ \citep{Gammie:2001}. Following \cite{Kurokawa:2018}, we set the dimensionless cooling time as $\beta=(m/0.1)^2$. We simulated the gas flow for at least $10^2$ Keplerian orbits, where the flow field seems to have reached a steady state \citepalias[see Sect. 2.3 of][for details]{kuwahara2022dust}.

%-----------------------------------
\begin{table}[htbp]
\caption{Parameters of hydrodynamical simulations. The following columns give the dimensionless planetary mass, the planetary mass in Earth mass units at 1 au, the planetary mass in Earth mass units at 50 au, and the Mach number of the headwind. The planetary masses with an asterisk were investigated in \citetalias{kuwahara2022dust}. See Appendix \ref{sec:Conversion to dimensional quantities} for the conversion from dimensionless planetary masses to dimensional ones.}
\centering
\begin{tabular}{lccc}\hline\hline
     $m$ & $M_{\rm p}\,[M_\oplus]$ (1 au) & $M_{\rm p}\,[M_\oplus]$ (50 au) & $\mathcal{M}_{\rm hw}$\\ \hline
    $0.03^{*}$ & 0.20 & 3.9 & 0.01, 0.03, 0.1 \\
    0.05       & 0.33 & 6.6 & 0.01, 0.03, 0.1 \\
    0.07       & 0.46 & 9.2 & 0.01, 0.03, 0.1 \\
    $0.1^*$    & 0.66 & 13  & 0.01, 0.03, 0.1 \\
    0.2        & 1.3  & 26  & 0.01, 0.03, 0.1 \\
    $0.3^*$    & 2.0  & 39  & 0.01, 0.03, 0.1 \\
    0.4        & 2.6  & 53  & 0.01, 0.03, 0.1 \\
    0.5        & 3.3  & 66  & 0.01, 0.03, 0.1 \\\hline
\end{tabular}
\label{tab:hydro simulations}
\end{table}

\subsection{Calculations of the radial radial drift velocity of dust perturbed by the gas flow}\label{sec:Calculations of the radial radial drift velocity of dust perturbed by the gas flow}
\add{The radial drift velocity of dust perturbed by the planet-induced gas flow was calculated by the same method as \citetalias{kuwahara2022dust}. (1) We first numerically integrated the equation of motion of dust in a local domain co-rotating with the planet (the local Cartesian coordinates $(x,y,z)$ centered at the planet), in which we used the gas velocity obtained from the hydrodynamical simulation to calculate the gas drag force acting on dust \citep{Kuwahara:2020a}. Hereafter we denote the $x$-, $y$-, and $z$-directions as the radially outward, azimuthal and vertical directions to the disk, respectively. (2) We then sampled the positions and the velocities of dust at fixed small time intervals in the local domain of orbital integration of dust, obtaining \addthi{in this way} the spatial distribution of dust. (3) We assumed the uniform and Gaussian distributions of dust in the azimuthal and vertical directions outside the local domain of orbital integration of dust, in which dust has the unperturbed steady-state drift velocity, $v_{\rm drift}\bm{e}_x$. \addthi{(4)} Finally, we computed the radial drift velocity of dust perturbed by the planet-induced gas flow, $\langle v_{\rm d}\rangle$, by averaging the $x$-component of the dust velocity in the vertical and full azimuthal directions in a disk \citepalias[see Sects. 2.4--2.5 of][for details]{kuwahara2022dust}.}

%---------------------------------------------------------
%--------------------------------------------------------
%-----------------------------------------
\subsection{Dust surface density calculation}\label{sec:Dust surface density calculation}
We computed the dust surface density by incorporating the perturbed radial drift velocity of dust into a 1D {advection-diffusion} equation:
\begin{align}
    \frac{\partial \Sigma_{\rm d}}{\partial t}+\frac{\partial }{\partial x}\bigg(\langle v_{\rm d}\rangle \Sigma_{\rm d}-\mathcal{D}\frac{\partial \Sigma_{\rm d}}{\partial x}\bigg)=0,\label{eq:1D advection diffusion equation}
\end{align}
where $\Sigma_{\rm d}$ is the dust surface density, $\mathcal{D}=\alpha_{\rm diff}/(1+{\rm St}^2)$ is the diffusion coefficient for the dust \citep{Youdin:2007}, and $\alpha_{\rm diff}$ is \addthi{a} dimensionless turbulent parameter describing turbulent diffusion of dust. \add{Because \citetalias{kuwahara2022dust} found that the dust rings and gaps do not appear when $\alpha_{\rm diff}\gtrsim10^{-3}$, in this study we only} assumed $\alpha_{\rm diff}=10^{-4}$ (hereafter referred to as the moderate-turbulence \addthi{case}) and $10^{-5}$ (hereafter referred to as the low-turbulence \addthi{case}). \addthi{In \Equref{eq:1D advection diffusion equation}, we neglect the effect of the disk curvature by focusing on a radial range sufficiently narrow compared to the orbital radius of the planet.} \rev{We did not consider the backreaction of dust on gas.}

While \citetalias{kuwahara2022dust} assumed a steady state in \Equref{eq:1D advection diffusion equation}, we computed the time evolution of the dust surface density in this study. %The following sections (Sect. XX—XX) describe the detailed method of the dust surface density calculation.
We assumed that the gas surface density is constant for simplicity, so that \Equref{eq:1D advection diffusion equation} does not contain the gas surface density. %This assumption is justified because we focus on a radial range sufficiently narrow compared to the orbital radius of the planet. %This assumption is justified by the following reasons. First, we focus on a radial range sufficiently narrow compared to the orbital radius of the planet. Second, we consider the planet mass below the so-called pebble isolation mass \citep[][]{Bitsch:2018}, at which a planet opens a shallow gas gap and creates a pressure bump at the gas gap edge. In our dimensionless unit, the pebble isolation mass can be described by $m\simeq0.6\equiv m_{\rm iso}$. %We discuss this assumption in Sect. \ref{sec:Discussion}.  
%In \citetalias{kuwahara2022dust}, we assumed a steady state. In this study, we calculate the time evolution of $\Sigma_{\rm d}$. 
We integrated \Equref{eq:1D advection diffusion equation} using \addthi{a} finite-volume method. The size of the calculation domain of dust surface density simulation was $x\in[x_{\rm in},x_{\rm out}]$. We set $x_{\rm in}=-100$ and $x_{\rm out}=100$. %unless otherwise noted. For the purpose of visualization, we extend the domain size to $x\in[-20,20]$ in some cases. We confirmed the changes of the domain size do not affect the results.
We used a fixed spatial interval $\Delta x=0.01$. A zero-diffusive flux condition was adopted at $x=x_{\rm in}$. At $x=x_{\rm out}$, we set the constant advective flux, $v_{\rm drift}\Sigma_{\rm d,0}$, where $\Sigma_{\rm d,0}=1$. The time step was calculated by:
\begin{align}
    \displaystyle\Delta t=\text{CFL}\times\min\Bigg(\frac{1}{\underset{i}{\max}\big(|\langle v_{\rm d}\rangle_i|/\Delta x\big)},\frac{(\Delta x)^2}{\mathcal{D}}\Bigg),
\end{align}
where \addthi{Courant–Friedrichs–Lewy (CFL) number was set to} $\text{CFL}=0.5$ and $\langle v_{\rm d}\rangle_i$ is the dust velocity at the $i$-th grid.

%------------------------------
%------------------------------
\subsection{\add{Analytic formulae for analyzing the numerical results}}
\add{In the following sections (Sects. \ref{sec:Drift and diffusion timescales of dust}--\ref{sec:Definition of the depth of the dust gap}), we introduce the analytic formulae which will be used to analyze the results of numerical simulations.}

%------------------------------
%------------------------------
\subsubsection{Drift and diffusion timescales of dust}\label{sec:Drift and diffusion timescales of dust}
We \addthi{identified two key timescales: the drift timescale of dust:}
\begin{align}
    &t_{\rm drift}=\frac{\mathcal{L}}{|v_{\rm drift}|}\simeq\frac{\mathcal{L}}{2{\rm St}\mathcal{M}_{\rm hw}}\simeq1.67\times10^4\,\Bigg(\frac{\rm St}{10^{-3}}\Bigg)^{-1}\Bigg(\frac{\mathcal{M}_{\rm hw}}{0.03}\Bigg)^{-1}\Bigg(\frac{\mathcal{L}}{H}\Bigg),\label{eq:tdrift}
\end{align}
\addthi{and the diffusion timescale:}
\begin{align}
    &t_{\rm diff}=\frac{\mathcal{L}^2}{\mathcal{D}}\simeq\frac{\mathcal{L}^2}{\alpha_{\rm diff}}=10^4\,\Bigg(\frac{\alpha_{\rm diff}}{10^{-4}}\Bigg)^{-1}\Bigg(\frac{\mathcal{L}}{H}\Bigg)^2.\label{eq:tdiff}
\end{align}
\addthi{Here,} $\mathcal{L}$ is the characteristic length and we assumed $1+{\rm St}^2\simeq1$ (${\rm St}\ll1$). The drift timescale coincides with the diffusion timescale when:
\begin{align}
    \mathcal{L}=\frac{\alpha_{\rm diff}}{2{\rm St}\mathcal{M}_{\rm hw}}\equiv\mathcal{L}_{\rm eq}\simeq1.67\,\Bigg(\frac{\alpha_{\rm diff}}{10^{-4}}\Bigg)\Bigg(\frac{\rm St}{10^{-3}}\Bigg)^{-1}\Bigg(\frac{\mathcal{M}_{\rm hw}}{0.03}\Bigg)^{-1}.\label{eq:L eq}
\end{align}

%------------------------------
%------------------------------
\subsubsection{Definition of the width of the outflow region}

\begin{figure}
    \centering
    \includegraphics[width=1\linewidth]{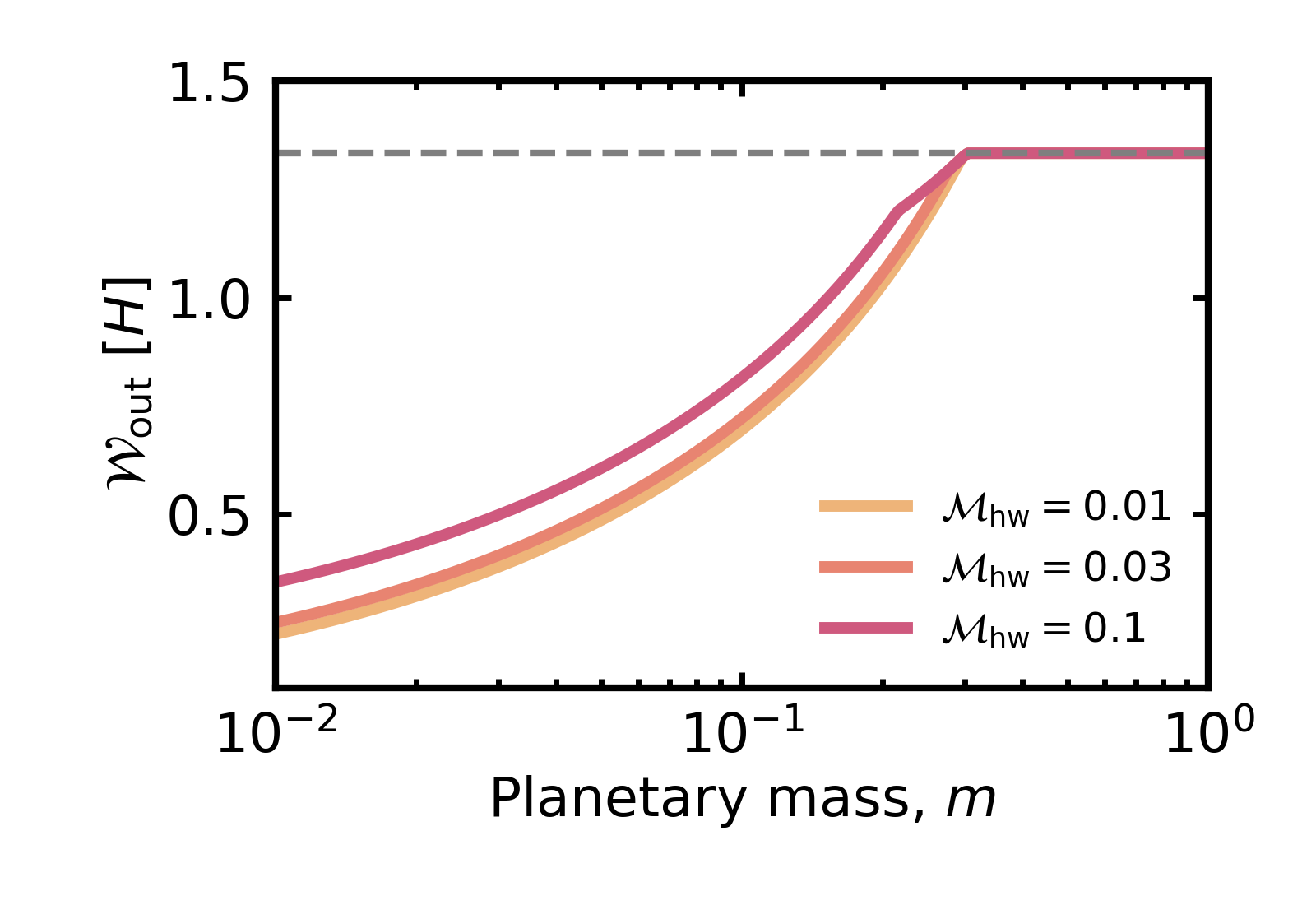}
    \caption{Width of the outflow region as a function of the planetary mass (Eq. \ref{eq:width of outflow region}).}
    \label{fig:wout}
\end{figure}

\add{The dust velocity is significantly perturbed at the edges of the outflow region. Following \cite{kuwahara2024analytic}, we refer to the outflow region as the region where the $x$-component of the gas velocity is dominantly perturbed.} The \addthi{dimensionless} $x$-coordinate of the edge of the outflow region is given by \citep{kuwahara2024analytic}:
\begin{align}
    w_{\rm out}^\pm=\pm\min\left(\frac{2}{3}\bigg(1\mp\mathcal{M}_{\rm hw}\bigg),w_{\rm HS}+\frac{2}{3}\mathcal{M}_{\rm hw}\right)\label{eq:w_out}
\end{align}
where 
\begin{align}
    w_{\rm HS}=\frac{1.05\sqrt{m}+3.4m^{7/3}}{1+2m^2},\label{eq:wHS}
\end{align}
is the half-width of the horseshoe region \citep{jimenez2017improved}. From \Equref{eq:w_out}, the width of the outflow region is given by:
\begin{align}
    \mathcal{W}_{\rm out}&\equiv w_{\rm out}^+-w_{\rm out}^-=\min\bigg(2w_{\rm HS}+\frac{4}{3}\mathcal{M}_{\rm hw},\,\frac{4}{3}\bigg).\label{eq:width of outflow region}
\end{align}
We plotted \Equref{eq:width of outflow region} in \Figref{fig:wout}. The width of the outflow region increases with the planetary mass when $m\lesssim0.3$ and \add{converges at} $m\gtrsim0.3$.

%------------------------------
%------------------------------

\subsubsection{Definition of the width of the dust ring and gap}\label{sec:Definition of the width of the dust ring and gap}

\begin{figure}
    \centering
    \includegraphics[width=1\linewidth]{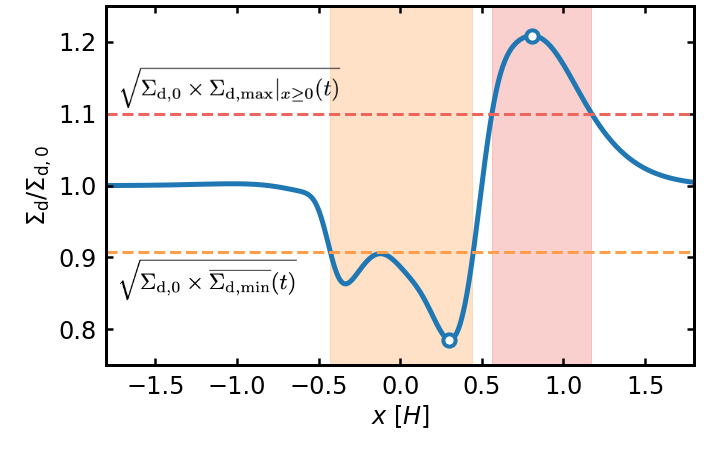}
    \caption{\add{Definition of the widths of the dust ring and gap. The red and yellow shaded regions show the numerically-calculated widths of the dust ring and gap. The circles mark $\Sigma_{\rm d,max}$ and $\Sigma_{\rm d,min}$.}}
    \label{fig:geometric_mean}
\end{figure}

Following \citetalias{kuwahara2022dust}, hereafter we refer to the regions where dust is depleted and accumulated as “dust gap” and “dust ring”, respectively. We numerically calculated the dust gap width by:
\begin{align}
    \mathcal{W}_{\rm gap}^{\rm num}(t)=x_{\rm gap,out}^{\rm num}(t)-x_{\rm gap,in}^{\rm num}(t),
\end{align}
\addsec{where the superscript \addthi{"num"} represents the value obtained from numerical simulations.} In the above equation, $x_{\rm gap,in}^{\rm num}(t)$ and $x_{\rm gap,out}^{\rm num}(t)$ are the edges of the dust gap where $\Sigma_{\rm d}(x,t)$ reaches the following value \citep{dong2017mass}\footnote{We used a slightly different definition of the dust gap width with respect to \cite{dong2017mass}, in which the authors calculated the geometric mean of the minimum and initial dust surface densities,  $\sqrt{\Sigma_{\rm d,0}\times\Sigma_{\rm d,min}}$.}:
\begin{align}
    &\Sigma_{\rm d}(x_{\rm gap,in}^{\rm num}(t))=\Sigma_{\rm d}(x_{\rm gap,out}^{\rm num}(t))=\sqrt{\Sigma_{\rm d,0}\times\overline{\Sigma_{\rm d,min}}(t)}.\label{eq:sigma mean}
\end{align}
We defined $\overline{\Sigma_{\rm d,min}}(t)$ as the average of the minimum dust surface density inside and outside the planetary orbit:
\begin{align}
    \overline{\Sigma_{\rm d,min}}(t)\equiv\frac{\Sigma_{\rm d,min}(t)|_{x<0}+\Sigma_{\rm d,min}(t)|_{x\geq0}}{2}.
\end{align}
%In our simulations, we found that an asymmetric depletion of dust inside and outside the planetary orbit in some cases (Sect. \ref{sec:Numerical results}). Thus in \Equref{eq:sigma mean}, we used the average of the minimum value of $\Sigma_{\rm d}$ at $x<0$ and $x\geq0$.
We numerically calculated the dust ring width by:
\begin{align}
    \mathcal{W}_{\rm ring}^{\rm num}(t)=x_{\rm ring,out}^{\rm num}(t)-x_{\rm ring,in}^{\rm num}(t).\label{eq:wring}
\end{align}
In \Equref{eq:wring}, $x_{\rm ring,in}^{\rm num}(t)$ and $x_{\rm ring,out}^{\rm num}(t)$ are the edges of the dust ring where $\Sigma_{\rm d}(x,t)$ reaches the geometric mean of the maximum and initial values:
\begin{align}
    \Sigma_{\rm d}(x_{\rm ring,in}^{\rm num}(t))=\Sigma_{\rm d}(x_{\rm ring,out}^{\rm num}(t))=\sqrt{\Sigma_{\rm d,0}\times\Sigma_{\rm d,max}|_{x\geq0}(t)}.
\end{align}
For the definition of the dust ring, we only focus on the dust accumulation outside the planetary orbit. \add{The definitions of the widths of the dust ring and gap were plotted in \Figref{fig:geometric_mean}.}

\subsubsection{Definition of the depth of the dust gap}\label{sec:Definition of the depth of the dust gap}
We defined the dust gap depth as the contrast between the minimum and maximum dust surface densities \citep[\Figref{fig:geometric_mean};][]{huang2018-DSHARP2,zhang2018-DSHARP7}. We numerically calculated the dust gap depth by:
\begin{align}
    \delta_{\rm gap}^{\rm num}(t)\equiv\frac{\Sigma_{\rm d,min}(t)}{\Sigma_{\rm d,max}(t)}.\label{eq:delta gap}
\end{align}
It should be noted that \Equref{eq:delta gap} may represent the amplitude of the dust ring if turbulent diffusion smooths the dust gap and then only the dust ring forms. Although it can differ from the intuitive definition of the dust gap depth, we consistently use \Equref{eq:delta gap} for measuring the dust gap depth.%, because it has been widely used in observational studies \citep[e.g.,][]{huang2018-DSHARP2}.
%only the dust gap formation occurs in some cases (Sect. \ref{sec:Numerical results}). In these cases, \Equref{eq:delta gap} represents the amplitude of the dust ring. However, we consistently use \Equref{eq:delta gap} as the definition of the dust gap in this study.

%==========================================

\begin{figure}[htbp]
    \centering
    \includegraphics[width=\linewidth]{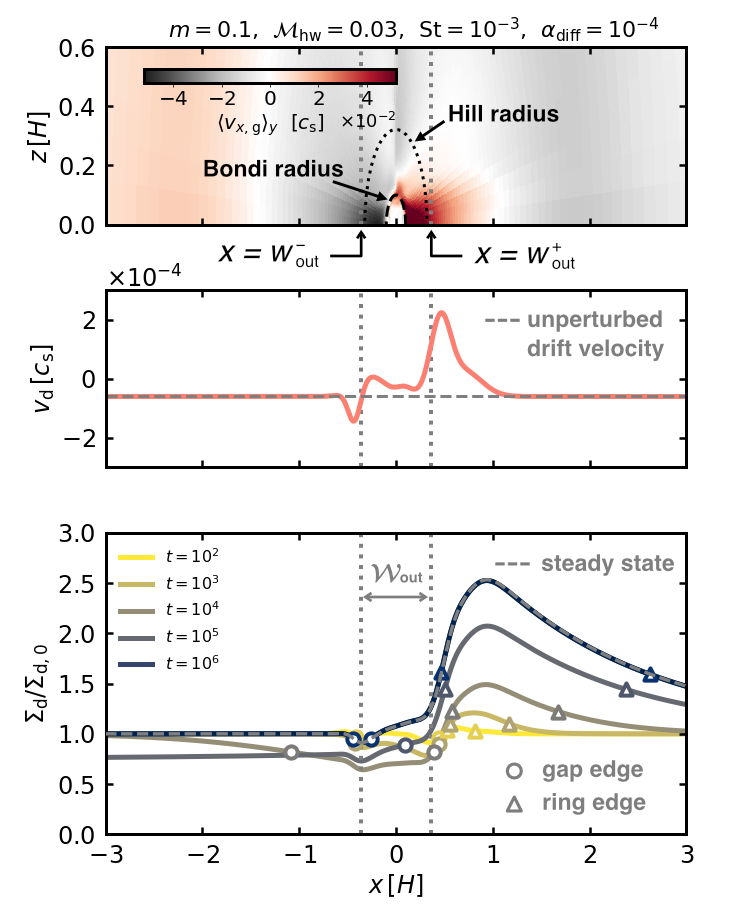}
    \caption{Perturbation of the planet-induced gas flow on the radial velocity of dust and the dust surface density. \add{We set $m=0.1,\,\mathcal{M}_{\rm hw}=0.03,\,{\rm St}=10^{-3}$, and $\alpha_{\rm diff}=10^{-4}$.} \textit{Top:} Gas flow structure at the meridian plane. The color contour represents the gas velocity in the $x$-direction averaged in the $y$-direction within the calculation domain of hydrodynamical simulation, $\langle v_{x,{\rm g}}\rangle_y$. The vertical dotted lines represent the $x$-coordinates of the edges of the outflow region, $w_{\rm out}^\pm$. \textit{Middle:} Perturbed radial drift velocity of dust. The horizontal dashed line represents $v_{\rm drift}$. \textit{Bottom:} Time evolution of the dust surface density. The gray dashed line corresponds to the steady-state dust surface density. The circle and triangle symbols denote the location of the numerically-calculated edges of the dust gap and ring.}
    \label{fig:gas_vd_sigmad}
\end{figure}

\begin{figure}[htbp]
    \centering
    \includegraphics[width=\linewidth]{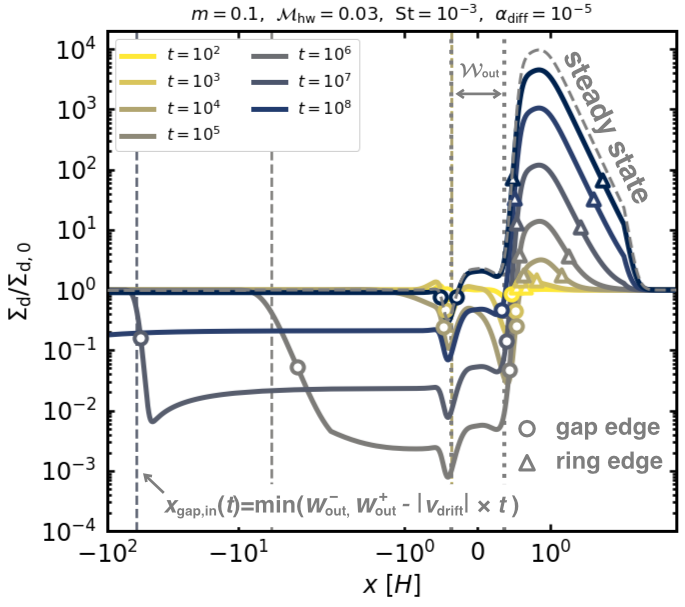}
    \caption{Time evolution of the dust surface density in low-turbulence disks. \add{We set $m=0.1,\,\mathcal{M}_{\rm hw}=0.03,\,{\rm St}=10^{-3}$, and $\alpha_{\rm diff}=10^{-5}$. The horizontal axis is on a log scale, whose range is extended to $x\in[-100,5]$.} The vertical dashed lines show the analytic model of the location of the inner edge of the dust gap, which moves with $v_{\rm drift}$ (Sect. \ref{sec:Semi-analytic models of dust rings and gaps}).}
    \label{fig:alpha1e-5}
\end{figure}

\begin{figure*}
    \centering
    \includegraphics[width=1\linewidth]{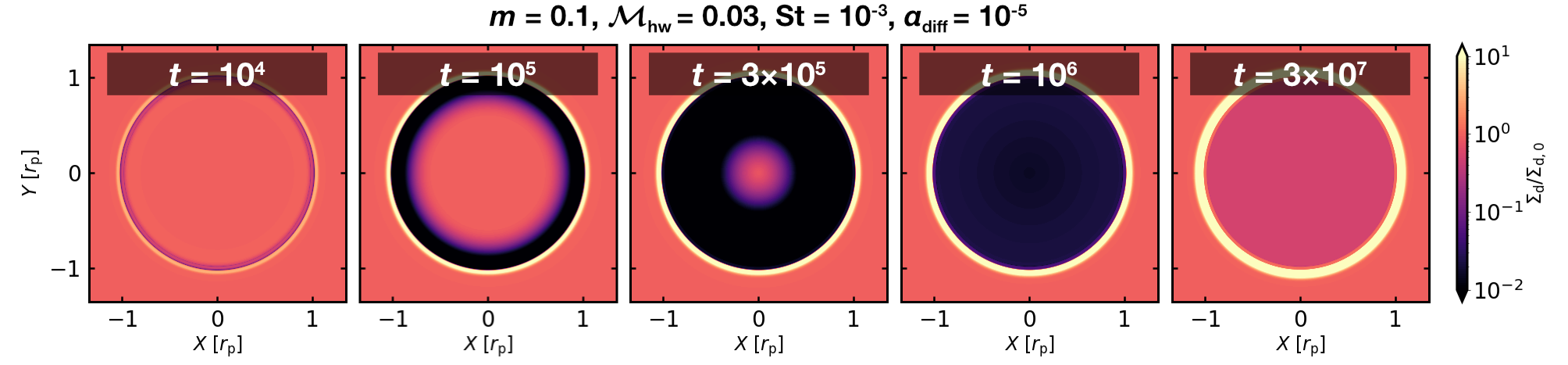}
    \caption{\add{Time evolution of $\Sigma_{\rm d}(t)$ in the two-dimensional plane. We set $m=0.1,\,\mathcal{M}_{\rm hw}=0.03,\,{\rm St}=10^{-3},$ and $\alpha_{\rm diff}=10^{-5}$. These images were generated based on the results of 1D calculations assuming an axisymmetric dust distribution, \addthi{neglecting disk curvature}. The axes are normalized by the planet location, $r_{\rm p}$, calculated by $X=Y=(r-r_{\rm p})/h_{\rm p}$, where $r$ is the radial coordinate centered at the star and $h_{\rm p}$ is the disk aspect ratio at $r=r_{\rm p}$. We set $h_{\rm p}=0.05$.}}
    \label{fig:disk_time}
\end{figure*}

\begin{figure}
    \centering
    \includegraphics[width=1\linewidth]{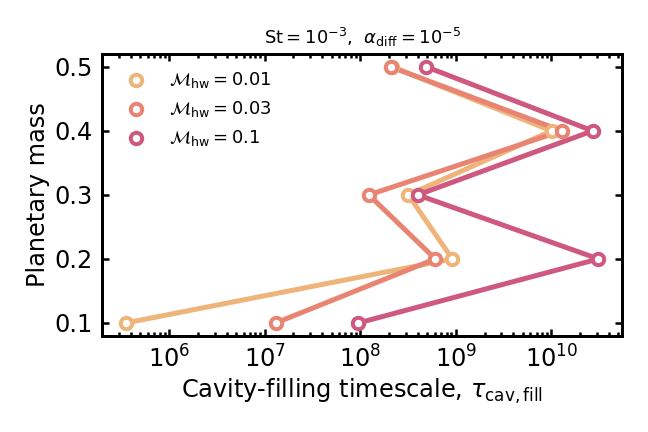}
    \caption{\rev{Cavity-filling timescale. We only focus $m\geq0.1$ in which the dust cavities form.}}
    \label{fig:tau_ring}
\end{figure}

\begin{figure*}[htbp]
    \centering
    \includegraphics[width=\linewidth]{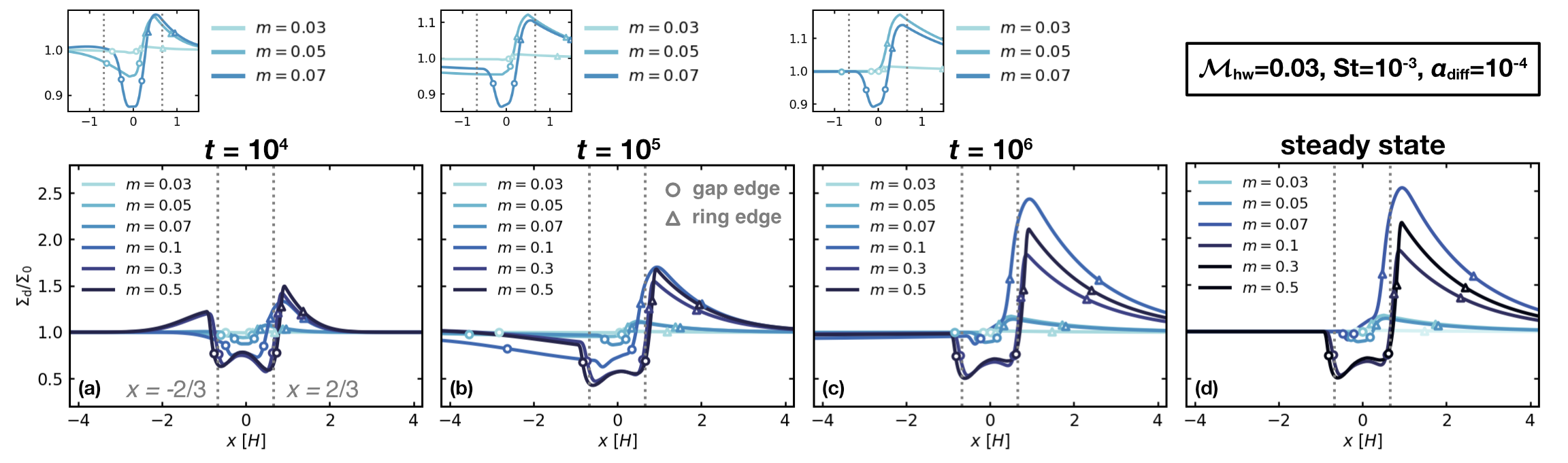}
    \caption{Dependence of $\Sigma_{\rm d}(t)$ on the planetary mass. \add{We set} $\mathcal{M}_{\rm hw}=0.03,\,{\rm St}=10^{-3}$, and $\alpha_{\rm diff}=10^{-5}$. The vertical dotted lines correspond to $|x|=4/3$ (the $x$-coordinate of the edge of the outflow region for $m\gtrsim0.3$; \Equref{eq:width of outflow region}). The figures on the upper left corners of the panels a--c are the zoom-in views for $m=0.03,\,0.05,$ and $0.07$.}
    \label{fig:planet mass dependence}
\end{figure*}

\begin{figure}[htbp]
    \centering
    \includegraphics[width=\linewidth]{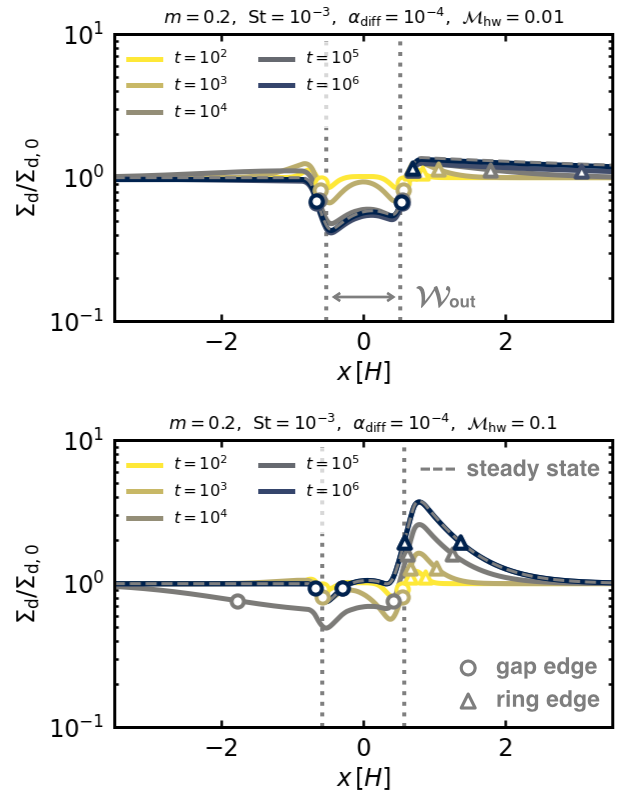}    
    \caption{Dependence of $\Sigma_{\rm d}(t)$ on the Mach number of the headwind. \add{We set \addthi{$m=0.2,\,{\rm St}=10^{-3}$}, and $\alpha_{\rm diff}=10^{-4}$. We varied the the Mach number of the headwind in each panel, $\mathcal{M}_{\rm hw}=0.01$ (\textit{top}) and $\mathcal{M}_{\rm hw}=0.1$ (\textit{bottom}).}}
    \label{fig:Mhw dependence}
\end{figure}

\begin{figure*}[htbp]
    \centering
    \includegraphics[width=\linewidth]{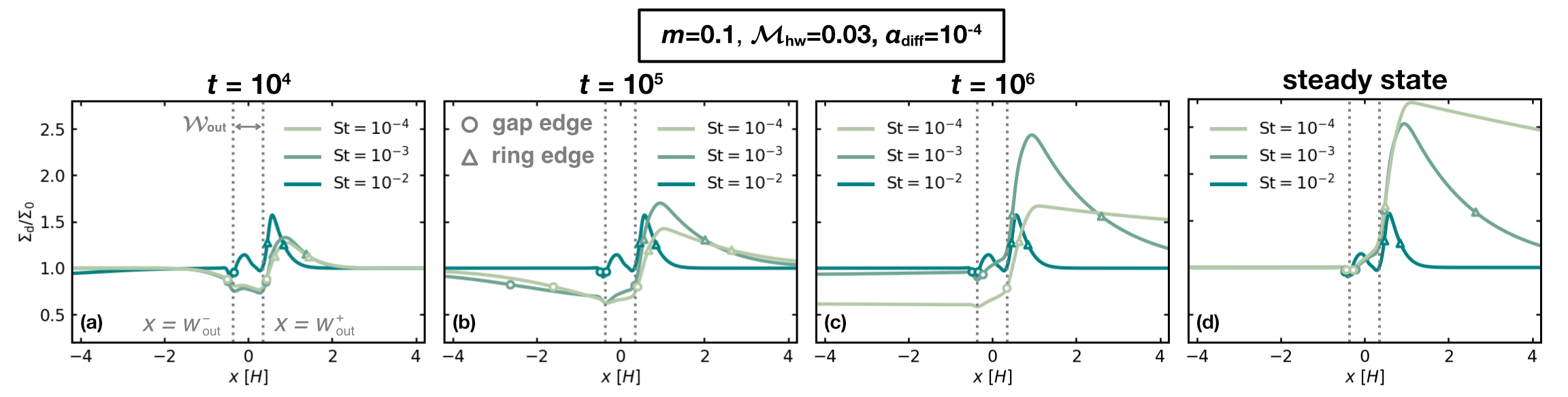}    
    \caption{Dependence of $\Sigma_{\rm d}(t)$ on the Stokes number. \add{We set} $m=0.1,\,\mathcal{M}_{\rm hw}=0.03$, and $\alpha_{\rm diff}=10^{-5}$. The vertical dotted lines correspond to $x=w_{\rm out}^\pm$.}
    \label{fig:St dependence}
\end{figure*}

\begin{figure}
    \centering
    \includegraphics[width=1\linewidth]{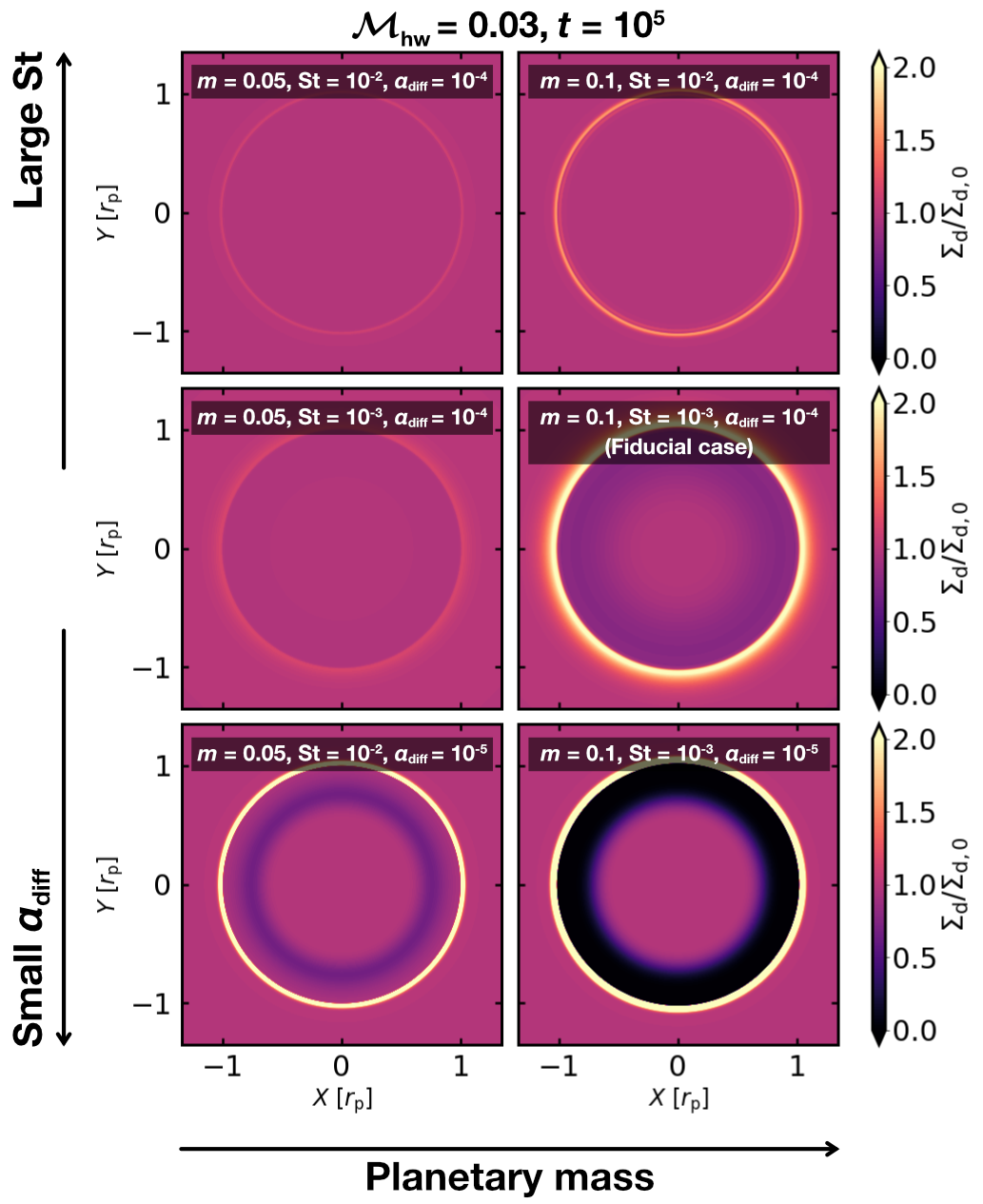}
    \caption{\add{Summary of the parameter dependence of $\Sigma_{\rm d}(t)$. We set $\mathcal{M}_{\rm hw}=0.03$ and $t=10^{5}$. We varied the planetary mass, the Stokes number, and the turbulent parameter.}}
    \label{fig:param_summary}
\end{figure}

\begin{figure}
    \centering
    \includegraphics[width=1\linewidth]{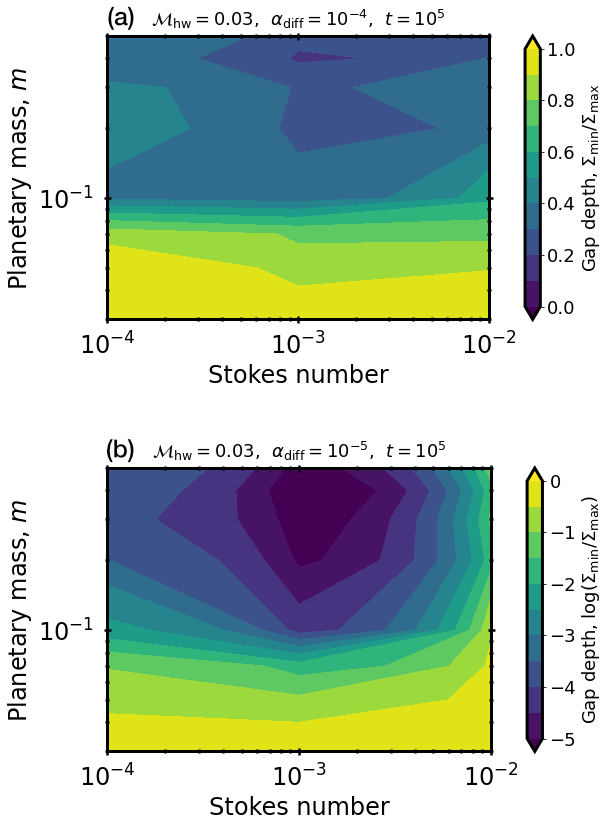}
    \caption{\add{Contour plot of the dust gap depth as a function of the planetary mass and the Stokes number. We set $\mathcal{M}_{\rm hw}=0.03$ and $t=10^5$. We varied the turbulent parameter in each panel, $\alpha_{\rm diff}=10^{-4}$ (\textit{panel a}) and $\alpha_{\rm diff}=10^{-5}$ (\textit{panel b}).}}
    \label{fig:heatmap}
\end{figure}

\begin{figure}
    \centering
    \includegraphics[width=1\linewidth]{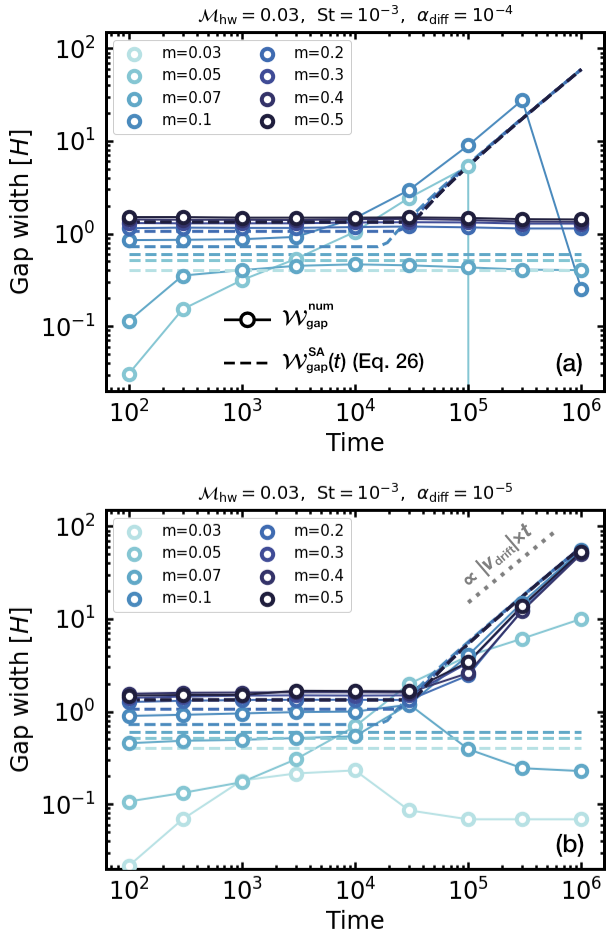}
    \caption{Time evolution of the dust gap width for different planetary masses. We fixed the Stokes number ${\rm St}=10^{-3}$ and the Mach number $\mathcal{M}_{\rm hw}=0.03$, and set $\alpha_{\rm diff}=10^{-4}$ in \textit{panel a} and $\alpha_{\rm diff}=10^{-5}$ in \textit{panel b}. \add{The solid lines with the circle symbols and the dashed lines are the numerically-calculated and the semi-analytic dust gap widths, respectively} (Eq. \ref{eq:W gap SA}; Sect. \ref{sec:Semi-analytic models of dust rings and gaps}). We note that in \textit{panels a and b}, the numerically-calculated dust gap width for $m=0.03$ is not shown because we obtained $\mathcal{W}_{\rm gap}^{\rm num}=0$.}
    \label{fig:gapwidth_time}
\end{figure}

% 1colのとき0.9
\begin{figure}
    \centering
    \includegraphics[width=\linewidth]{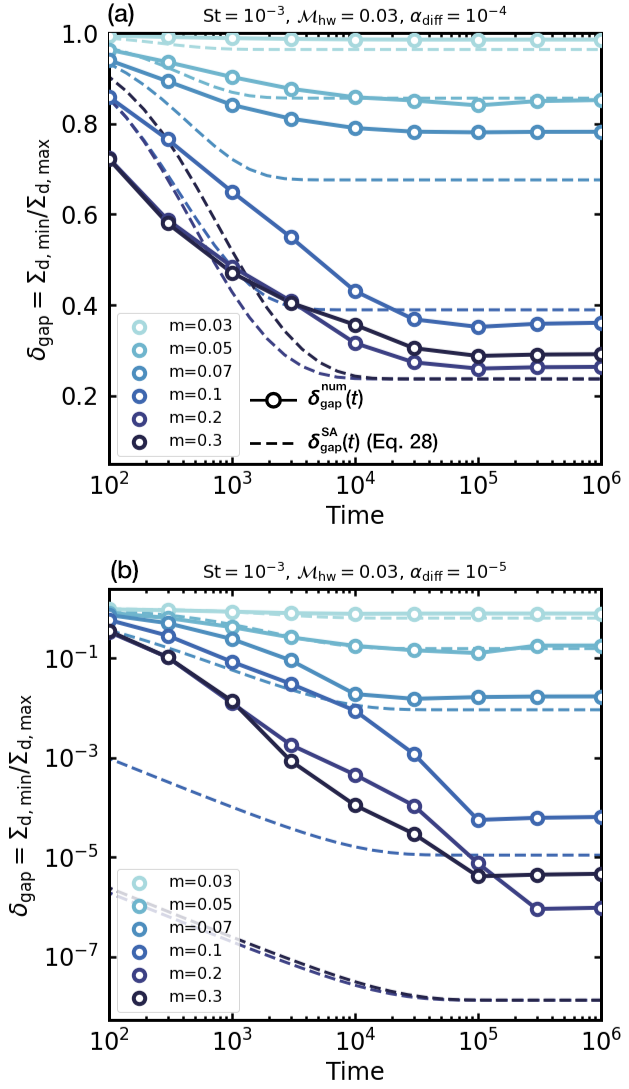}
    \caption{Time evolution of the dust gap depth for different planetary masses. We fixed the Stokes number ${\rm St}=10^{-3}$ and the Mach number $\mathcal{M}_{\rm hw}=0.03$, and set $\alpha_{\rm diff}=10^{-4}$ in \textit{panel a} and $\alpha_{\rm diff}=10^{-5}$ in \textit{panel b}. \addsec{The solid lines with the circle symbols and the dashed lines are the numerically-calculated and the semi-analytic dust gap depths, respectively} (Eq. \ref{eq:delta SA}; Sect. \ref{sec:Semi-analytic models of dust rings and gaps}).}
    \label{fig:gapdepth_time}
\end{figure}

\begin{figure}
    \centering
    \includegraphics[width=1\linewidth]{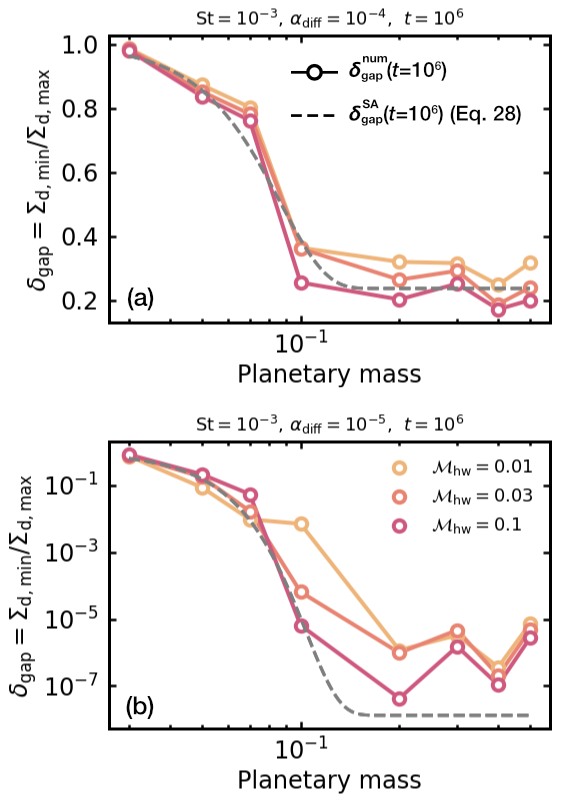}
    \caption{Dust gap depth as a function of the planetary mass at $t=10^6$. We fixed the Stokes number ${\rm St}=10^{-3}$ and set $\alpha_{\rm diff}=10^{-4}$ in \textit{panel a} and $\alpha_{\rm diff}=10^{-5}$ in \textit{panel b}. \addsec{The solid lines with the circle symbols and the dashed lines are the numerically-calculated and the semi-analytic dust gap depths, respectively} (Eq. \ref{eq:delta SA}; Sect. \ref{sec:Semi-analytic models of dust rings and gaps}).}
    \label{fig:delta_m}
\end{figure}

% 2colのときは0.87
% one colのときは0.7
\begin{figure}
    \centering
    \includegraphics[width=0.87\linewidth]{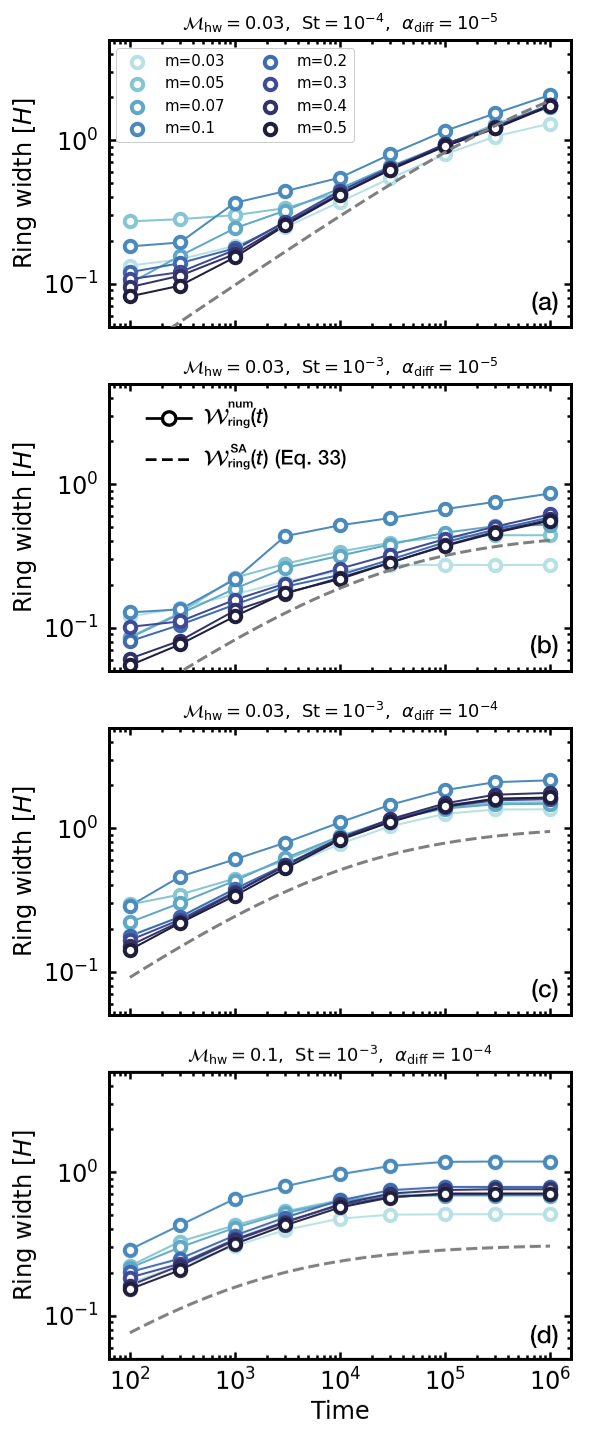}
    \caption{\add{Time evolution of the dust ring width for different planetary masses.} The assumed parameters ($\mathcal{M}_{\rm hw},\,{\rm St}$, and $\alpha_{\rm diff}$) are shown at the top of each panel. \add{The solid lines with the circle symbols and the dashed lines are the numerically-calculated and the semi-analytic dust ring widths, respectively} (Eq. \ref{eq: W ring SA}; Sect. \ref{sec:Semi-analytic models of dust rings and gaps}).}
    \label{fig:ringwidth}
\end{figure}

\begin{figure}[htbp]
    \centering
    \includegraphics[width=\linewidth]{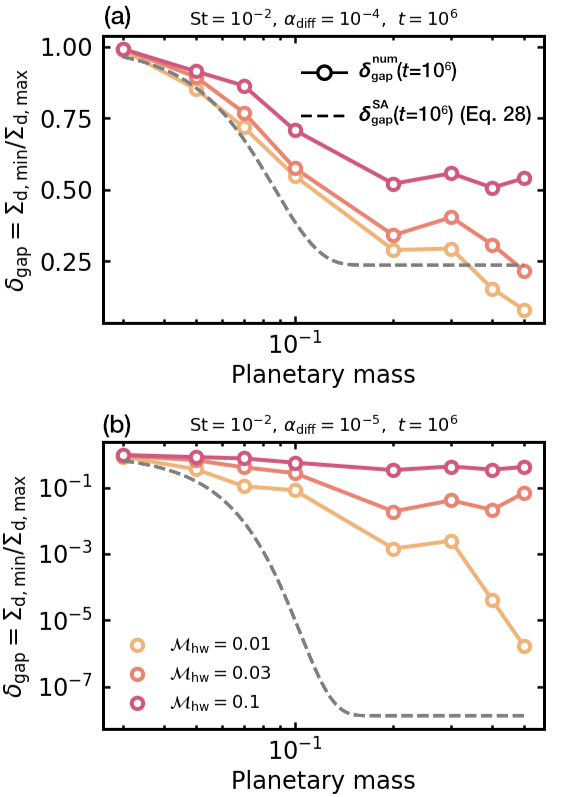}    
    \caption{Same as \Figref{fig:delta_m}, but we set ${\rm St}=10^{-2}$.}
    \label{fig:delta_m_St1e-2}
\end{figure}

%==========================================
\section{Numerical results}\label{sec:Numerical results}
In this section, we present the numerical results of this study, \addthi{comparing them with the steady-state solutions \addthi{\citepalias{kuwahara2022dust}}.} Section \ref{sec:Physical processes of the time evolution of Sigma_d} describe the physical processes of the time evolution of the dust surface density. Section \ref{sec:Properties of dust rings and gaps} shows the properties of the dust rings and gaps, such as the widths and the depths. Based on the results presented in this section, we introduce the semi-analytic models of the widths of the dust ring and gap and the depth of the dust gap in Sect. \ref{sec:Semi-analytic models of dust rings and gaps}.

%----------------------------------
%----------------------------------
%----------------------------------
\subsection{\addthi{Time evolution of $\Sigma_{\rm d}(t)$}}\label{sec:Physical processes of the time evolution of Sigma_d}

We describe the behavior of the time evolution of $\Sigma_{\rm d}(t)$ influenced by the planet-induced gas flow. %In the following sections, we used the dimensionless quantities introduced in Sect. \ref{sec:Non-dimensionalization}. 
\add{Given the wide parameter spaces in our study, we first show the numerical results for a fiducial parameter set with $\alpha_{\rm diff}=10^{-4},\,m=0.1,\,\mathcal{M}_{\rm hw}=0.03$, and ${\rm St}=10^{-3}$ (Sect. \ref{sec:Fiducial case}). We then show the dependence of $\Sigma_{\rm d}(t)$ on the turbulent parameter (Sect. \ref{sec:Dependence on the turbulence}), the planetary mass (Sect. \ref{sec:Dependence on the planetary mass}), the Mach number of the headwind (Sect. \ref{sec:Dependence on the headwind}), and the Stokes number (Sect. \ref{sec:Dependence on the Stokes number}), respectively. } %We show the results in moderate-turbulence disks ($\alpha_{\rm diff}=10^{-4}$; Sect. \ref{sec:Results in a moderate-turbulence disk}), and then show the results in low-turbulence disks ($\alpha_{\rm diff}=10^{-5}$; Sect. \ref{sec:Results in a low-turbulence disk}). 

\subsubsection{\add{Fiducial case}}\label{sec:Fiducial case}
\add{The outflow of the gas at the midplane induced by low-mass planets perturbs the radial drift velocity of dust, causing the dust rings and gaps (\Figref{fig:gas_vd_sigmad}).} Figure \ref{fig:gas_vd_sigmad} shows the perturbations of low-mass planets on the gas and dust, where we \add{set $\alpha_{\rm diff}=10^{-4},\,m=0.1,\,\mathcal{M}_{\rm hw}=0.03$, and ${\rm St}=10^{-3}$, as the fiducial parameter set}. In \Figref{fig:gas_vd_sigmad}, the top, the middle, and the bottom panels show the velocity field of the gas at the meridian plane, the radial drift velocity of dust influenced by the gas flow, and the dust surface density, respectively. 

The dust surface density, which has initially a flat profile, changes with time, and then reaches a steady state within \add{$t\lesssim10^6$} (\Figref{fig:gas_vd_sigmad}). The outflow of the gas at the midplane perturbs the radial drift velocity of dust, causing positive and negative peaks in the profile of $\langle v_{\rm d}\rangle$ (the middle panel of \Figref{fig:gas_vd_sigmad}). The radially-outward (inward) outflow of the gas inhibits (enhances) the radial drift of dust. The dust surface density decreases with time around the planetary orbit \addthi{creating a dust gap}, while it increases with time outside the planetary orbit \addthi{creating a dust ring}. %More dust accumulates outside the planetary orbit when the larger $\mathcal{M}_{\rm hw}$ is assumed, because the flow speed of the radially-outward outflow increase with $\mathcal{M}_{\rm hw}$ \citep{kuwahara2024analytic}. 
\add{The dust surface density only changes by a factor of $\sim2$ in \Figref{fig:gas_vd_sigmad} due to efficient dust diffusion. Given a characteristic length of a perturbation is set by $\mathcal{L}=\mathcal{W}_{\rm out}$, the drift and diffusion timescales are given by $t_{\rm drift}\sim1.2\times10^4$ and $t_{\rm diff}\sim5.2\times10^3$, \addthi{resulting in a diffusion-dominated regime} $t_{\rm diff}<t_{\rm drift}$.}

The \add{locations} of the edges of the dust gap  %, in particular the outer edge of the dust gap, 
\add{are determined by those of} the edges of the outflow region, $x=w_{\rm out}^\pm$ (Eq. \ref{eq:w_out}; the vertical dotted lines in \Figref{fig:gas_vd_sigmad}). \addsec{Thus, the dust gap widths can be estimated by $\mathcal{W}_{\rm out}$ (Eq. \ref{eq:width of outflow region}).} %The outer edge of the dust gap hardly changes with time. 
The location of the inner edge of the dust ring \add{is} set by $x=w_{\rm out}^+$ and hardly changes with time. The outer edge of the dust ring moves with time due to diffusion (\Figref{fig:gas_vd_sigmad}).

\subsubsection{\add{Dependence of $\Sigma_{\rm d}(t)$ on the turbulent parameter}}\label{sec:Dependence on the turbulence}
\add{A perturbation to $\Sigma_{\rm d}(t)$ due to the planet-induced gas flow strongly depends on the turbulent parameter. Figure \ref{fig:alpha1e-5} shows the time-dependent $\Sigma_{\rm d}(t)$ in the low-turbulence disk ($\alpha_{\rm diff}=10^{-5}$). The planetary mass, the Mach number of the headwind, and the Stokes number are the same as in the fiducial case. Compared to the fiducial case, $\Sigma_{\rm d}(t)$ changes more significantly, because a steep gradient of $\Sigma_{\rm d}$ needs to achieve a steady state due to inefficient dust diffusion ($t_{\rm diff}>t_{\rm drift}$ in \Figref{fig:alpha1e-5}). In  \Figref{fig:alpha1e-5}, $\Sigma_{\rm d}(t)$ increases (decreases) by $\sim3\text{--}4$ orders of magnitude. The dust surface density reaches the steady state after $t>10^8$.}
% We found that $\Sigma_{\rm d}(t)$ changes by orders of magnitude. The profile of $\Sigma_{\rm d}(t)$ deviates significantly from the steady-state solution during the time evolution. 
%While the dust surface density reaches a steady state in moderate-turbulence disks within $t\sim10^{4}\text{--}10^5$ (corresponding to $\sim0.01\text{--}0.1$ Myr at 1 au), 
%We found that it took a long time % (longer than the typical disk lifetime)
%for $\Sigma_{\rm d}(t)$ to reach a steady state
%Figure \ref{fig:alpha1e-5} shows the time evolution of $\Sigma_{\rm d}(t)$ for $\alpha_{\rm diff}=10^{-5}$. We fixed $\mathcal{M}_{\rm hw}=0.03$  and ${\rm St}=10^{-3}$ and assumed $m=0.07$  (\Figref{fig:alpha1e-5}a) and $m=0.1$ (\Figref{fig:alpha1e-5}b). 
% (corresponding to $>$1 Myr at 1 au). 

During the time evolution $\Sigma_{\rm d}(t)$ changes in a complex manner. The profile of $\Sigma_{\rm d}(t)$ deviates significantly from the steady-state solution. The dust accumulates over time at $x\gtrsim w_{\rm out}^+$, which is similar to the \add{fiducial case}. At $x\lesssim w_{\rm out}^+$, the dust surface density drops significantly in the early stage of the time evolution, $t\lesssim10^4\text{--}10^5$. Due to inefficient dust diffusion, the dust \add{hardly} leak out to the inside of the planetary orbit. As a result, at the early stage of the time evolution, the minimum value of $\Sigma_{\rm d}(t)$ can be orders of magnitude smaller than that of the steady-state solution.

Moreover, we found that the dust gap expands with time, and then the dust is depleted in a wide range of $x\lesssim w_{\rm out}^+$ \add{when $m\gtrsim0.1$} (\Figref{fig:alpha1e-5}; see also \Figref{fig:planet mass dependence appendix} for different planetary masses). We found that the inner edge of the dust gap moves with $|v_{\rm drift}|$. In this case, the dust gap width can no longer estimated by $\mathcal{W}_{\rm out}$ (Eq. \ref{eq:width of outflow region}). Because we assumed a constant supply of dust from the outer region of the disk, the dust slowly leaks to the inside of the planetary orbit, and then $\Sigma_{\rm d}(t)$ at $x\lesssim w_{\rm out}^+$ \add{increased} in the late stage of the time evolution, $t\gtrsim10^8$. 

\addthi{For illustrative purposes, we also display these time sequences in a two-dimensional (2D) plane in \Figref{fig:disk_time}}, which were generated from the results of 1D calculations shown in \Figref{fig:alpha1e-5}. \revsec{The dust cavity, the gap as wide as the planet’s orbital distance, forms after $t\geq10^6$. The cavity-opening timescale can be estimated by \Equref{eq:tdrift} with $\mathcal{L}=r_{\rm p}$.} \addsec{We note that our model for the dust surface density evolution ignores the effect of curvature (Eq. \ref{eq:1D advection diffusion equation}).} %We describe quantitatively the width and the depth of dust gaps in Sect. \ref{sec:Properties of dust rings and gaps}, and then introduce the semi-analytic models of them in Sect. \ref{sec:Semi-analytic models of dust rings and gaps}.

%時間的に拡大するギャップとキャビティの観測への示唆は、5.3説で追って議論する。

\revsec{A disk with a dust cavity formed by the gas-flow mechanism could be observed as a transition disk \citep{francis2020dust}, although the dependence of the cavity evolution time on $m,\,\mathcal{M}_{\rm hw},\,{\rm St},$ and $\alpha_{\rm diff}$ remains unclear. We speculate that the dust cavity is filled on long timescale ($\gtrsim10^6\text{--}10^7$, corresponding to $>5\text{--}50$ Myr at 10 au; \Figref{fig:tau_ring}). Since the cavity is filled by diffusion of dust at the ring, we considered that the cavity-filling timescale would be proportional to the formation timescale of the dust ring \citepalias[Eq. 35 of][]{kuwahara2022dust}: $\tau_{\rm cav,fill}\sim M_{\rm ring}/\dot{M}_{\rm dust}$, where $M_{\rm ring}$ is the mass of the steady-state dust ring and $\dot{M}_{\rm dust}$ is the radial inward mass flux of dust. We discuss the implications for transition disks in Sect. \ref{sec:Dust gap width Discussion}. }

%------------------------------------
%------------------------------------
%------------------------------------
\subsubsection{Dependence of $\Sigma_{\rm d}(t)$ on the planetary mass}\label{sec:Dependence on the planetary mass}
\addthi{When higher-mass planets are assumed, we find deeper dust gaps and higher concentrations of dust in a ring.} %\add{The deeper dust gaps and higher concentrations of dust into a ring can be seen when the higher-mass planets are assumed, which is consistent with the conclusion of \citetalias{kuwahara2022dust} (\Figref{fig:planet mass dependence}).} 
\add{In \Figref{fig:planet mass dependence}, the Mach number of the headwind, the Stokes number, and the turbulent parameter are the same as in the fiducial case: $\mathcal{M}_{\rm hw}=0.03$, ${\rm St}=10^{-3}$, and $\alpha_{\rm diff}=10^{-4}$}. %Overall trends of \Figref{fig:planet mass dependence} indicate that  %, in which the more drastic changes of $\Sigma_{\rm d}(t)$ can be seen when the higher-mass planets were assumed. 
The location of the outer edge of the dust gap, $x_{\rm gap,out}^{\rm num}\sim w_{\rm out}^+$, hardly changes when $m\gtrsim0.3$, \add{at which point $w_{\rm out}^{+}$ converges} (Eq. \ref{eq:w_out}). The dust gap depths \add{converge at} $m\gtrsim0.3$. \add{This is because} the outflow speed in the $x$-direction has a peak at $m\sim0.3$ and then the influence of the gas flow on the dust motion saturates  \citep{kuwahara2024analytic}. In \citetalias{kuwahara2022dust}, \add{where we only considered $m=0.03,\,0.1,$ and 0.3,} we set the condition of the dust ring and gap formation as $m\gtrsim0.1$. \add{However,} we found that even low-mass planets with $m=0.05$ can generate dust rings and (or) gaps (zoom-in panels of \Figref{fig:planet mass dependence}, see also \Figref{fig:planet mass dependence appendix} for $\alpha_{\rm diff}=10^{-5}$).

%------------------------------------
%------------------------------------
\subsubsection{\add{Dependence of $\Sigma_{\rm d}(t)$ on the Mach number of the headwind}}\label{sec:Dependence on the headwind}
\add{The amplitude of the dust ring becomes higher when the larger $\mathcal{M}_{\rm hw}$ is assumed (\Figref{fig:Mhw dependence}). The flow speed of the radially-outward outflow increases with $\mathcal{M}_{\rm hw}$ \citep{kuwahara2024analytic}, which leads to higher concentrations of dust into a ring outside the planetary orbit.}

%------------------------------------
\subsubsection{Dependence of $\Sigma_{\rm d}(t)$ on the Stokes number}\label{sec:Dependence on the Stokes number}
In this section, we focus on the dependence of $\Sigma_{\rm d}(t)$ on the Stokes number. In \Figref{fig:St dependence}, \add{the planetary mass, the Mach number of the headwind, and the turbulent parameter are the same as in the fiducial case: $m=0.1,\,\mathcal{M}_{\rm hw}=0.03$, and $\alpha_{\rm diff}=10^{-4}$.}

\citetalias{kuwahara2022dust} found that the deeper dust gaps and higher concentrations of dust into a ring can be seen for the smaller Stokes numbers in a steady state, because smaller dust particles are more sensitive to the gas flow. \add{This is successfully reproduced in \Figref{fig:St dependence}d.} %\citetalias{kuwahara2022dust} set the condition for an apparent dust ring and gap formation occurs when ${\rm St}\lesssim10^{-3}$ (Sect.\ref{sec:Short summary of Paper1}).
\add{The time required to reach the steady state is shorter when the larger Stokes number is assumed (Figs. \ref{fig:St dependence}a--c, see also \Figref{fig:St dependence appendix} for $\alpha_{\rm diff}=10^{-5}$),}
%Same as the conclusion of \citetalias{kuwahara2022dust}, \add{a perturbation to $\Sigma_{\rm d}(t)$ is weak when ${\rm St}=10^{-2}$.} During the time evolution of $\Sigma_{\rm d}(t)$, \add{dust gaps are deeper and dust concentration is higher in ${\rm St}=10^{-3}$ than ${\rm St}=10^{-4}$} (Figs. \ref{fig:St dependence}b and c). %- This is caused by differences in the advection speed of dust. 
because the drift timescale of dust is \add{shorter for the larger Stokes numbers} ($t_{\rm drift}\propto{\rm St}^{-1}$; \Equref{eq:tdrift}).

%---------------------------------------------------------
\subsubsection{Summary of the parameter dependence}
\add{We summarize the dependence of $\Sigma_{\rm d}(t)$ on $\alpha_{\rm diff},\,m$ and St for a fixed $\mathcal{M}_{\rm hw}$ and the time in \Figref{fig:param_summary}. We set $\mathcal{M}_{\rm hw}=0.03$ and $t=10^5$ in \Figref{fig:param_summary}. A perturbation to $\Sigma_{\rm d}(t)$ is stronger when the smaller $\alpha_{\rm diff}$, the higher-mass planets, or the smaller St are assumed.}

\add{Figure \ref{fig:heatmap} is a contour plot of the dust gap depth as a function of the planetary mass and the Stokes number for a fixed $\mathcal{M}_{\rm hw}$ and time ($\mathcal{M}_{\rm hw}=0.03$ and $t=10^5$), showing that the dust gap forms when $m\gtrsim0.05$ and the dust gap deepens as the planetary mass increases. At $t=10^5$, the dust gap depth has a peak at ${\rm St}=10^{-3}$.} %Since the influences of gas flow are weaker for the large St, the dust gaps become shallow as St increases. Since the drift timescale of dust is longer for smaller St, the dust gap depth for ${\rm St}=10^{-4}$ is shallower than that for ${\rm St}=10^{-3}$ at $t=10^5$.}

%---------------------------------------------------------
%---------------------------------------------------------
\section{\rev{Properties of dust rings and gaps}}\label{sec:Properties of dust rings and gaps}
\rev{This section shows the widths of the ring and gap and depth of the gap. We first show the numerical results in Sects. \ref{sec:Dust gap width}--\ref{sec:Dust ring width}. We then introduce semi-analytic models of dust rings and gaps in Sect. \ref{sec:Semi-analytic models of dust rings and gaps} based on the obtained numerical results.}

\subsection{\rev{Numerically-calculated dust gap width}}\label{sec:Dust gap width}
\add{We found that the dust gap \addthi{width either stays constant or expands} with time (\Figref{fig:gapwidth_time}).} %Figure \ref{fig:gapwidth_time} shows the time evolution of the dust gap width for different planetary masses. %The outer edge of the dust gap is hardly change with time since it is set by the outer edge of the outflow region, $x=w_{\rm out}^+$. Thus, the dust gap width depends on the location of the inner edge of the dust gap. 
In general, \add{once the dust gaps form, their} widths do not change significantly with time in the moderate-turbulence disks ($\alpha_{\rm diff}=10^{-4}$; \Figref{fig:gapwidth_time}a), because $\Sigma_{\rm d}(t)$ decreases only within the outflow region (\Figref{fig:gas_vd_sigmad}). The dust gap widths increase with the planetary mass when $m\lesssim0.3$, \add{and converge at $m\gtrsim0.3$}. These numerical results can be explained by the dependence of $\mathcal{W}_{\rm out}$ on the planetary mass, %These numerical results suggest that when $\alpha_{\rm diff}=10^{-4}$ the dust gap width can be scaled by the width of the outflow region, $\mathcal{W}_{\rm out}$ (\Equref{eq:width of outflow region}). %, even though there are few exceptions such as the lines of $m=0.05,\,0.1$ in \Figref{fig:gapwidth_time}a and $m=0.1$ in \Figref{fig:gapwidth_time}b.
which is independent of time. \add{It increases with the planetary mass when $m\lesssim0.3$, and has a constant value at $m\gtrsim0.3$} (\Figref{fig:wout}). 

The dust gap \add{keeps expanding} with time when $m\gtrsim0.1$ after $t\gtrsim10^4$ in the low-turbulence disk ($\alpha_{\rm diff}=10^{-5}$; \Figref{fig:gapwidth_time}b). Since the inner edge of the dust gap moves with $|v_{\rm drift}|$ (\Figref{fig:alpha1e-5}), the width of the expanding dust gap is independent of the planetary mass. \add{We note that the semi-analytic model of the dust gap width (the dashed lines in \Figref{fig:gapwidth_time}) will be introduced in Sect. \ref{sec:Semi-analytic models of dust rings and gaps}.}

%---------------------------------
%---------------------------------
\subsection{\rev{Numerically-calculated dust gap depth}}\label{sec:Dust gap depth}
\add{The dust gaps deepen initially with time, and then their depths $\delta_{\rm gap}^{\rm num}(t)$ converge after $t\gtrsim10^3\text{--}10^4$ (\Figref{fig:gapdepth_time}).} Initially, $\delta_{\rm gap}^{\rm num}(t)$ decreases because the dust surface density decreases at $x\lesssim w_{\rm out}^+$. 
As $\Sigma_{\rm d}(t)$ stops decreasing at $x\lesssim w_{\rm out}^+$ or a decrease in $\Sigma_{\rm d}(t)$ at $x\lesssim w_{\rm out}^+$ balances an increase in $\Sigma_{\rm d}(t)$ at $x\gtrsim w_{\rm out}^+$ after $t\gtrsim10^3\text{--}10^4$, the dust gap depth eventually becomes constant (\Figref{fig:alpha1e-5}). %The dust gap depths depend strongly on $\alpha_{\rm diff}$. %The dust gap depths are deeper for smaller turbulent parameters. Under weak turbulence, the \add{steep} gradient of $\Sigma_{\rm d}$ needs to achieve a steady state, which results in a significant decrease in $\delta_{\rm gap}^{\rm num}(t)$. 
The dust gaps \add{deepen} with the planetary mass when $m\lesssim0.3$ and \add{their depths} converge \add{at} $m\gtrsim0.3$ (\Figref{fig:delta_m}), because the outflow speed has a peak \add{at} $m\sim0.3$ and, \add{consequently,} the perturbation of the gas flow on the dust motion saturates \citep{kuwahara2024analytic}. \add{We note that the semi-analytic model of the dust gap depth (the dashed lines in \Figref{fig:gapdepth_time}) will be introduced in Sect. \ref{sec:Semi-analytic models of dust rings and gaps}.}

%---------------------------------
%---------------------------------
\subsection{\rev{Numerically-calculated dust ring width}}\label{sec:Dust ring width}
\add{The dust ring widths increase with time due to diffusion and then reach a steady state (\Figref{fig:ringwidth}).} The dust ring widths have the radial extent of $\lesssim1\text{--}10$ times the gas scale height, \add{which is} weakly dependent on the planetary mass. \add{Figure \ref{fig:ringwidth} summarizes the parameter dependence of the dust ring width at a certain time, showing the following trends. (1) The dust ring width decreases as St increases (Figs. \ref{fig:ringwidth}a and b). (2) The dust ring width increases as $\alpha_{\rm diff}$ increases (Figs. \ref{fig:ringwidth}b and c). (3) The dust ring width decreases as $\mathcal{M}_{\rm hw}$ increases} (Figs. \ref{fig:ringwidth}c and d). These trends suggest that the dust ring width is proportional to the length where the drift timescale coincides with the diffusion timescale: $\mathcal{L}_{\rm eq}\propto\alpha_{\rm diff}{\rm St}^{-1}\mathcal{M}_{\rm hw}^{-1}$. \add{We note that the semi-analytic model of the dust ring width (the dashed lines in \Figref{fig:ringwidth}) will be introduced in Sect. \ref{sec:Semi-analytic models of dust rings and gaps}.}%As described in Sect. \ref{sec:Results in a moderate-turbulence disk}, the inner edge of the dust ring is set by $x=w_{\rm out}^+$ and it hardly changes with time. Thus, the ring width is determined by the location of the outer edge of the dust ring. 

%------------------------------------
%------------------------------------
%------------------------------------
\subsection{\rev{Semi-analytic models of dust rings and gaps}}\label{sec:Semi-analytic models of dust rings and gaps}

Based on the obtained numerical results in Sect. \ref{sec:Numerical results}, we introduce semi-analytic models of the width of the dust ring and gap and the depth of the dust gap \add{as functions of $m,\,\mathcal{M}_{\rm hw},\,{\rm St},\,\alpha_{\rm diff}$, and $t$. Since a significant perturbation to $\Sigma_{\rm d}(t)$ due to the planet-induced gas flow appears only when the smaller Stokes numbers were assumed, we restrict our attention to the limited range of the Stokes number, ${\rm St}\lesssim10^{-3}$.}

\add{Section \ref{sec:Dust gap width sect 4} introduces the semi-analytic model of the dust gap width. By fitting of the numerical results, we derived a criterion for the dust gap width which distinguish between the temporally constant and expanding gaps, and then described the dust gap widths in each case. In Sect. \ref{sec:Dust gap depth sect 4}, we considered that the time evolution of the dust gap depth is described by a logistic differential equation. Using an analytical solution to the logistic equation combined with the fitting of numerical results, we obtained the semi-analytic model of the dust gap depth.  In Sect. \ref{sec:Dust ring width sect 4}, we considered the time evolution of the dust ring width is described by a sigmoid curve with a steady-state dust ring width as an asymptote. By fitting the sigmoid curve to the numerical results, we obtained the semi-analytic model of the dust ring width.}

%------------------------------------
\subsubsection{\rev{Dust gap width}}\label{sec:Dust gap width sect 4}
As mentioned in Sect. \ref{sec:Dust gap width}, \add{the dust \addthi{gap is either} constant or expanding with time. The temporally constant dust gap width can be modeled by the width of the outflow region, $\mathcal{W}_{\rm out}$ (Eq. \ref{eq:width of outflow region}). When the dust gap expands with time, the inner edge of the dust gap is set by $x_{\rm gap,in}(t)=-|v_{\rm drift}|t$ or $w_{\rm out}^-$, whichever smaller (\Figref{fig:alpha1e-5}).} %addthi{Based on these numerical results, we derive semi-analytic models for dust gap widths, $\mathcal{W}_{\rm gap}^{\rm SA}(t)$.}

\addthi{Considering the dust surface density within the outflow region, we construct a semi-analytic model for the dust gap width, $\mathcal{W}_{\rm gap}^{\rm SA}(t)$. We expect that the dust gap width keeps constant, $\mathcal{W}_{\rm gap}^{\rm SA}(t)=\mathcal{W}_{\rm out}$, when the diffusive flux of dust dominates the time evolution of the dust surface density, while we assume that the dust gap expands with time when the advective flux dominates. We determine the diffusion-dominated or advection-dominated regime by comparing \addfor{$\Sigma_{\rm d}(x,t)$} at the gap location with a certain critical value, $\Sigma_{\rm crit}$. Given the balance between the advective and diffusive flux of dust, we derive the critical dust surface density:
\begin{align}
    \langle v_{\rm d}\rangle\Sigma_{\rm d}=\mathcal{D}\frac{\partial \Sigma_{\rm d}}{\partial x}.\label{eq:flux balance} 
\end{align}
Here we focus on the limited region, $w_{\rm out}^-<x<w_{\rm out}^+$, in which the radial drift velocity of dust is approximately given by $\langle v_{\rm d}\rangle\sim v_{\rm dfrit}$ (the middle panel of \Figref{fig:gas_vd_sigmad}). We set $\langle v_{\rm d}\rangle=v_{\rm drift}$ for simplicity in \Equref{eq:flux balance}. We then integrate \Equref{eq:flux balance} over a range of $\mathcal{W}_{\rm out}=w_{\rm out}^+-w_{\rm out}^-$ (Eq. \ref{eq:width of outflow region}) and obtain:
\addfif{\begin{align}
    \ln\Bigg(\frac{\Sigma_{\rm d}(w_{\rm out}^+)}{\Sigma_{\rm d}(w_{\rm out}^-)}\Bigg)=\frac{v_{\rm drift}}{\mathcal{D}}\int_{w_{\rm out}^-}^{w_{\rm out}^+}\mathrm{d}x.\label{eq:integration}
\end{align}
Equation (\ref{eq:integration}) gives:
\begin{align}
    \Sigma_{\rm d}(w_{\rm out}^+)=\Sigma_{\rm d,0}\exp\Bigg[-\frac{2{\rm St}\mathcal{M}_{\rm hw}\mathcal{W}_{\rm out}}{\alpha_{\rm diff}}\Bigg]\equiv\Sigma_{\rm crit},\label{eq:Sigma crit}
\end{align}
where we set $\Sigma_{\rm d}(w_{\rm out}^-)=\Sigma_{\rm d,0}$ and assume $1+{\rm St}^2\simeq1$.} The diffusive (advective) flux of dust dominates the time evolution of the dust gap when \addfor{$\Sigma_{\rm d}(x,t)>\Sigma_{\rm crit}$ ($\Sigma_{\rm d}(x,t)<\Sigma_{\rm crit}$)} within the limited region, ${w_{\rm out}^-<x<w_{\rm out}^+}$.} \addfor{We compared the time evolution of the dust surface density with $\Sigma_{\rm crit}$ in \Figref{fig:Smin_time}.}

\addthi{Practically, we consider that the dust gap stays constant when $\Sigma_{\rm min}^{\rm global}\geq\Sigma_{\rm crit}$ and expands with time when $\Sigma_{\rm min}^{\rm global}<\Sigma_{\rm crit}$, where $\Sigma_{\rm min}^{\rm global}$ is the global minimum of the time-dependent dust surface density at the gap location:}
\begin{align}
    \displaystyle\Sigma_{\rm min}^{\rm global}\equiv\min_{t>0}\Sigma_{\rm d,min}(t).
\end{align}
We \addfif{find} that $\Sigma_{\rm min}^{\rm global}$ can be fitted by (Appendix \ref{sec:Appendix Fitting formulae}):
\begin{align}
    \Sigma_{\rm min}^{\rm global}=\min(1,\,\Sigma_{\rm min}^{\rm fit}),
\end{align}
where,\rev{
\begin{align}
    &\Sigma_{\rm min}^{\rm fit}=10^{\mathcal{S}_{\rm min}^{\rm fit}(\alpha_{\rm diff},m)},\label{eq:Sigma min fit}\\
    &\mathcal{S}_{\rm min}^{\rm fit}(\alpha_{\rm diff},m)=-0.37\,\bigg(\frac{\alpha_{\rm diff}}{10^{-4}}\bigg)^{-1.1}\times\mathrm{erf}\Bigg(3.2\times10^2\,\bigg(\frac{\alpha_{\rm diff}}{10^{-4}}\bigg)^{-0.17}m^{2.8}\Bigg),
\end{align}
\add{with} $\mathrm{erf}$} \add{being} the error function \rev{(${\rm erf}(m)\rightarrow1$ when $m\rightarrow\infty$)}. %\erase{In \Equref{eq:Sigma min fit}, $\mathcal{S}_{\rm min}^{\rm fit}$ determines the shape of the function and $\Sigma_{\rm low,lim}^{\rm fit}$ sets the lower limit.}

In summary, the semi-analytic formula of the dust gap width is given by:
\begin{mdframed}
\begin{align}
    \mathcal{W}_{\rm gap}^{\rm SA}(t)=
    \begin{cases}
        \mathcal{W}_{\rm out}\quad\left(\Sigma_{\rm min}^{\rm global}\geq\Sigma_{\rm crit}\right),\\
        \max\left(\mathcal{W}_{\rm out},w_{\rm out}^+-|v_{\rm drift}|t\right)\quad\left(\Sigma_{\rm min}^{\rm global}<\Sigma_{\rm crit}\right).\label{eq:W gap SA}
    \end{cases}
\end{align}
\end{mdframed}
We plotted \Equref{eq:W gap SA} in \Figref{fig:gapwidth_time} with the dashed line (see also \Figref{fig:gapwidth_time_appendix}). In \Figref{fig:gapwidth_time}, \Equref{eq:W gap SA} predicts that \add{the dust gaps keep expanding with time when $m\gtrsim0.1\text{--}0.2$ at $t\gtrsim10^4$, which is consistent with the numerical results in the low-turbulence disk ($\alpha_{\rm diff}=10^{-5}$; \Figref{fig:gapwidth_time}b). When $\alpha_{\rm diff}=10^{-4}$, \Equref{eq:W gap SA} \addfif{fails} to reproduce the numerical results for $m\geq0.2$ which are constant with time (\Figref{fig:gapwidth_time}a). We speculate that this deviation \addfif{is} caused by the assumption of $\langle v_{\rm d}\rangle\sim v_{\rm dfrit}$ in \Equref{eq:flux balance}. \addsec{Nevertheless, we use \Equref{eq:flux balance} as it reproduces an overall trend in the planetary-mass dependence.}}%Although \Figref{fig:gapwidth_time} (and also \Figref{fig:gapwidth_time_appendix}) shows the limited applicability of our semi-analytic model, \Equref{eq:W gap SA} can describe the dependence of the dust gap width on the planetary mass. %the dust gap widths are constant when $\alpha_{\rm diff}=10^{-4}$. Equation (\ref{eq:W gap SA}) predicts that the dust gap expands with time when $m\gtrsim0.1$ at $t\gtrsim10^3\text{--}10^4$ in a low-turbulence disk ($\alpha_{\rm diff}=10^{-5}$), which  is consistent with the numerical results. 

%------------------------------------
%------------------------------------
%------------------------------------
\subsubsection{\rev{Dust gap depth}}\label{sec:Dust gap depth sect 4}
As described in Sect. \ref{sec:Dust gap depth}, the dust gap depth deepens with time and has a lower limit. We \addfif{assume} that the time evolution of the dust gap depth obeys the following equation\footnote{\addfif{Equation (\ref{eq:delta SA}) is an analytic solution to the following logistic equation:
\begin{align}
    \frac{\partial \delta_{\rm gap}^{\rm SA}(t)}{\partial t}=-\frac{\delta_{\rm gap}^{\rm SA}(t)}{\tau_{\rm gap}}\Bigg(\frac{\delta_{\rm gap}^{\rm SA}(t)-\delta_{\infty}}{\delta_{\infty}}\Bigg).\label{eq:delta differential eq}
\end{align}}}:
\begin{mdframed}
\begin{align}
    \delta_{\rm gap}^{\rm SA}(t)=\delta_\infty\Bigg[1-\bigg(1-\frac{\delta_\infty}{\delta_0}\bigg)\,e^{-t/\tau_{\rm gap}}\Bigg]^{-1},\label{eq:delta SA}
\end{align}
\end{mdframed}
where $\delta_{\rm gap}^{\rm SA}(t)$ is the semi-analytic model of the dust gap depth, $\delta_{\infty}$ is the steady-state dust gap depth, \addsec{$\delta_0=\delta_{\rm gap}^{\rm SA}(0)\equiv1$}, and $\tau_{\rm gap}$ is the characteristic timescale. \addfif{Equation (\ref{eq:delta SA}) shows that} $\delta_{\rm gap}^{\rm SA}(t)$ decreases with time and then approaches \addfif{the steady-state value}, $\delta_{\rm gap}^{\rm SA}(t)\rightarrow\delta_\infty$, \addfif{at $t\gg\tau_{\rm gap}$}. We \addfif{define}
\begin{align}
    \tau_{\rm gap}\equiv\min(t_{\rm drift},t_{\rm diff}),\label{eq:tau}
\end{align}
where we set \add{$\mathcal{L}=0.41\times\mathcal{W}_{\rm out}$} in both $t_{\rm drift}$ and $t_{\rm diff}$ (Eqs. \ref{eq:tdrift} and \ref{eq:tdiff}). The coefficient of \add{0.41} \addfif{is} determined by the least-squares \add{fitting of numerical results}. \add{The characteristic timescale $\tau_{\rm gap}$ is a function of $m,\,\mathcal{M}_{\rm hw},\,{\rm St},$ and $\alpha_{\rm diff}$, having on the order of $\sim10^3\text{--}10^4$ in our parameter sets}. 

We \addfif{find} that the steady-state dust gap depth can be fitted by (Appendix \ref{sec:Appendix Fitting formulae}):\rev{
\begin{align}
    \delta_\infty=\min(1,\,\delta_{\infty}^{\rm fit}),\label{eq:gap depth SA}
\end{align}
where
\begin{align}
    &\delta_{\infty}^{\rm fit}=10^{\mathcal{S}_{\infty}^{\rm fit}(\alpha_{\rm diff},m)},\label{eq:delta fit}\\
    &\mathcal{S}_{\infty}^{\rm fit}(\alpha_{\rm diff},m)=-0.63\,\bigg(\frac{\alpha_{\rm diff}}{10^{-4}}\bigg)^{-1.1}\times\mathrm{erf}\Bigg(4.2\times10^2\,\bigg(\frac{\alpha_{\rm diff}}{10^{-4}}\bigg)^{0.022}\,m^{2.8}\Bigg).
\end{align}}
%In \Equref{eq:delta fit}, $\mathcal{S}_{\infty}^{\rm fit}$ determines the shape of the function and $\delta_{\rm low,lim}^{\rm fit}$ sets the lower limit.

We plotted \Equref{eq:delta SA} in Figs. \ref{fig:gapdepth_time} and \ref{fig:delta_m} (see also \Figref{fig:gapdepth_time_appendix}) with the dashed line. \add{Although \Equref{eq:delta SA} does not completely reproduce the numerical results, it \addsec{shows good agreement} with the numerical result,} in particular when $t\gtrsim\tau_{\rm gap}\sim10^3\text{--}10^4$. When $t\lesssim\tau_{\rm gap}$, \Equref{eq:delta SA} only agrees with the numerical results of $m<0.1$. We speculate that the deviation \addfif{is} caused by the assumption of $\tau_{\rm gap}$, which \addfif{is} set by the drift or the diffusion timescale with $\mathcal{L}\propto\mathcal{W}_{\rm out}$, whichever smaller (Eq. \ref{eq:tau}). However, the radial drift speed of dust deviates from the unperturbed value within the outflow region, which changes $t_{\rm drift}$.

%---------------------------------------------------------
%---------------------------------------------------------
\subsubsection{\rev{Dust ring width}}\label{sec:Dust ring width sect 4}
As described in Sect. \ref{sec:Dust ring width}, the dust ring width increases with time and then reaches a steady state. We \addfif{assume} that the time evolution of the dust ring width is described by the following sigmoid curve:
\begin{mdframed}
\begin{align}
    \mathcal{W}_{\rm ring}^{\rm SA}(t)=\mathcal{W}_{\rm ring,\infty}^{\rm fit}\Bigg(1-\frac{1}{1+(t/\tau_{\rm ring})^q}\Bigg),\label{eq: W ring SA}
\end{align}
\end{mdframed}
where $\mathcal{W}_{\rm ring}^{\rm SA}(t)$ is the semi-analytic model of the dust ring width, $\mathcal{W}_{\rm ring,\infty}^{\rm fit}$ is the \add{fitting formula for the} steady-state dust ring width, $\tau_{\rm ring}$ is the characteristic timescale, and \addsec{$q=0.42$} (Appendix \ref{sec:Appendix Fitting formulae}). Numerical results showed that $\mathcal{W}_{\rm ring,\infty}^{\rm fit}$ would be proportional to $\mathcal{L}_{\rm eq}$ (Eq. \ref{eq:L eq}), \addsec{and, consequently, $\alpha_{\rm diff}$} (Sect. \ref{sec:Dust ring width}), \add{but we \addfif{find} that the dependence is weaker than \addsec{$\mathcal{W}_{{\rm ring},\infty}^{\rm fit}\propto\alpha_{\rm diff}$}. We \addfif{find} that} $\mathcal{W}_{\rm ring,\infty}^{\rm fit}$ can be fitted by (Appendix \ref{sec:Appendix Fitting formulae}):
\begin{align}
    \mathcal{W}_{\rm ring,\infty}^{\rm fit}=0.63\,\Bigg(\frac{\alpha_{\rm diff}}{10^{-4}}\Bigg)^{-0.65}\times\mathcal{L}_{\rm eq}.
\end{align}
The dust rings expand due to dust diffusion. Thus, we \addfif{define}
\begin{align}
    \tau_{\rm ring}\equiv\frac{\left(\mathcal{W}_{\rm ring,\infty}^{\rm fit}\right)^2}{\alpha_{\rm diff}}.
\end{align}
We plotted \Equref{eq: W ring SA} in \Figref{fig:ringwidth} (see also \Figref{fig:ringwidth_appendix}) with the dotted lines, which \addsec{show} good agreement with the numerical results.

\subsubsection{\rev{Caveat}}
\add{So far we have considered the regime in which the dust is tightly coupled with the gas, ${\rm St}\lesssim10^{-3}$. Since we developed our semi-analytic models by fitting the numerical results with ${\rm St}=10^{-3}$ (Appendix \ref{sec:Appendix Fitting formulae}), our semi-analytic models would be invalid when ${\rm St}\gtrsim10^{-2}$. Figure \ref{fig:delta_m_St1e-2} compares our semi-analytic model for the dust gap depth with the numerical result when ${\rm St}=10^{-2}$, showing a significant deviation appears in particular when $\alpha_{\rm diff}=10^{-5}$.} %The time evolution of the widths of the dust ring and gap and the depth of the dust gap for ${\rm St}=10^{-2}$ are shown in Figs. \ref{fig:gapwidth_time_appendix}--\ref{fig:ringwidth_appendix}, which also show the limited applicability of our semi-analytic formulae for the large Stokes numbers, ${\rm St}=10^{-2}$.}

%---------------------------------------------------------
%---------------------------------------------------------
%---------------------------------------------------------
\section{Discussion}\label{sec:Discussion}

%---------------------------------------------------------
\subsection{Time evolution of $\Sigma_{\rm d}(t)$ \addthi{with} dimensional units}\label{sec:Time evolution of Sigma in the dimensional units}

\addthi{To facilitate the interpretation of our results, }
%In this section, to provide a better understanding of our study for readers, 
we rewrite our numerical results \addthi{with dimensional units}. Assuming that \addthi{a} typical steady accretion disk model \citep{Ida:2016}, we \addthi{convert} the dimensionless quantities into the dimensional ones. Appendix \ref{sec:Conversion to dimensional quantities} describes the method for the conversion. %\erase{, in which we assumed that the temperature of the disk gas is determined by viscous heating and irradiation heating from \add{a} solar-mass star. The temperature profile of the disk gas determines the gas scale height and the aspect ratio of the disk, which allows us to convert the units of the length from $H$ to au. We then calculated the dimensional planet mass at a certain orbital radius for a given dimensionless thermal mass, $M_{\rm p}=mM_\ast h^3$.} 
We set the orbital radius of the planet as $r_{\rm p}=1$ au or 50 au, at which the Mach number of the headwind has the value of $\mathcal{M}_{\rm hw}=0.03$ or $0.1$ \addthi{(Eq. \ref{eq:Mhw value})}. %\erase{We also assumed the gas surface density profile, and then calculated the physical radius of dust, $s$, for a given Stokes number. }
%So far we described the physical process of the time evolution of $\Sigma_{\rm d}(t)$ in our dimensionless unit. 
%In this section, by assuming a disk model and converting dimensionless quantities into dimensional ones, we discuss the implications of our results for planet formation and disk observations. As described in Sect. \ref{sec:Conversion to dimensional quantities}, we can convert the dimensionless parameters into dimensional ones. 

%In this section, we consider the gas gap formation by the planet using an analytic model \citep[Eqs. 9 and 16--19 in][]{duffell2020empirically} because even a low-mass planet can open a shallow gas gap in a low-turbulence disk. Although we did not consider the effect of gas gap on the dust motion in our simulations, we assumed that the dust particles can pass through the gas gap because the pressure bump formed at the gas gap edge may be permeable for the small dust particles \citep[${\rm St}\lesssim10^{-3}$;][]{Bitsch:2018}.

\addthi{Here we} show the time-dependent dust surface density $\Sigma_{\rm d}(t)$ \addthi{with sizes of} $s\simeq4$ mm\addthi{-sized particles} (${\rm St}=10^{-3}$) perturbed by \add{gas flow induced by} an Earth-like planet at 1 au  (Sect. \ref{sec:Earth-like planet at 1 au}). \addthi{We also show solids with sizes of} $s\simeq0.2$ mm (${\rm St}=3\times10^{-3}$) perturbed by a Neptune-like planet at 50 au (Sect \ref{sec:Neptune-like planet at 50 au}). The dust size was chosen to be consistent with \addthi{nonsticky slicate grains inside the \ce{H2O} snow ($\lesssim$ a few au) and with nonsticky icy \ce{CO2}-covered grains outside the \ce{CO2} snow line \addfif{located approximately} outside 10 au \citep{Musiolik:2016a,Musiolik:2016b}. This limits particle sizes to $\sim2$ mm at $\sim1$ au and $\sim0.4$ mm at 50 au \citep{Okuzumi:2019}.}%the results of a theoretical model in which the realistic stickness of the dust grains were considered \citep{Okuzumi:2019}. The dust grains exist as nonsticky silicate grains inside the \ce{H2O} snow line ($\lesssim$ a few au) and the maximum dust size is $\sim2$ mm at $\sim1$ au \citep{Okuzumi:2019}. The dust grains covered with the \ce{CO2} ice mantle outside the \ce{CO2} snow line ($\gtrsim10$ au) is as nonsticky as silicate grains \citep{Musiolik:2016a,Musiolik:2016b} and the obtained maximum dust size is $\sim0.4$ mm at 50 au in \cite{Okuzumi:2019}. 

\subsubsection{Earth-like planet at 1 au}\label{sec:Earth-like planet at 1 au}
Figure \ref{fig:earth_case} shows \addthi{the evolution of the solid surface density $\Sigma_{\rm d}(t)$ of $\sim4$ mm-sized particles around an Earth-like planet with a mass of $\sim0.7\,M_\oplus$ } %\add{shows a situation where an Earth-like planet, $\sim0.7\,M_\oplus$,} is 
embedded at 1 au in a \addthi{disk with low midplane turbulence} ($\alpha_{\rm diff}=10^{-5}$; \Figref{fig:earth_case}a) and \addthi{in a disk with moderater midplane turbulence} ($\alpha_{\rm diff}=10^{-4}$; \Figref{fig:earth_case}b). %showing the time-dependent $\Sigma_{\rm d}(t)$ of $\sim4$ mm-sized dust. 
\addthi{In} the low-turbulence disk, the dust is depleted by \addsec{$\sim2$ orders of magnitude} within 1 Myr at $<$1 au (\Figref{fig:earth_case}b). A significant amount of dust concentrates into a very narrow ring whose width is less than 0.1 au at the planet location. In the narrow ring, $\Sigma_{\rm d}(t)$ increases by $\sim10^2$ times the initial value. %The profile of $\Sigma_{\rm d}(t)$ does not significantly change when we increase the planet mass, because the influence of the planet-induced gas flow on $\Sigma_{\rm d}(t)$ saturates (see \Figref{fig:planet mass dependence}c).  %The outer edge of the gas gap is located outside the outer edge of the dust gap. These trends do not change significantly when we increase the planet mass (\Figref{fig:planet mass dependence}), except for the expected gas gap profile. 
\addthi{In contrast, in} the moderate-turbulence disk ($\alpha_{\rm diff}=10^{-4}$; \Figref{fig:earth_case}a), \addthi{the Earth-mass planets make} neither a significant dust depletion nor concentration, for the assumed \addthi{parameters}. \addthi{A} shallow dust gap appears within $0.1$ Myr at $<1$ au, but it is smoothed within $1$ Myr. Only a narrow dust ring whose width is $\sim0.1$ au remains outside the planet location at 1 Myr. %The maximum contrast of the dust ring is $\Sigma_{\rm d}/\Sigma_{\rm d,0}\sim2.5$. When we assume higher-mass planets, the narrow dust gap whose width is $\sim0.1$ au would appear around $1$ au (Figs. \ref{fig:St_Mhw003_alpha1e-5.png}b-1--b-3). 

\addthi{The assumed planetary mass ($M_{\rm p}\simeq0.7\,M_\oplus$) falls below the pebble isolation mass \citep[$M_{\rm iso}\simeq3\,M_\oplus$;][]{Lambrechts:2014,Bitsch:2018} at which a growing planet opens a shallow gas gap and then the pebble flux is highly reduced inside the planetary orbit.} \addthi{Even planets with masses below the pebble isolation mass can affect significantly $\Sigma_{\rm d}(t)$. When the planetary mass exceeds $m\gtrsim0.1$ ($M_{\rm p}\gtrsim0.7\,M_\oplus$ at 1 au), the subsequent growth of the planet would be suppressed and planets remain small in the terrestrial planet forming-region.}
%From \Figref{fig:earth_case}, we conclude that even planets with masses below the pebble isolation mass can affect the dust surface density, which has an impact on terrestrial planet formation in a low-turbulence disk, $\alpha_{\rm diff}\lesssim10^{-5}$ (Sect. \ref{sec:Implications for planet formation}).

%-------------------------
%-------------------------
%-------------------------
\subsubsection{Neptune-like planet at 50 au}\label{sec:Neptune-like planet at 50 au}
Figure \ref{fig:neptune_case} \add{shows a situation where a Neptune-like planet, $\sim13\,M_{\oplus}$,} is located at 50 au. In both the low- and moderate-turbulence disks ($\alpha_{\rm diff}=10^{-5}$ and $10^{-4}$), the Neptune-like planet can generate the dust ring and gap whose widths are a few au in the distribution of $\sim0.2$ mm-sized dust within $1$ Myr. %At 1 Myr, the obtained dust gap depths are $\delta_{\rm gap}^{\rm num}=6.2\times10^{-3}$ when $\alpha_{\rm diff}=10^{-5}$ and $\delta_{\rm gap}^{\rm num}=0.37$ when $\alpha_{\rm diff}=10^{-4}$. 

%-------------------------
%-------------------------
%-------------------------
\subsection{Implications for planet formation \add{via pebble accretion}}\label{sec:Implications for planet formation}
%Previous sections showed that even planets with masses below the pebble isolation mass can affect the dust surface density and generate the dust ring and gap, suggesting that the planet-induced gas flow affects the processes of planet formation.
%\erase{The dust ring would be sites for the dust growth. In particular, a significant accumulation of dust outside the planetary orbit in low-turbulence disks may lead to the subsequent formation of planetesimals via streaming instability \citep{Youdin:2005,Johansen:2007}. These planetesimals may rapidly grow to planets via efficient accretion of $\sim$mm-sized dust particles (pebbles) at the dust rings \citep{morbidelli2020planet,jiang2023efficient}.}

\begin{figure}
    \centering
    \includegraphics[width=\linewidth]{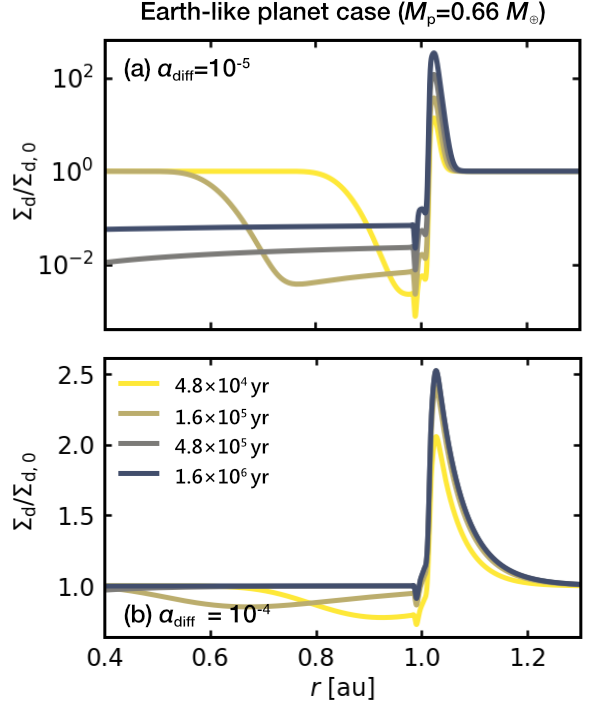}
    \caption{Time-dependent dust surface density \addthi{with sizes of $s=3.7$ mm-sized particles} perturbed by an Earth-like planet \addthi{($M_{\rm p}=0.66\,M_\oplus$)} at 1 au. Numerical simulations were conducted in the dimensionless unit. \addthi{We set $m=0.1,\,\mathcal{M}_{\rm hw}=0.03,\,{\rm St}=10^{-3}$, and varied the turbulent parameter in each panel, $\alpha_{\rm diff}=10^{-5}$ (\textit{panel a}) and $\alpha_{\rm diff}=10^{-4}$ (\textit{panel b}). For the assumed parameter set, the pebble isolation masses are given by $M_{\rm iso}=2.8\,M_\oplus$ (\textit{panel a}) and $3\,M_\oplus$ (\textit{panel b}; Eq. \ref{eq:Miso}).}}
    \label{fig:earth_case}
\end{figure}

\begin{figure}
    \centering
    \includegraphics[width=\linewidth]{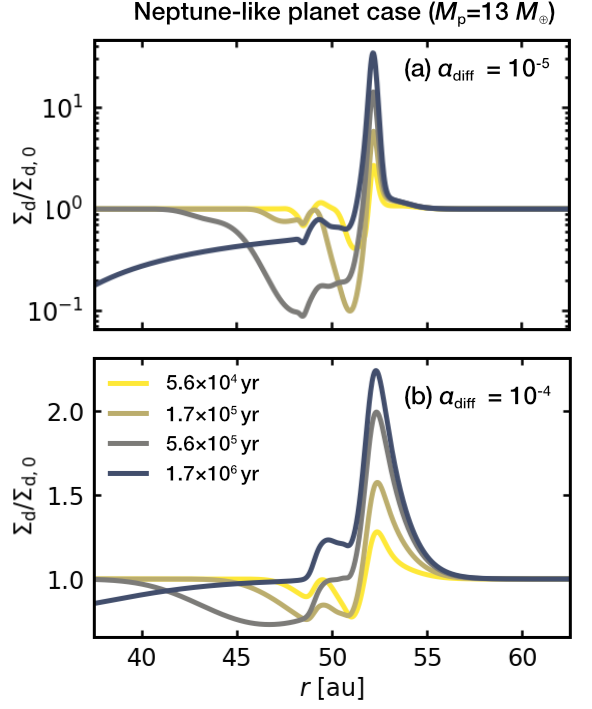}
    \caption{\addthi{Time-depenendt dust surface density with sizes of $s=0.23$ mm-sized particles for a Neptune-like planet ($M_{\rm p}=13\,M_\oplus$) at $50$ au. Numerical simulations were conducted in the dimensionless unit. We set $m=0.1,\,\mathcal{M}_{\rm hw}=0.1,\,{\rm St}=3\times10^{-3}$, and varied the turbulent parameter in each panel, $\alpha_{\rm diff}=10^{-5}$ (\textit{panel a}) and $\alpha_{\rm diff}=10^{-4}$ (\textit{panel b}). For the assumed parameter set, the pebble isolation masses are given by $M_{\rm iso}=56\,M_\oplus$ (\textit{panel a}) and $60\,M_\oplus$ (\textit{panel b}; Eq. \ref{eq:Miso}).}}
    \label{fig:neptune_case}
\end{figure}

Once the gap forms in the distribution of pebbles, it reduces the accretion rate of pebbles onto protoplanets. As shown in \Figref{fig:earth_case}a, even planets with masses below the pebble isolation mass can affect significantly $\Sigma_{\rm d}(t)$. When the planetary mass exceeds $m\gtrsim0.1$ ($M_{\rm p}\gtrsim0.7\,M_\oplus$ at 1 au), the subsequent growth of the planet would be suppressed and planets remain small in the terrestrial planet forming-region. %\erase{These remaining small planets would experience the giant impacts during disk dispersal, which may lead to the formation of super-Earth and sub-Neptune masses planets.} 
\addsec{The suppression of pebble accretion in the terrestrial planet forming-region due to the dust-gap-opening effect by the gas-flow mechanism may be helpful in explaining the current observed period-mass diagram of exoplanets, in which a large fraction of low-mass planets ($\lesssim10\,M_\oplus$) has been found at short-period orbits \citep[$\lesssim100$ days; e.g.,][\rev{see Sect. 4.4.2 of \citetalias{kuwahara2022dust} for further discussion}]{Fressin:2013,Weiss:2014}.}

%-------------------------------
%-------------------------------
\subsection{Comparison to disk observations}\label{sec:Comparison to disk observations}

% 1colのとき0.8
\begin{figure}
    \centering
    \includegraphics[width=\linewidth]{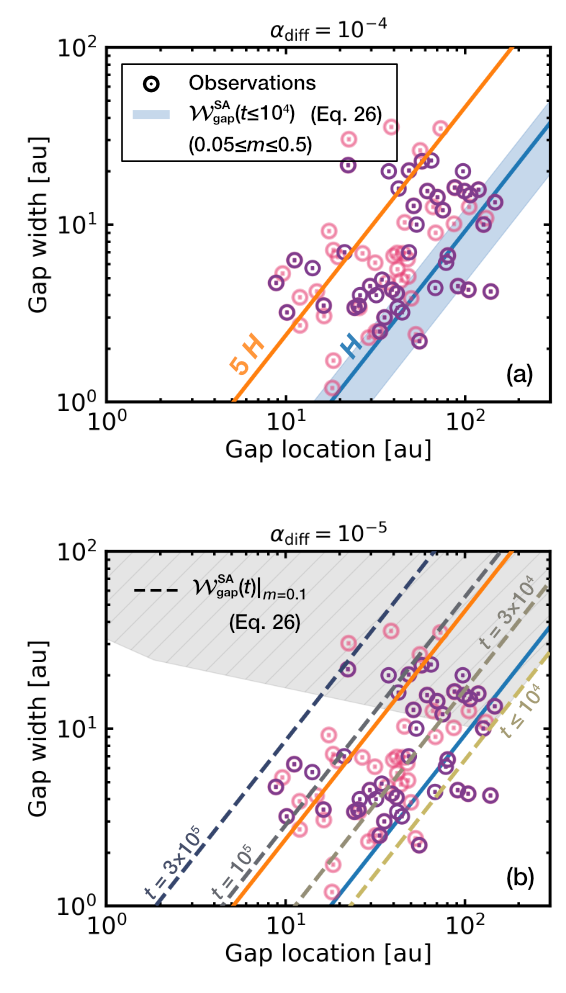}
    \caption{\add{Dust gap width as a function of the dust gap location. We set ${\rm St}=10^{-3}$ and $\mathcal{M}_{\rm hw}=0.03$, and varied the turbulent parameter in each panel, $\alpha_{\rm diff}=10^{-4}$ (\textit{panel a}) and $\alpha_{\rm diff}=10^{-5}$ (\textit{panel b}). The blue and orange solid lines denote $H$ and $5\,H$, respectively (Eq. \ref{eq:H gas}). \textit{Panel a}: The \addthi{blue} shaded region is given by $\mathcal{W}_{\rm gap}^{\rm SA}(t)$ (Eq. \ref{eq:W gap SA}) at $t\leq10^4$. The lower and upper limits were set by $m=0.05$ and 0.5, respectively. \textit{Panel b}: The dashed lines are given by \Equref{eq:W gap SA} with a fixed planetary mass, $m=0.1$. \addsec{Different colors correspond to different times, $t\leq10^4$, $t=3\times10^4,\,10^5$, and $3\times10^5$, respectively.} \addthi{We hatched the region in which \rev{the time required for gap formation} exceeds \rev{3 Myr} for the assumed dimensionless time, $t$.} The observational data indicated with pink and purple markers show observed gaps that are accompanied by a ring, taken from 
    \cite{zhang2023substructures} (compiled from \cite{huang2018-DSHARP2,long2018gaps}, and \cite{zhang2023substructures}) \addthi{and \cite{huang2018-DSHARP2} (DSHARP sample), respectively. \rev{These samples do not include disks with inner dust cavities.}}}}
    \label{fig:gap_width_vs_obs}
\end{figure}

\begin{figure}
    \centering
    \includegraphics[width=1\linewidth]{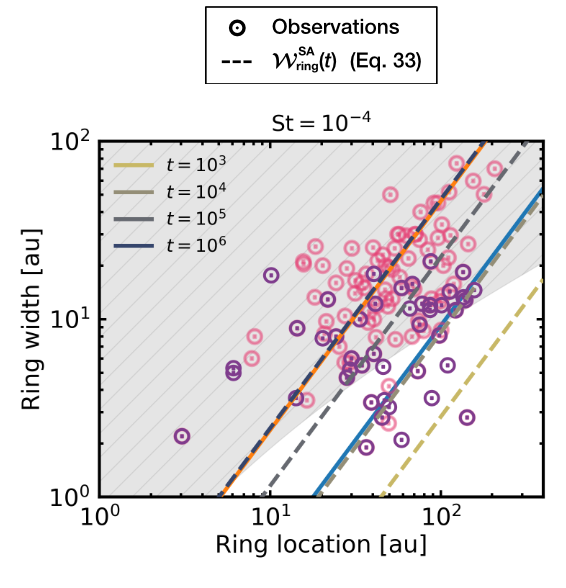}
    \caption{\add{Dust ring width as a function of the ring location. We set $\mathcal{M}_{\rm hw}=0.03$ and $\alpha_{\rm diff}=10^{-4}$. The dashed and dotted lines are given by \Equref{eq: W ring SA} with ${\rm St}=10^{-3}$ and $10^{-4}$ in which we converted the units of $\mathcal{W}_{\rm ring}^{\rm SA}$ from $H$ to au using \Equref{eq:H gas}. The blue and orange solid lines denote $H$ and $5\,H$, respectively. \addthi{We hatched the region in which \rev{the time required for ring formation} exceeds \rev{3 Myr} for the assumed dimensionless time, $t$.} The observational data \addthi{with pink and purple markers} from \cite{bae2023structured} and \cite{huang2018-DSHARP2} (DSHARP sample), respectively.}}
    \label{fig:ring_width_vs_obs}
\end{figure}

\add{We compared our semi-analytic models of the widths of the dust ring and gap with the observational data, finding that a fraction of the observed dust rings and gaps could be explained by \addthi{the gas-flow driven by low-mass planets}. \addsec{We considered \addthi{here} a single planet embedded in a disk}. \addsec{Provided} that the thermal emission of the dust is optically thin, and the opacity and the temperature are constant within a dust substructure, the dust surface density is proportional to the observed intensity profile, $\Sigma_{\rm d}\propto I_{\nu}$.} \add{In the following paragraphs, we only compared our semi-analytic models with the observed widths of the dust ring and gap. A direct comparison of our semi-analytic model of the dust gap depth with the observational data is difficult because the optically thin assumption would be invalid at the dust ring locations \rev{\citep{guerra2024into,ribas2024inner}} \addthi{and, consequently, the observed intensity does not correspond to a unique dust surface density.}} \revsec{In the highly optically thick regime the rings may not be observed, because strong optical depth effects lead to flat-topped radial intensity profiles \citep{dullemond2018-DSHARP6}. Our semi-analytic model of the dust ring width is valid as long as a ring with a moderate optical depth is detectable in the radial intensity profile. The uncertainty in the optical depth of the observed no-flat-topped ring does not significantly affect the observed value of the dust ring width \citep{dullemond2018-DSHARP6}.}

\addsec{To compare with the observational data, we converted the units of $\mathcal{W}_{\rm gap}^{\rm SA}$ and $\mathcal{W}_{\rm ring}^{\rm SA}$ from $H$ to au using \Equref{eq:H gas}. Same as in Sect. \ref{sec:Time evolution of Sigma in the dimensional units}, we considered the typical steady accretion disk model with the fixed stellar mass, stellar luminosity, and the mass accretion rate (Appendix \ref{sec:Conversion to dimensional quantities}): $M_\ast=1\,M_\odot,\,L_\ast=1\,L_\odot$, and $\dot{M}_\ast=10^{-8}\,M_\odot\text{/yr}$. We note that rings and gaps have been observed around various types of stars, so that, in reality, these values vary in disks \citep{huang2018-DSHARP2,long2018gaps,bae2023structured}: $M_\ast\sim0.2\text{--}2\,M_\odot$, $L_\ast\sim0.1\text{--}20\,L_\odot$, and $\dot{M}_{\ast}\sim10^{-10}\text{--}10^{-7}\,M_\odot\text{/yr}$.}

\subsubsection{Dust gap width}\label{sec:Dust gap width Discussion}
\add{Figure \ref{fig:gap_width_vs_obs} compares our semi-analytic model of the dust gap width, $\mathcal{W}_{\rm gap}^{\rm SA}$, with the observational data, $\mathcal{W}_{\rm gap}^{\rm obs}$, in which the dust gap widths are plotted as a function of the gap \addsec{location}, $r_{\rm gap}$. \addthi{The observational data were obtained from the Atacama Large Millimeter/submillimeter Array (ALMA) surveys \citep{huang2018-DSHARP2,long2018gaps,zhang2023substructures}, including  the Disk Substructures at High Angular Resolution Project \citep[DSHARP;][]{huang2018-DSHARP2}.} The assumed Stokes number and the Mach number of the headwind to plot $\mathcal{W}_{\rm gap}^{\rm SA}$ were the same values as in the fiducial case: ${\rm St}=10^{-3},$ and $\mathcal{M}_{\rm hw}=0.03$.}

\add{About $20\%$ of the observed dust gaps, whose widths are comparable to the gas scale height $\mathcal{W}_{\rm gap}^{\rm obs}\sim H$, could be explained by \addsec{our gas-flow mechanism in the moderate-turbulence disks} ($\alpha_{\rm diff}=10^{-4}$; \Figref{fig:gap_width_vs_obs}a). \addsec{The gas-flow mechanism has the potential to explain the observed wide dust gaps with $\mathcal{W}_{\rm gap}^{\rm obs}\gtrsim H$ by assuming \addthi{weaker midplane} turbulence ($\alpha_{\rm diff}=10^{-5}$) and a long time \addthi{evolution exceeding} $t\gtrsim10^4$.}} \addthi{Given that an upper limit for the time required for gap formation is \rev{3 Myr, $\sim65\%$} of the observed gaps can be explained (\Figref{fig:gap_width_vs_obs}b).} %Given a typical disk lifetime is $1$ Myr, the observed wide dust gaps at $\lesssim30$ au would be explained by our temporally expanding dust gaps, because $t=10^4$ corresponds to $\sim1.6$\,Myr at 30 au (\Equref{eq:time}).}

\add{These comparisons may suggest the existence of low-mass planets ($m\gtrsim0.05$) at wide orbits as an origin of the observed dust gaps, which could be consistent with a large population of such planets inferred from a population synthesis model \citep{drazkowska2023planet}. However, it is difficult to constrain the masses of unseen planets, because the dust gap widths in our model converge when $m\gtrsim0.3$ or $t\gtrsim10^4$ (\Figref{fig:gapwidth_time}).} %To constrain the masses of unseen planets from the observed dust gaps, we concluded that it is necessary to consider the dust gap widths as well as the dust gap depths and even the times after the planet formation

\revsec{Figure \ref{fig:gap_width_vs_obs} suggests that low-mass planets within $\lesssim10$ au could carve gaps as wide as their location, which are entering the transition disk regime \citep{francis2020dust}. However, since the dependence of the evolution time of the dust cavities on the parameters ($m,\,\mathcal{M}_{\rm hw},\,{\rm St},$ and $\alpha_{\rm diff}$) is unclear (\Figref{fig:tau_ring}), further investigations are needed to link the transition disks with our gas-flow mechanism.}

\subsubsection{Dust ring width}
\add{The observed dust ring widths range from a few to few tens of au, which are predominantly wider than those predicted by our semi-analytic model. Figure \ref{fig:ring_width_vs_obs} compares our semi-analytic model of the dust ring width with the observational data, $\mathcal{W}_{\rm ring}^{\rm obs}$, in which the dust ring widths are plotted as a function of the ring locations. \addthi{The observational data were compiled from \cite{huang2018-DSHARP2} and \cite{bae2023structured}.} %\rev{In \Figref{fig:ring_width_vs_obs}, the rings in the observed radial intensity profiles do not show flat-topped shapes.} 
The assumed Mach number of the headwind and the turbulent parameter to plot $\mathcal{W}_{\rm ring}^{\rm SA}$ were the same values as in the fiducial case: $\mathcal{M}_{\rm hw}=0.03$ and $\alpha_{\rm diff}=10^{-4}$. We considered ${\rm St}=10^{-4}$ in \Figref{fig:ring_width_vs_obs}.} \addsec{We note that we do not show $\alpha_{\rm diff}=10^{-5}$ \addthi{or ${\rm St}=10^{-3}$} because it leads to narrower gaps and poor agreement with observed rings.}

\add{About \rev{$15\%$} of the observed rings, whose widths are less than $\mathcal{W}_{\rm ring}^{\rm obs}\lesssim H$, could be explained by \addsec{the gas-flow mechanism}, \addthi{given that an upper limit of the time required for ring formation is \rev{3 Myr}} (\Figref{fig:ring_width_vs_obs}). \addthi{We found that only $\sim3\%$ of the observed dust rings could be explained by the gas-flow mechanism when ${\rm St}=10^{-3}$ and $\alpha_{\rm diff}=10^{-4}$.} %\erase{The most of the observed rings with $\mathcal{W}_{\rm ring}^{\rm obs}>H$ are reproduced only when we assume ${\rm St}=10^{-4}$ at $t\gtrsim10^4$,} 
\addfif{This may suggest that the dust particles at rings are small due to collisional fragmentation or bouncing \citep[][]{blum2008growth,guttler2010outcome,zsom2010outcome}.} %(e.g., fragments produced by collisional fragmentation). 
\addfif{The inferred Stokes numbers at the dust rings in this study are consistent with the results of dust growth simulations considering the fragility of porous dust \citep[${\rm St}\sim10^{-4}\text{--}10^{-3}$;][]{ueda2024support}. However, when compact dust was considered, larger Stokes numbers were inferred from the multi-wavelength analysis of the continuum emission, the modeling of dust rings at gas pressure maxima, and the dust growth simulations \citep[${\rm St}\sim10^{-3}\text{--}10^{-1}$;][]{sierra2021molecules,doi2023constraints,jiang2024grain}.} Further \addsec{discussion is} difficult, because the observed wide rings might not sufficiently resolved, and thus they could be composed of multiple narrow rings \citep{bae2023structured}.}

\subsubsection{Potential of multiple planets}
\addsec{So far we have been only considered the dust ring and gap formation by a single planet. The observed wide dust gaps with $\mathcal{W}_{\rm gap}^{\rm obs}\gtrsim H$ may also be explained by the radially-outward gas flows induced by the multiple planets. When the multiple planets are embedded in the disks with an orbital separation of $\Delta a\lesssim\mathcal{W}_{\rm gap}^{\rm SA}$, the wide dust gaps may form which are shared by the multiple planets, although the orbital stability of planets is beyond the scope of this study. In this case, a single dust ring forms outside the orbit of the outermost planet.}

\addsec{The observed wide rings may consist of the narrow rings. When the multiple planets are embedded in the disks with the orbital separation of $\Delta a>\mathcal{W}_{\rm gap}^{\rm SA}\sim H$, the multiple rings may form with a separation of $\sim\Delta a$. If the spatial resolution of the observations is low ($\gtrsim H$), these multiple narrow rings may be observed as a single wide ring.}

%-----------------------------
%-----------------------------
\subsection{Implications for future disk observations}\label{sec:Implications for disk observations}
The \addthi{relatively} deep and wide dust gaps formed by the \addthi{gas flows driven by low-mass planets in the process of their formation} in the inner few au of low-turbulence disks could be detected by future observations (\Figref{fig:earth_case}a). Due to limitations of the angular resolution, it is difficult to detect the dust substructures in the inner few au of disks by current observations. A possible future extension would improve the angular resolution of the current ALMA by several times, which could lead to further detection of dust substructure in the inner few au of disks \citep{carpenter2020alma,burrill2022investigating}. A next-generation Very Large Array (ngVLA) is expected to capture the dust thermal emission at $\sim$mm\text{--}cm wavelengths with the best angular resolution of $\sim$0.001 arcsec \citep{selina2018ngvla}, which will also provide the capability to probe the inner few au of disks. Simulations of ngVLA observations suggest that the dust gaps with the widths of $\sim$2\text{--}3 au are detectable at $\sim$5 au under weak turbulence \citep[$\alpha_{\rm diff}\lesssim10^{-5}$;][]{ricci2018investigating,harter2020imaging}. \addthi{Future high-angular-resolution observations may also detect narrow dust rings and gaps with widths comparable to or less than the gas scale height, $H$.} Thus, our gas-flow mechanism needs to be compared with these future observations.

\rev{Future observations may detect the outflow region produced by our gas-flow mechanism. The spatial scale of the outflow region is on the order of the disk gas scale height, $\sim H\simeq8.9\,{\rm au}\,(r/100\,{\rm au})^{9/7}$ (Eq. \ref{eq:aspect ratio}), in the radial and azimuthal directions \citep{kuwahara2024analytic}. The maximum amplitude of the velocity perturbation of the outflow is $\sim0.3\,c_{\rm s}=0.03\,v_{\rm K}\,(h/0.1)\simeq0.09 \,{\rm km/s}\,(r/100\,{\rm au})^{-1/2}(M_\ast/M_\odot)^{1/2}$ \citep{kuwahara2024analytic}. Given that the distance to the disk is 100 parsecs and a planet is embedded in the disk at 100 au, the capability to resolve the disk at $\sim0.1$ arcsec and $\sim0.1$ km/s is required. The current ALMA has the angular resolution of $\gtrsim0.1$ arcsec and the velocity resolution of $\gtrsim0.01$ km/s for the gas. The kinematic features of the outflow induced by low-mass planets may be detectable. However, as discussed in Sect. 4.5.2 of \citetalias{kuwahara2022dust}, the kinematic features of the outflow can only be appeared in the region close to the midplane ($z<H$; \Figref{fig:gas_vd_sigmad}). Therefore, molecules that can trace low heights in disks, such as \ce{C^17O}, \ce{HCN}, and \ce{C_2H}, should be used as tracers.}% Thus, we expect that our model for the dust ring and gap formation can be tested by these future observations. 

%-------------------------------
%-------------------------------
\subsection{Comparison to previous studies}
%\textcolor{red}{We compare our results with  \cite{dong2018multiple,yang2020morphological}}

\add{Previous studies have \addthi{mostly focused on} \addsec{gap-opening planets} \addthi{to explain} the observed dust gap widths (the gas-gap mechanism), \addthi{often} using empirical relations between the planetary mass and the dust gap width obtained from the disk-planet interaction simulations. \cite{zhang2018-DSHARP7} performed 2D hydrodynamical simulations of gas and dust with gas-gap-opening planets. The authors defined the dust gap width by $\Delta^{\rm Z18}\equiv(r_{\rm out}-r_{\rm in})/r_{\rm out}$, where $r_{\rm in}$ and $r_{\rm out}$ are the edges of the dust gap normalized by the planet location $r_{\rm p}$, and then obtained $\Delta^{\rm Z18}\sim0.1\text{--}1$ for different disk models with $h=0.05,\,0.07,$ and 0.1. In our dimensionless unit, the dust gap width defined in \cite{zhang2018-DSHARP7} can be described by $\mathcal{W}_{\rm gap}^{\rm Z18}=\Delta^{\rm Z18}r_{\rm out}/h$. Assuming $r_{\rm out}\sim r_{\rm p}\equiv1$, we obtained $\mathcal{W}_{\rm gap}^{\rm Z18}\sim1\text{--}20\,H$.}

\add{An empirical relation in which the dust gap width is assumed to be proportional to the Hill radius has been obtained by the hydrodynamical simulations with gas-gap-opening planets \citep[$\mathcal{W}_{\rm gap}^{\rm Hill}\sim4\text{--}7.5\,R_{\rm Hill}$;][]{rosotti2016minimum,lodato2019newborn,wang2021architecture}. In our parameter sets, the Hill radius ranges from $R_{\rm Hill}\simeq0.22$ to 0.55, which leads to $\mathcal{W}_{\rm gap}^{\rm Hill}\sim1\text{--}4\,H$.}

\add{\cite{dong2018multiple} performed 2D hydrodynamical simulations of gas and dust with embedded planets, investigating the dust gap formation due to the shallow gas-gap opening by a planet under the weak turbulence condition ($\alpha_{\rm diff}\lesssim10^{-5}$). The authors considered the planets with $m=0.04\text{--}0.8$, obtaining the empirical relations between the dust gap width and the planetary mass, $\mathcal{W}_{\rm gap}^{\rm D18}\simeq3.6\,l_{\rm sh}$, where $l_{\rm sh}$ is the so-called shocking length of the density waves launched by an embedded planet \citep{goodman2001planetary}:
\begin{align}
    l_{\rm sh}\simeq0.8\,\Bigg(\frac{\gamma+1}{12/5}m\Bigg)^{-2/5}\,H\simeq2\,\Bigg(\frac{m}{0.1}\Bigg)^{-2/5}\bigg|_{\gamma=1.4}\,H.\label{eq:shocking length}
\end{align}
Here, $\gamma=1.4$ is the adiabatic index. Thus, in our dimensionless unit, we obtained $\mathcal{W}_{\rm gap}^{\rm D18}\sim3\text{--}10\,H$.}

\add{The dust gap widths obtained in this study are narrower than those obtained in the previous studies mentioned above as long as a temporally constant dust gap was considered, $\mathcal{W}_{\rm gap}\lesssim4/3\,H$. The difference in dust gap widths obtained in the previous studies and this study is attributed to different physics and the different parameter range of the Stokes number assumed in each study. The models for dust ring and gap formation \addthi{driven by higher-mass planets carving gas gaps }%due to deep or shallow gas gaps considered in the previous studies 
(\addsec{the gas-gap mechanisms}) are more susceptible to occur when larger Stokes numbers are assumed \citep[${\rm St}\gtrsim10^{-3}\text{--}10^{-2}$;][]{zhu2012dust,zhu2014particle,rosotti2016minimum,weber2018characterizing}. \addthi{Then} the dust particles are trapped at $x\gtrsim H$. %The dust particles are trapped at the outer edge of the gas gap which forms at $x\gtrsim l_{\rm sh}$ through nonlinear density wave steepening \citep{goodman2001planetary,rafikov2002planet}. Equation (\ref{eq:shocking length}) gives $l_{\rm sh}\sim1\text{--}3\,H$ in our parameter sets. 
On the other hand, our \addsec{gas-flow mechanism} works well \addthi{for smaller solids with} ${\rm St}\lesssim10^{-2}$. \addthi{Such} dust particles are trapped by the outflow of the gas at $x\lesssim2/3\,H$.} %\addsec{These differences are useful in distinguishing the mechanisms of dust ring and gap formation.}%, and then they are trapped by the radially-outward gas flow.} Although we did not consider the gas gap formation in this study, we would expect that the dust gap width in the populations of the large and small dust do not match each other even when we consider the shallow gas-gap opening effect, because the small dust particles can pass through the gas gap due to diffusion or the viscous accretion flow \citep[e.g.,][]{rice2006dust}.

\addsec{Although we did not consider the (shallow) gas-gap formation in this study, we would expect that the gas-flow mechanism can coexist with the gas-gap-opening mechanism. The gas-gap mechanism generates the dust gaps with $\mathcal{W}_{\rm gap}\gtrsim H$ in the distribution of large dust (${\rm St}\gtrsim10^{-3}\text{--}10^{-2}$). Because the small dust particles can pass through the gas gap due to diffusion or the viscous accretion flow \citep[e.g.,][]{rice2006dust}, the dust gaps also form in the distribution of small dust (${\rm St}\lesssim10^{-2}$) by the gas-flow mechanism, whose widths depend on the assumed parameters. The locations of the outer edge of the dust gaps should differ between the gas-flow and gas-gap-opening mechanisms.}

\rev{Several studies have shown that low-mass planets, which do not form gas gaps or pressure bumps, can create rings and gaps. The gravitational torque exerted by embedded planets can carve gaps that appear only in the dust distribution \citep{Muto:2009,dipierro2017opening}. \cite{Muto:2009} showed that, when the global pressure gradient of the disk gas is neglected, a planet with a mass of $2\,M_\oplus$ can open a gap in the distribution of large dust grains (${\rm St}\gtrsim0.1$). \cite{dipierro2017opening} derived a grain-size-dependent criterion for dust gap opening in disks, showing that a planet with a mass of $\lesssim10\,M_\oplus$ can carve a dust gap when ${\rm St}\gtrsim1$.}

\rev{Compared to these previous studies that work well for the dust with large Stokes numbers, our results show the opposite trend. In our study, dust rings and gaps form when the smaller Stokes numbers are considered (${\rm St}\lesssim10^{-2}$), as we focused on the potential effect of the gas flow driven by an embedded planet, which was not considered in the previous studies.}

%-------------------------------
%-------------------------------
\subsection{Model limitations}\label{sec:Model limitations}
\add{Although we did not consider the evolution of the gas disk, \addsec{the outer part of a disk evolving with viscous angular momentum transport could spread} outward due to the conservation of the angular momentum \citep{lynden1974evolution}. The disk gas at $r>r_{\rm exp}$ expands outward, where $r_{\rm exp}=r_{0}/2(1+t/t_{\nu})$, $r_{0}$ is an initial disk radius, and $t_{\nu}$ is a characteristic viscous timescale at $r=r_0$. The small dust considered in this study could move outward together with the resulting outward flow of the background disk gas \citep{liu2022natural}. When a planet is embedded at $r>r_{\rm exp}$, the dust particles will be trapped by the radially-inward outflow of the gas induced by an embedded planet. Then the dust ring could form inside the planetary orbit and the dust could deplete outside the planetary orbit. \addsec{However, the direction of the flow of the background disk gas at the midplane is a controversial issue. The disk gas evolution may be driven by magnetic disk winds \citep[e.g.,][]{bai2013wind}. The winds extract angular momentum from the disk surface and drive inward gas accretion at the midplane.}}

\add{Low-mass planets would undergo inward migration \citep[so-called type $\rm I$ migration;][]{ward1986density,tanaka2002three}. The timescale of the type $\rm I$ migration is described by:\rev{
\begin{align}
    t_{\rm typeI}\simeq\frac{M_\ast}{M_{\rm p}}\frac{M_\ast}{\Sigma_{\rm g}r^2}h^3\,\Omega^{-1}=10^4\,\Bigg(\frac{m}{0.1}\Bigg)^{-1}\Bigg(\frac{\Sigma_{\rm g}r^2/M_\ast}{10^{-3}}\Bigg)^{-1}\,\Omega^{-1},
\end{align}
}where $\Sigma_{\rm g}$ is the gas surface density. The type $\rm I$ migration timescale is comparable or shorter than the timescale for the dust ring and gap formation by the \addsec{gas-flow mechanism.} When the migration timescale is shorter than the timescale for the dust ring and gap formation, the positions of the dust ring and gap do not necessarily coincide with the planet location \citep{kanagawa2021dust}.
}

\add{We fixed the planetary masses in this study, whereas planets grow by pebble accretion. A perturbation to the dust surface density becomes strong as the planetary mass increases. \rev{The dust gap widens and deepens as the planetary mass increases if the growth timescale of the planet is shorter than the timescale for the dust gap formation. When the planetary mass exceeds approximately $m\sim0.1\text{--}0.3$, the gap width is independent of the planetary mass (\Figref{fig:gapwidth_time}). Thus, we could constrain the lower bound of the mass of the embedded planet from the observed dust gap width.} We also fixed the Stokes number, even though it could be varied by dust growth and fragmentation at the dust rings. These additional processes would further complicate the time evolution of $\Sigma_{\rm d}(t)$ and should be included in further studies.}

\rev{An eccentricity of an embedded planet could alter our results. A planet on eccentric orbit induces a time-dependent gas flow field \citep{bailey2021three}, in which the radially-outward gas flow would disappear. However, the eccentricity of the planet can be damped by the disk-planet interaction. The eccentricity damping timescale is given by \citep{tanaka2004three}:
\begin{align}
    t_{\rm damp}&\simeq1.282\frac{M_\ast}{M_{\rm p}}\frac{M_\ast}{\Sigma_{\rm g}r^2}h^4\,\Omega^{-1}\nonumber\\
    &\simeq6.4\times10^2\,\Bigg(\frac{m}{0.1}\Bigg)^{-1}\Bigg(\frac{\Sigma_{\rm g}r^2/M_\ast}{10^{-3}}\Bigg)^{-1}\Bigg(\frac{h}{0.05}\Bigg)\,\Omega^{-1},
\end{align}
Thus, the dust ring and gap formation by the gas-flow mechanism is valid when the eccentricity damping timescale is shorter than the timescale for the dust ring and gap formation.}

\add{\addthi{We finally note that} the backreaction of dust on gas, which is not considered in this study, would be important at the dust rings. When the backreaction is included, \rev{the axisymmetric dust rings form without planets due to the self-induced dust trap mechanism \citep{gonzalez2017self,vericel2020self,vericel2021dust}. When a planet embedded in a disk, the dust ring outside the planetary orbit} \addsec{with the high local dust-to-gas density ratio ($\gtrsim1$)} could be unstable and broken into small-scale dust-gas vortices \citep{pierens2019vortex,yang2020morphological}, which could change an axisymmetric morphology of the dust rings considered in this study.}

%---------------------------------------------------------
%---------------------------------------------------------
%---------------------------------------------------------
%%%%%%%%%%%%%%%%%%%%%%%%%%%%%%%%%%%%%%%%%%%%%%%%%%%%%%%%%%
%%%%%%%%%%%%%%%%%%%%%%%%%%%%%%%%%%%%%%%%%%%%%%%%%%%%%%%%%%
%%%%%%%%%%%%%%%%%%%%%%%%%%%%%%%%%%%%%%%%%%%%%%%%%%%%%%%%%%
\section{Conclusions} \label{sec:Conclusions}

\add{We investigated the \addsec{time evolution of dust rings and gaps formed} by low-mass planets \addsec{inducing \addthi{a} radially-outward gas flow}. By fitting \addthi{our} numerical results, we developed semi-analytic models describing the widths of the dust ring and gap and the depth of the dust gap. Our main results are as follows:}
\begin{enumerate}
    \item Under weak turbulence ($\alpha_{\rm diff}\lesssim10^{-4}$), low-mass planets with $m\gtrsim0.05$ (corresponding to $\gtrsim0.33\,M_\oplus$ at 1 au \addthi{or $\gtrsim1.7\,M_\oplus$ at 10 au}) can generate dust rings and gaps in the distribution of small dust, ${\rm St}\lesssim10^{-2}$. %The properties of the dust rings and gaps depend on the planetary mass, the Stokes number, the turbulent parameter, and the time.
    \item \addthi{Dust gaps have a width comparable to the gas scale height, but can expand further in size when $m\gtrsim0.1$ and $\alpha_{\rm diff}
    \lesssim10^{-5}$, at a rate set by the dust drift speed \rev{(Eq. \ref{eq:W gap SA})}.}%The dust gap widths keep constant or expanding with time, depending on the assumed parameters such as the planetary mass, the turbulent parameter, and the time. The temporally constant dust gap has the width of $\lesssim H$, whose width increases with the planetary mass when $m\lesssim0.3$, and converges at $m\gtrsim0.3$. The width of the temporally expanding dust gap is determined by the drift speed of dust, independent of the planetary mass. %In moderate-turbulence disks ($\alpha_{\rm diff}=10^{-4}$), the dust gap width is constant with time, increases with the planetary mass when $m\lesssim0.3$, and converges when $m\gtrsim0.3$. In a low-turbulence disk $\alpha_{\rm diff}=10^{-5}$, the dust gap width expands with time when $m\gtrsim0.1$.
    \item The dust gap depth deepens as the planetary mass increases when $m\lesssim0.3$, \addthi{but} converges at $m\gtrsim0.3$ \addthi{to a depletion factor of $\delta_{\rm gap}\sim0.2$ when $\alpha_{\rm diff}=10^{-4}$ ($\delta_{\rm gap}\sim10^{-7}$ when $\alpha_{\rm diff}=10^{-5}$;  \rev{Eq. \ref{eq:delta SA}})}. \addthi{Deeper} dust gaps form when smaller turbulent parameters are assumed. 
    \item The dust rings \addthi{outside of the planetary orbit} widen with time due to diffusion and then reach a steady state, whose widths range from $\sim0.1\,H$ to 10 $H$ depending on ${\rm St},\,\alpha_{\rm diff}$, and $\mathcal{M}_{\rm hw}$ \rev{(Eq. \ref{eq: W ring SA})}.
    %\item Our semi-analytic models of the dust rings and gaps describe the widths of the dust ring and gap, and the depth of the dust gap as functions of the planetary mass, the drift speed of dust, the turbulent parameter, and the time, reproducing the numerical results. \erase{\item Our semi-analytic models can reproduce the above mentioned numerical results, describing the widths of the dust ring and gap and the depth of the dust gap as functions of $m,\,{\rm St},\,\alpha_{\rm diff},\,\mathcal{M}_{\rm hw}$, and $t$.}
\end{enumerate}
\add{By comparing our semi-analytic models of the dust ring and gap with the observational data, we found that \addthi{up to approximately \rev{$65\%$ ($15\%$)} of the observed dust gaps (rings)} could be generated by \addsec{\addthi{the gas-flow driven by a single low-mass planet}. When ${\rm St}=10^{-3}$ and $\alpha_{\rm diff}=10^{-4}$ are considered as the fiducial values, \addthi{low-mass planets} could explain approximately $20\%$ ($3\%$) of the observed dust gaps (rings) with the radial widths of $\sim H$ within $t\leq10^4\text{--}10^5$\addthi{, corresponding to \rev{$\lesssim0.05\text{--}0.5$ Myr} at 10 au}. \addthi{On longer times} ($t\gtrsim10^4\text{--}10^5$), the gas-flow mechanism also has the potential to explain approximately \rev{$65\%$ ($15\%$)} of the observed wide gaps (rings) with widths \addthi{exceeding the gas scale height $H$. Wide gaps require a low level of midplane turbulence ($\alpha_{\rm diff}\lesssim10^{-5}$) and wide rings require the very small Stokes numbers (${\rm St}\lesssim10^{-4}$)}.}}

\addfif{Our model for the dust ring and gap formation favors low values of St (${\rm St}\lesssim10^{-4}\text{--}10^{-3}$), which may suggest the existence of fragile dust grains in protoplanetary disks \citep{Okuzumi:2019,jiang2024grain,ueda2024support}.} %\erase{Our findings reveal that even low-mass planets, which do not form gas gaps or pressure bumps, can create rings and gaps.} 
\addthi{A fraction of the observed disk substructures may already be consistent with such low-mass planets in wide orbits where they may be ubiquitous during planet formation, before migrating into their final orbits \citep{drazkowska2023planet}.}%Our gas-flow mechanism has the potential to explain the observed wide gaps and rings whose widths are $>H$, depending on the assumed parameters. The observed wide gaps could be explained by the gas-flow mechanism when the weak turbulence ($\alpha_{\rm diff}\lesssim10^{-5}$) and the long timescale ($t\gtrsim10^4$) are assumed. The observed wide rings could be explained when the very small Stokes number (${\rm St}\lesssim10^{-4}$) and the long timescale ($t\gtrsim10^4$) are assumed.

%-------------------------------------------------------------------------------------------
\begin{acknowledgements}
\rev{We would like to thank an anonymous referee for constructive comments that have improved the quality of this manuscript.} We thank Athena++ developers. Numerical computations were in part carried out on Cray XC50 at the Center for Computational Astrophysics at the National Astronomical Observatory of Japan. \add{A.K. and M.L. acknowledge the ERC starting grant 101041466-EXODOSS. \addsec{H.K. is supported by JSPS KAKENHI Grant No. 20KK0080, 21H04514, 21K13976, 22H01290, 22H05150.}}
\end{acknowledgements}
%

%% references
%%\raggedright              %% only for adsaa with dvips, not for pdflatex

%\ref{Ref}
%\bibliography{Ref}

\begin{thebibliography}{89}
\expandafter\ifx\csname natexlab\endcsname\relax\def\natexlab#1{#1}\fi

\bibitem[{{ALMA Partnership} {et~al.}(2015){ALMA Partnership}, {Brogan},
  {P{\'e}rez}, {Hunter}, {Dent}, {Hales}, {Hills}, {Corder}, {Fomalont},
  {Vlahakis}, {Asaki}, {Barkats}, {Hirota}, {Hodge}, {Impellizzeri}, {Kneissl},
  {Liuzzo}, {Lucas}, {Marcelino}, {Matsushita}, {Nakanishi}, {Phillips},
  {Richards}, {Toledo}, {Aladro}, {Broguiere}, {Cortes}, {Cortes}, {Espada},
  {Galarza}, {Garcia-Appadoo}, {Guzman-Ramirez}, {Humphreys}, {Jung}, {Kameno},
  {Laing}, {Leon}, {Marconi}, {Mignano}, {Nikolic}, {Nyman}, {Radiszcz},
  {Remijan}, {Rod{\'o}n}, {Sawada}, {Takahashi}, {Tilanus}, {Vila Vilaro},
  {Watson}, {Wiklind}, {Akiyama}, {Chapillon}, {de Gregorio-Monsalvo}, {Di
  Francesco}, {Gueth}, {Kawamura}, {Lee}, {Nguyen Luong}, {Mangum}, {Pietu},
  {Sanhueza}, {Saigo}, {Takakuwa}, {Ubach}, {van Kempen}, {Wootten},
  {Castro-Carrizo}, {Francke}, {Gallardo}, {Garcia}, {Gonzalez}, {Hill},
  {Kaminski}, {Kurono}, {Liu}, {Lopez}, {Morales}, {Plarre}, {Schieven},
  {Testi}, {Videla}, {Villard}, {Andreani}, {Hibbard}, \&
  {Tatematsu}}]{ALMA:2015}
{ALMA Partnership}, {Brogan}, C.~L., {P{\'e}rez}, L.~M., {et~al.} 2015, The
  Astrophysical Journal, 808, L3 \csname ALMA:2015link\endcsname~\csname
  ALMA:2015note\endcsname

\bibitem[{Andrews {et~al.}(2018)Andrews, Huang, P{\'e}rez, Isella, Dullemond,
  Kurtovic, Guzm{\'a}n, Carpenter, Wilner, Zhang,
  {et~al.}}]{andrews2018-DSHARP1}
Andrews, S.~M., Huang, J., P{\'e}rez, L.~M., {et~al.} 2018, The Astrophysical
  Journal Letters, 869, L41 \csname andrews2018-DSHARP1link\endcsname~\csname
  andrews2018-DSHARP1note\endcsname

\bibitem[{Bae {et~al.}(2023)Bae, Isella, Zhu, Martin, Okuzumi, \&
  Suriano}]{bae2023structured}
Bae, J., Isella, A., Zhu, Z., {et~al.} 2023, in Astronomical Society of the
  Pacific Conference Series, Vol. 534, 423 \csname
  bae2023structuredlink\endcsname~\csname bae2023structurednote\endcsname

\bibitem[{Bai \& Stone(2013)}]{bai2013wind}
Bai, X.-N. \& Stone, J.~M. 2013, The Astrophysical Journal, 769, 76 \csname
  bai2013windlink\endcsname~\csname bai2013windnote\endcsname

\bibitem[{Bailey {et~al.}(2021)Bailey, Stone, \& Fung}]{bailey2021three}
Bailey, A., Stone, J.~M., \& Fung, J. 2021, The Astrophysical Journal, 915, 113
  \csname bailey2021threelink\endcsname~\csname bailey2021threenote\endcsname

\bibitem[{Bitsch {et~al.}(2018)Bitsch, Morbidelli, Johansen, Lega, Lambrechts,
  \& Crida}]{Bitsch:2018}
Bitsch, B., Morbidelli, A., Johansen, A., {et~al.} 2018, Astronomy \&
  Astrophysics, 612, A30 \csname Bitsch:2018link\endcsname~\csname
  Bitsch:2018note\endcsname

\bibitem[{Blum \& Wurm(2008)}]{blum2008growth}
Blum, J. \& Wurm, G. 2008, Annu. Rev. Astron. Astrophys., 46, 21 \csname
  blum2008growthlink\endcsname~\csname blum2008growthnote\endcsname

\bibitem[{Burrill {et~al.}(2022)Burrill, Ricci, Harter, Zhang, \&
  Zhu}]{burrill2022investigating}
Burrill, B.~P., Ricci, L., Harter, S.~K., Zhang, S., \& Zhu, Z. 2022, The
  Astrophysical Journal, 928, 40 \csname
  burrill2022investigatinglink\endcsname~\csname
  burrill2022investigatingnote\endcsname

\bibitem[{Carpenter {et~al.}(2020)Carpenter, Iono, Kemper, \&
  Wootten}]{carpenter2020alma}
Carpenter, J., Iono, D., Kemper, F., \& Wootten, A. 2020, arXiv preprint
  arXiv:2001.11076 \csname carpenter2020almalink\endcsname~\csname
  carpenter2020almanote\endcsname

\bibitem[{Dipierro \& Laibe(2017)}]{dipierro2017opening}
Dipierro, G. \& Laibe, G. 2017, Monthly Notices of the Royal Astronomical
  Society, 469, 1932 \csname dipierro2017openinglink\endcsname~\csname
  dipierro2017openingnote\endcsname

\bibitem[{Dipierro {et~al.}(2016)Dipierro, Laibe, Price, \&
  Lodato}]{dipierro2016two}
Dipierro, G., Laibe, G., Price, D.~J., \& Lodato, G. 2016, Monthly Notices of
  the Royal Astronomical Society: Letters, 459, L1 \csname
  dipierro2016twolink\endcsname~\csname dipierro2016twonote\endcsname

\bibitem[{Doi \& Kataoka(2023)}]{doi2023constraints}
Doi, K. \& Kataoka, A. 2023, The Astrophysical Journal, 957, 11 \csname
  doi2023constraintslink\endcsname~\csname doi2023constraintsnote\endcsname

\bibitem[{Dong \& Fung(2017)}]{dong2017mass}
Dong, R. \& Fung, J. 2017, The Astrophysical Journal, 835, 146 \csname
  dong2017masslink\endcsname~\csname dong2017massnote\endcsname

\bibitem[{Dong {et~al.}(2018)Dong, Li, Chiang, \& Li}]{dong2018multiple}
Dong, R., Li, S., Chiang, E., \& Li, H. 2018, The Astrophysical Journal, 866,
  110 \csname dong2018multiplelink\endcsname~\csname
  dong2018multiplenote\endcsname

\bibitem[{Drazkowska {et~al.}(2023)Drazkowska, Bitsch, Lambrechts, Mulders,
  Harsono, Vazan, Liu, Ormel, Kretke, \& Morbidelli}]{drazkowska2023planet}
Drazkowska, J., Bitsch, B., Lambrechts, M., {et~al.} 2023, in Astronomical
  Society of the Pacific Conference Series, Vol. 534, 717 \csname
  drazkowska2023planetlink\endcsname~\csname drazkowska2023planetnote\endcsname

\bibitem[{Dullemond {et~al.}(2018)Dullemond, Birnstiel, Huang, Kurtovic,
  Andrews, Guzm{\'a}n, P{\'e}rez, Isella, Zhu, Benisty,
  {et~al.}}]{dullemond2018-DSHARP6}
Dullemond, C.~P., Birnstiel, T., Huang, J., {et~al.} 2018, The Astrophysical
  Journal Letters, 869, L46 \csname dullemond2018-DSHARP6link\endcsname~\csname
  dullemond2018-DSHARP6note\endcsname

\bibitem[{Emsenhuber {et~al.}(2021)Emsenhuber, Mordasini, Burn, Alibert, Benz,
  \& Asphaug}]{emsenhuber2021new}
Emsenhuber, A., Mordasini, C., Burn, R., {et~al.} 2021, Astronomy \&
  Astrophysics, 656, A69 \csname emsenhuber2021newlink\endcsname~\csname
  emsenhuber2021newnote\endcsname

\bibitem[{Fernandes {et~al.}(2019)Fernandes, Mulders, Pascucci, Mordasini, \&
  Emsenhuber}]{Fernandes:2019}
Fernandes, R.~B., Mulders, G.~D., Pascucci, I., Mordasini, C., \& Emsenhuber,
  A. 2019, The Astrophysical Journal, 874, 81 \csname
  Fernandes:2019link\endcsname~\csname Fernandes:2019note\endcsname

\bibitem[{Francis \& van~der Marel(2020)}]{francis2020dust}
Francis, L. \& van~der Marel, N. 2020, The Astrophysical Journal, 892, 111
  \csname francis2020dustlink\endcsname~\csname francis2020dustnote\endcsname

\bibitem[{{Fressin} {et~al.}(2013){Fressin}, {Torres}, {Charbonneau}, {Bryson},
  {Christiansen}, {Dressing}, {Jenkins}, {Walkowicz}, \&
  {Batalha}}]{Fressin:2013}
{Fressin}, F., {Torres}, G., {Charbonneau}, D., {et~al.} 2013, The
  Astrophysical Journal, 766, 81 \csname Fressin:2013link\endcsname~\csname
  Fressin:2013note\endcsname

\bibitem[{Fulton {et~al.}(2021)Fulton, Rosenthal, Hirsch, Isaacson, Howard,
  Dedrick, Sherstyuk, Blunt, Petigura, Knutson,
  {et~al.}}]{fulton2021california}
Fulton, B.~J., Rosenthal, L.~J., Hirsch, L.~A., {et~al.} 2021, The
  Astrophysical Journal Supplement Series, 255, 14 \csname
  fulton2021californialink\endcsname~\csname fulton2021californianote\endcsname

\bibitem[{{Fung} {et~al.}(2015){Fung}, {Artymowicz}, \& {Wu}}]{Fung:2015}
{Fung}, J., {Artymowicz}, P., \& {Wu}, Y. 2015, The Astrophysical Journal, 811,
  101 \csname Fung:2015link\endcsname~\csname Fung:2015note\endcsname

\bibitem[{{Gammie}(2001)}]{Gammie:2001}
{Gammie}, C.~F. 2001, The Astrophysical Journal, 553, 174 \csname
  Gammie:2001link\endcsname~\csname Gammie:2001note\endcsname

\bibitem[{Garaud \& Lin(2007)}]{garaud2007effect}
Garaud, P. \& Lin, D. 2007, The Astrophysical Journal, 654, 606 \csname
  garaud2007effectlink\endcsname~\csname garaud2007effectnote\endcsname

\bibitem[{Gonzalez {et~al.}(2017)Gonzalez, Laibe, \&
  Maddison}]{gonzalez2017self}
Gonzalez, J.-F., Laibe, G., \& Maddison, S.~T. 2017, Monthly Notices of the
  Royal Astronomical Society, 467, 1984 \csname
  gonzalez2017selflink\endcsname~\csname gonzalez2017selfnote\endcsname

\bibitem[{Goodman \& Rafikov(2001)}]{goodman2001planetary}
Goodman, J. \& Rafikov, R. 2001, The Astrophysical Journal, 552, 793 \csname
  goodman2001planetarylink\endcsname~\csname goodman2001planetarynote\endcsname

\bibitem[{Guerra-Alvarado {et~al.}(2024)Guerra-Alvarado, Carrasco-Gonz{\'a}lez,
  Mac{\'\i}as, van~der Marel, Houge, Maud, Pinilla, Villenave, Asaki, \&
  Humphreys}]{guerra2024into}
Guerra-Alvarado, O.~M., Carrasco-Gonz{\'a}lez, C., Mac{\'\i}as, E., {et~al.}
  2024, Astronomy \& Astrophysics, 686, A298 \csname
  guerra2024intolink\endcsname~\csname guerra2024intonote\endcsname

\bibitem[{G{\"u}ttler {et~al.}(2010)G{\"u}ttler, Blum, Zsom, Ormel, \&
  Dullemond}]{guttler2010outcome}
G{\"u}ttler, C., Blum, J., Zsom, A., Ormel, C.~W., \& Dullemond, C.~P. 2010,
  Astronomy \& Astrophysics, 513, A56 \csname
  guttler2010outcomelink\endcsname~\csname guttler2010outcomenote\endcsname

\bibitem[{{Haisch} {et~al.}(2001){Haisch}, {Lada}, \& {Lada}}]{Haisch2001ApJ}
{Haisch}, Karl~E., J., {Lada}, E.~A., \& {Lada}, C.~J. 2001, The Astrophysical
  Journal, 553, L153 \csname Haisch2001ApJlink\endcsname~\csname
  Haisch2001ApJnote\endcsname

\bibitem[{Harter {et~al.}(2020)Harter, Ricci, Zhang, \&
  Zhu}]{harter2020imaging}
Harter, S.~K., Ricci, L., Zhang, S., \& Zhu, Z. 2020, The Astrophysical
  Journal, 905, 24 \csname harter2020imaginglink\endcsname~\csname
  harter2020imagingnote\endcsname

\bibitem[{Huang {et~al.}(2018)Huang, Andrews, Dullemond, Isella, P{\'e}rez,
  Guzm{\'a}n, {\"O}berg, Zhu, Zhang, Bai, {et~al.}}]{huang2018-DSHARP2}
Huang, J., Andrews, S.~M., Dullemond, C.~P., {et~al.} 2018, The Astrophysical
  Journal Letters, 869, L42 \csname huang2018-DSHARP2link\endcsname~\csname
  huang2018-DSHARP2note\endcsname

\bibitem[{{Ida} {et~al.}(2016){Ida}, {Guillot}, \& {Morbidelli}}]{Ida:2016}
{Ida}, S., {Guillot}, T., \& {Morbidelli}, A. 2016, Astronomy \& Astrophysics,
  591, A72 \csname Ida:2016link\endcsname~\csname Ida:2016note\endcsname

\bibitem[{Jiang {et~al.}(2024)Jiang, Mac{\'\i}as, Guerra-Alvarado, \&
  Carrasco-Gonz{\'a}lez}]{jiang2024grain}
Jiang, H., Mac{\'\i}as, E., Guerra-Alvarado, O.~M., \& Carrasco-Gonz{\'a}lez,
  C. 2024, Astronomy \& Astrophysics, 682, A32 \csname
  jiang2024grainlink\endcsname~\csname jiang2024grainnote\endcsname

\bibitem[{Jiang {et~al.}(2022)Jiang, Zhu, \& Ormel}]{jiang2022no}
Jiang, H., Zhu, W., \& Ormel, C.~W. 2022, The Astrophysical Journal Letters,
  924, L31 \csname jiang2022nolink\endcsname~\csname jiang2022nonote\endcsname

\bibitem[{Jim{\'e}nez \& Masset(2017)}]{jimenez2017improved}
Jim{\'e}nez, M.~A. \& Masset, F.~S. 2017, Monthly Notices of the Royal
  Astronomical Society, 471, 4917 \csname
  jimenez2017improvedlink\endcsname~\csname jimenez2017improvednote\endcsname

\bibitem[{Kanagawa {et~al.}(2021)Kanagawa, Muto, \& Tanaka}]{kanagawa2021dust}
Kanagawa, K.~D., Muto, T., \& Tanaka, H. 2021, The Astrophysical Journal, 921,
  169 \csname kanagawa2021dustlink\endcsname~\csname
  kanagawa2021dustnote\endcsname

\bibitem[{{Kurokawa} \& {Tanigawa}(2018)}]{Kurokawa:2018}
{Kurokawa}, H. \& {Tanigawa}, T. 2018, Monthly Notices of the Royal
  Astronomical Society, 479, 635 \csname Kurokawa:2018link\endcsname~\csname
  Kurokawa:2018note\endcsname

\bibitem[{{Kuwahara} \& {Kurokawa}(2020)}]{Kuwahara:2020a}
{Kuwahara}, A. \& {Kurokawa}, H. 2020, Astronomy \& Astrophysics, 633, A81
  \csname Kuwahara:2020alink\endcsname~\csname Kuwahara:2020anote\endcsname

\bibitem[{Kuwahara \& Kurokawa(2024)}]{kuwahara2024analytic}
Kuwahara, A. \& Kurokawa, H. 2024, Astronomy \& Astrophysics, 682, A14 \csname
  kuwahara2024analyticlink\endcsname~\csname kuwahara2024analyticnote\endcsname

\bibitem[{{Kuwahara} {et~al.}(2019){Kuwahara}, {Kurokawa}, \&
  {Ida}}]{Kuwahara:2019}
{Kuwahara}, A., {Kurokawa}, H., \& {Ida}, S. 2019, Astronomy \& Astrophysics,
  623, A179 \csname Kuwahara:2019link\endcsname~\csname
  Kuwahara:2019note\endcsname

\bibitem[{Kuwahara {et~al.}(2022)Kuwahara, Kurokawa, Tanigawa, \&
  Ida}]{kuwahara2022dust}
Kuwahara, A., Kurokawa, H., Tanigawa, T., \& Ida, S. 2022, Astronomy \&
  Astrophysics, 665, A122 \csname kuwahara2022dustlink\endcsname~\csname
  kuwahara2022dustnote\endcsname

\bibitem[{{Lambrechts} {et~al.}(2014){Lambrechts}, {Johansen}, \&
  {Morbidelli}}]{Lambrechts:2014}
{Lambrechts}, M., {Johansen}, A., \& {Morbidelli}, A. 2014, Astronomy \&
  Astrophysics, 572, A35 \csname Lambrechts:2014link\endcsname~\csname
  Lambrechts:2014note\endcsname

\bibitem[{Liu {et~al.}(2022)Liu, Johansen, Lambrechts, Bizzarro, \&
  Haugb{\o}lle}]{liu2022natural}
Liu, B., Johansen, A., Lambrechts, M., Bizzarro, M., \& Haugb{\o}lle, T. 2022,
  Science Advances, 8, eabm3045 \csname liu2022naturallink\endcsname~\csname
  liu2022naturalnote\endcsname

\bibitem[{Lodato {et~al.}(2019)Lodato, Dipierro, Ragusa, Long, Herczeg,
  Pascucci, Pinilla, Manara, Tazzari, Liu, {et~al.}}]{lodato2019newborn}
Lodato, G., Dipierro, G., Ragusa, E., {et~al.} 2019, Monthly Notices of the
  Royal Astronomical Society, 486, 453 \csname
  lodato2019newbornlink\endcsname~\csname lodato2019newbornnote\endcsname

\bibitem[{Long {et~al.}(2018)Long, Pinilla, Herczeg, Harsono, Dipierro,
  Pascucci, Hendler, Tazzari, Ragusa, Salyk, {et~al.}}]{long2018gaps}
Long, F., Pinilla, P., Herczeg, G.~J., {et~al.} 2018, The Astrophysical
  Journal, 869, 17 \csname long2018gapslink\endcsname~\csname
  long2018gapsnote\endcsname

\bibitem[{Luhman {et~al.}(2009)Luhman, Allen, Espaillat, Hartmann, \&
  Calvet}]{luhman2009disk}
Luhman, K., Allen, P., Espaillat, C., Hartmann, L., \& Calvet, N. 2009, The
  Astrophysical Journal Supplement Series, 186, 111 \csname
  luhman2009disklink\endcsname~\csname luhman2009disknote\endcsname

\bibitem[{Lynden-Bell \& Pringle(1974)}]{lynden1974evolution}
Lynden-Bell, D. \& Pringle, J.~E. 1974, Monthly Notices of the Royal
  Astronomical Society, 168, 603 \csname
  lynden1974evolutionlink\endcsname~\csname lynden1974evolutionnote\endcsname

\bibitem[{Mordasini {et~al.}(2018)Mordasini, Deeg, \&
  Belmonte}]{mordasini2018handbook}
Mordasini, C., Deeg, H., \& Belmonte, J. 2018, Cham: Springer, 143 \csname
  mordasini2018handbooklink\endcsname~\csname
  mordasini2018handbooknote\endcsname

\bibitem[{Mulders {et~al.}(2021)Mulders, Pascucci, Ciesla, \&
  Fernandes}]{mulders2021mass}
Mulders, G.~D., Pascucci, I., Ciesla, F.~J., \& Fernandes, R.~B. 2021, The
  Astrophysical Journal, 920, 66 \csname mulders2021masslink\endcsname~\csname
  mulders2021massnote\endcsname

\bibitem[{M{\"u}ller-Horn {et~al.}(2022)M{\"u}ller-Horn, Pichierri, \&
  Bitsch}]{muller2022emerging}
M{\"u}ller-Horn, J., Pichierri, G., \& Bitsch, B. 2022, Astronomy \&
  Astrophysics, 663, A163 \csname muller2022emerginglink\endcsname~\csname
  muller2022emergingnote\endcsname

\bibitem[{{Musiolik} {et~al.}(2016{\natexlab{a}}){Musiolik}, {Teiser},
  {Jankowski}, \& {Wurm}}]{Musiolik:2016a}
{Musiolik}, G., {Teiser}, J., {Jankowski}, T., \& {Wurm}, G.
  2016{\natexlab{a}}, The Astrophysical Journal, 818, 16 \csname
  Musiolik:2016alink\endcsname~\csname Musiolik:2016anote\endcsname

\bibitem[{{Musiolik} {et~al.}(2016{\natexlab{b}}){Musiolik}, {Teiser},
  {Jankowski}, \& {Wurm}}]{Musiolik:2016b}
{Musiolik}, G., {Teiser}, J., {Jankowski}, T., \& {Wurm}, G.
  2016{\natexlab{b}}, The Astrophysical Journal, 827, 63 \csname
  Musiolik:2016blink\endcsname~\csname Musiolik:2016bnote\endcsname

\bibitem[{{Muto} \& {Inutsuka}(2009)}]{Muto:2009}
{Muto}, T. \& {Inutsuka}, S.-i. 2009, The Astrophysical Journal, 695, 1132
  \csname Muto:2009link\endcsname~\csname Muto:2009note\endcsname

\bibitem[{Nakagawa {et~al.}(1986)Nakagawa, Sekiya, \& Hayashi}]{Nakagawa:1986}
Nakagawa, Y., Sekiya, M., \& Hayashi, C. 1986, Icarus, 67, 375 \csname
  Nakagawa:1986link\endcsname~\csname Nakagawa:1986note\endcsname

\bibitem[{Ndugu {et~al.}(2019)Ndugu, Bitsch, \& Jurua}]{ndugu2019observed}
Ndugu, N., Bitsch, B., \& Jurua, E. 2019, Monthly Notices of the Royal
  Astronomical Society, 488, 3625 \csname
  ndugu2019observedlink\endcsname~\csname ndugu2019observednote\endcsname

\bibitem[{Oka {et~al.}(2011)Oka, Nakamoto, \& Ida}]{oka2011evolution}
Oka, A., Nakamoto, T., \& Ida, S. 2011, The Astrophysical Journal, 738, 141
  \csname oka2011evolutionlink\endcsname~\csname oka2011evolutionnote\endcsname

\bibitem[{{Okuzumi} \& {Tazaki}(2019)}]{Okuzumi:2019}
{Okuzumi}, S. \& {Tazaki}, R. 2019, The Astrophysical Journal, 878, 132 \csname
  Okuzumi:2019link\endcsname~\csname Okuzumi:2019note\endcsname

\bibitem[{{Ormel} {et~al.}(2015){Ormel}, {Shi}, \& {Kuiper}}]{Ormel:2015b}
{Ormel}, C.~W., {Shi}, J.-M., \& {Kuiper}, R. 2015, Monthly Notices of the
  Royal Astronomical Society, 447, 3512 \csname
  Ormel:2015blink\endcsname~\csname Ormel:2015bnote\endcsname

\bibitem[{Paardekooper \& Mellema(2006)}]{paardekooper2006dust}
Paardekooper, S.-J. \& Mellema, G. 2006, Astronomy \& Astrophysics, 453, 1129
  \csname paardekooper2006dustlink\endcsname~\csname
  paardekooper2006dustnote\endcsname

\bibitem[{Pierens {et~al.}(2019)Pierens, Lin, \& Raymond}]{pierens2019vortex}
Pierens, A., Lin, M.-K., \& Raymond, S.~N. 2019, Monthly Notices of the Royal
  Astronomical Society, 488, 645 \csname
  pierens2019vortexlink\endcsname~\csname pierens2019vortexnote\endcsname

\bibitem[{Ribas {et~al.}(2024)Ribas, Clarke, \& Zagaria}]{ribas2024inner}
Ribas, {\'A}., Clarke, C.~J., \& Zagaria, F. 2024, Monthly Notices of the Royal
  Astronomical Society, 532, 1752 \csname ribas2024innerlink\endcsname~\csname
  ribas2024innernote\endcsname

\bibitem[{Ricci {et~al.}(2018)Ricci, Liu, Isella, \&
  Li}]{ricci2018investigating}
Ricci, L., Liu, S.-F., Isella, A., \& Li, H. 2018, The Astrophysical Journal,
  853, 110 \csname ricci2018investigatinglink\endcsname~\csname
  ricci2018investigatingnote\endcsname

\bibitem[{Rice {et~al.}(2006)Rice, Armitage, Wood, \& Lodato}]{rice2006dust}
Rice, W., Armitage, P.~J., Wood, K., \& Lodato, G. 2006, Monthly Notices of the
  Royal Astronomical Society, 373, 1619 \csname
  rice2006dustlink\endcsname~\csname rice2006dustnote\endcsname

\bibitem[{Rosotti {et~al.}(2016)Rosotti, Juhasz, Booth, \&
  Clarke}]{rosotti2016minimum}
Rosotti, G.~P., Juhasz, A., Booth, R.~A., \& Clarke, C.~J. 2016, Monthly
  Notices of the Royal Astronomical Society, 459, 2790 \csname
  rosotti2016minimumlink\endcsname~\csname rosotti2016minimumnote\endcsname

\bibitem[{Selina {et~al.}(2018)Selina, Murphy, McKinnon, Beasley, Butler,
  Carilli, Clark, Durand, Erickson, Grammer, {et~al.}}]{selina2018ngvla}
Selina, R.~J., Murphy, E.~J., McKinnon, M., {et~al.} 2018, Science with a Next
  Generation Very Large Array, 517, 15 \csname
  selina2018ngvlalink\endcsname~\csname selina2018ngvlanote\endcsname

\bibitem[{{Shakura} \& {Sunyaev}(1973)}]{Shakura:1973}
{Shakura}, N.~I. \& {Sunyaev}, R.~A. 1973, Astronomy \& Astrophysics, 24, 337
  \csname Shakura:1973link\endcsname~\csname Shakura:1973note\endcsname

\bibitem[{Sierra {et~al.}(2021)Sierra, P{\'e}rez, Zhang, Law, Guzm{\'a}n, Qi,
  Bosman, {\"O}berg, Andrews, Long, {et~al.}}]{sierra2021molecules}
Sierra, A., P{\'e}rez, L.~M., Zhang, K., {et~al.} 2021, The Astrophysical
  Journal Supplement Series, 257, 14 \csname
  sierra2021moleculeslink\endcsname~\csname sierra2021moleculesnote\endcsname

\bibitem[{Stone {et~al.}(2020)Stone, Tomida, White, \&
  Felker}]{stone2020athena++}
Stone, J.~M., Tomida, K., White, C.~J., \& Felker, K.~G. 2020, The
  Astrophysical Journal Supplement Series, 249, 4 \csname
  stone2020athena++link\endcsname~\csname stone2020athena++note\endcsname

\bibitem[{Tanaka {et~al.}(2002)Tanaka, Takeuchi, \& Ward}]{tanaka2002three}
Tanaka, H., Takeuchi, T., \& Ward, W.~R. 2002, The Astrophysical Journal, 565,
  1257 \csname tanaka2002threelink\endcsname~\csname
  tanaka2002threenote\endcsname

\bibitem[{Tanaka \& Ward(2004)}]{tanaka2004three}
Tanaka, H. \& Ward, W.~R. 2004, The Astrophysical Journal, 602, 388 \csname
  tanaka2004threelink\endcsname~\csname tanaka2004threenote\endcsname

\bibitem[{Tzouvanou {et~al.}(2023)Tzouvanou, Bitsch, \&
  Pichierri}]{tzouvanou2023all}
Tzouvanou, A., Bitsch, B., \& Pichierri, G. 2023, Astronomy \& Astrophysics,
  677, A82 \csname tzouvanou2023alllink\endcsname~\csname
  tzouvanou2023allnote\endcsname

\bibitem[{Ueda {et~al.}(2024)Ueda, Tazaki, Okuzumi, Flock, \&
  Sudarshan}]{ueda2024support}
Ueda, T., Tazaki, R., Okuzumi, S., Flock, M., \& Sudarshan, P. 2024, arXiv
  preprint arXiv:2406.07427 \csname ueda2024supportlink\endcsname~\csname
  ueda2024supportnote\endcsname

\bibitem[{van~der Marel \& Mulders(2021)}]{van2021stellar}
van~der Marel, N. \& Mulders, G.~D. 2021, The Astronomical Journal, 162, 28
  \csname van2021stellarlink\endcsname~\csname van2021stellarnote\endcsname

\bibitem[{Vericel \& Gonzalez(2020)}]{vericel2020self}
Vericel, A. \& Gonzalez, J.-F. 2020, Monthly Notices of the Royal Astronomical
  Society, 492, 210 \csname vericel2020selflink\endcsname~\csname
  vericel2020selfnote\endcsname

\bibitem[{Vericel {et~al.}(2021)Vericel, Gonzalez, Price, Laibe, \&
  Pinte}]{vericel2021dust}
Vericel, A., Gonzalez, J.-F., Price, D.~J., Laibe, G., \& Pinte, C. 2021,
  Monthly Notices of the Royal Astronomical Society, 507, 2318 \csname
  vericel2021dustlink\endcsname~\csname vericel2021dustnote\endcsname

\bibitem[{Villenave {et~al.}(2022)Villenave, Stapelfeldt, Duch{\^e}ne,
  M{\'e}nard, Lambrechts, Sierra, Flores, Dent, Wolff, Ribas,
  {et~al.}}]{villenave2022highly}
Villenave, M., Stapelfeldt, K., Duch{\^e}ne, G., {et~al.} 2022, The
  Astrophysical Journal, 930, 11 \csname
  villenave2022highlylink\endcsname~\csname villenave2022highlynote\endcsname

\bibitem[{Wang {et~al.}(2021)Wang, Kanagawa, \& Suto}]{wang2021architecture}
Wang, S., Kanagawa, K.~D., \& Suto, Y. 2021, The Astrophysical Journal, 923,
  165 \csname wang2021architecturelink\endcsname~\csname
  wang2021architecturenote\endcsname

\bibitem[{Ward(1986)}]{ward1986density}
Ward, W.~R. 1986, icarus, 67, 164 \csname ward1986densitylink\endcsname~\csname
  ward1986densitynote\endcsname

\bibitem[{Weber {et~al.}(2018)Weber, Ben{\'\i}tez-Llambay, Gressel, Krapp, \&
  Pessah}]{weber2018characterizing}
Weber, P., Ben{\'\i}tez-Llambay, P., Gressel, O., Krapp, L., \& Pessah, M.~E.
  2018, The Astrophysical Journal, 854, 153 \csname
  weber2018characterizinglink\endcsname~\csname
  weber2018characterizingnote\endcsname

\bibitem[{{Weidenschilling}(1977)}]{Weidenschilling:1977}
{Weidenschilling}, S.~J. 1977, Astrophysics and Space Science, 51, 153 \csname
  Weidenschilling:1977link\endcsname~\csname Weidenschilling:1977note\endcsname

\bibitem[{{Weiss} \& {Marcy}(2014)}]{Weiss:2014}
{Weiss}, L.~M. \& {Marcy}, G.~W. 2014, The Astrophysical Journal, 783, L6
  \csname Weiss:2014link\endcsname~\csname Weiss:2014note\endcsname

\bibitem[{Yang \& Zhu(2020)}]{yang2020morphological}
Yang, C.-C. \& Zhu, Z. 2020, Monthly Notices of the Royal Astronomical Society,
  491, 4702 \csname yang2020morphologicallink\endcsname~\csname
  yang2020morphologicalnote\endcsname

\bibitem[{{Youdin} \& {Lithwick}(2007)}]{Youdin:2007}
{Youdin}, A.~N. \& {Lithwick}, Y. 2007, Icarus, 192, 588 \csname
  Youdin:2007link\endcsname~\csname Youdin:2007note\endcsname

\bibitem[{Zhang {et~al.}(2021)Zhang, Booth, Law, Bosman, Schwarz, Bergin,
  {\"O}berg, Andrews, Guzm{\'a}n, Walsh, {et~al.}}]{zhang2021molecules}
Zhang, K., Booth, A.~S., Law, C.~J., {et~al.} 2021, The Astrophysical Journal
  Supplement Series, 257, 5 \csname zhang2021moleculeslink\endcsname~\csname
  zhang2021moleculesnote\endcsname

\bibitem[{Zhang {et~al.}(2023)Zhang, Kalscheur, Long, Zhang, Long, Bergin, Zhu,
  \& Trapman}]{zhang2023substructures}
Zhang, S., Kalscheur, M., Long, F., {et~al.} 2023, The Astrophysical Journal,
  952, 108 \csname zhang2023substructureslink\endcsname~\csname
  zhang2023substructuresnote\endcsname

\bibitem[{Zhang {et~al.}(2018)Zhang, Zhu, Huang, Guzm{\'a}n, Andrews,
  Birnstiel, Dullemond, Carpenter, Isella, P{\'e}rez,
  {et~al.}}]{zhang2018-DSHARP7}
Zhang, S., Zhu, Z., Huang, J., {et~al.} 2018, The Astrophysical Journal
  Letters, 869, L47 \csname zhang2018-DSHARP7link\endcsname~\csname
  zhang2018-DSHARP7note\endcsname

\bibitem[{Zhu {et~al.}(2012)Zhu, Nelson, Dong, Espaillat, \&
  Hartmann}]{zhu2012dust}
Zhu, Z., Nelson, R.~P., Dong, R., Espaillat, C., \& Hartmann, L. 2012, The
  Astrophysical Journal, 755, 6 \csname zhu2012dustlink\endcsname~\csname
  zhu2012dustnote\endcsname

\bibitem[{Zhu {et~al.}(2014)Zhu, Stone, Rafikov, \& Bai}]{zhu2014particle}
Zhu, Z., Stone, J.~M., Rafikov, R.~R., \& Bai, X.-n. 2014, The Astrophysical
  Journal, 785, 122 \csname zhu2014particlelink\endcsname~\csname
  zhu2014particlenote\endcsname

\bibitem[{Zsom {et~al.}(2010)Zsom, Ormel, G{\"u}ttler, Blum, \&
  Dullemond}]{zsom2010outcome}
Zsom, A., Ormel, C.~W., G{\"u}ttler, C., Blum, J., \& Dullemond, C. 2010,
  Astronomy \& Astrophysics, 513, A57 \csname
  zsom2010outcomelink\endcsname~\csname zsom2010outcomenote\endcsname

\end{thebibliography}

\clearpage
\begin{appendix}
%\appendix
%\setcounter{section}{1}
\def\thesection{A}
\setcounter{equation}{0}
\def\theequation{A.\arabic{equation}}
\setcounter{figure}{0}
\def\thefigure{A.\arabic{figure}}

\section{Conversion to dimensional quantities}\label{sec:Conversion to dimensional quantities}
It is practical to convert dimensionless quantities into dimensional ones for discussion. The following sections describe the method of the conversion for a given disk model.

\begin{figure}
    \centering
    \includegraphics[width=1\linewidth]{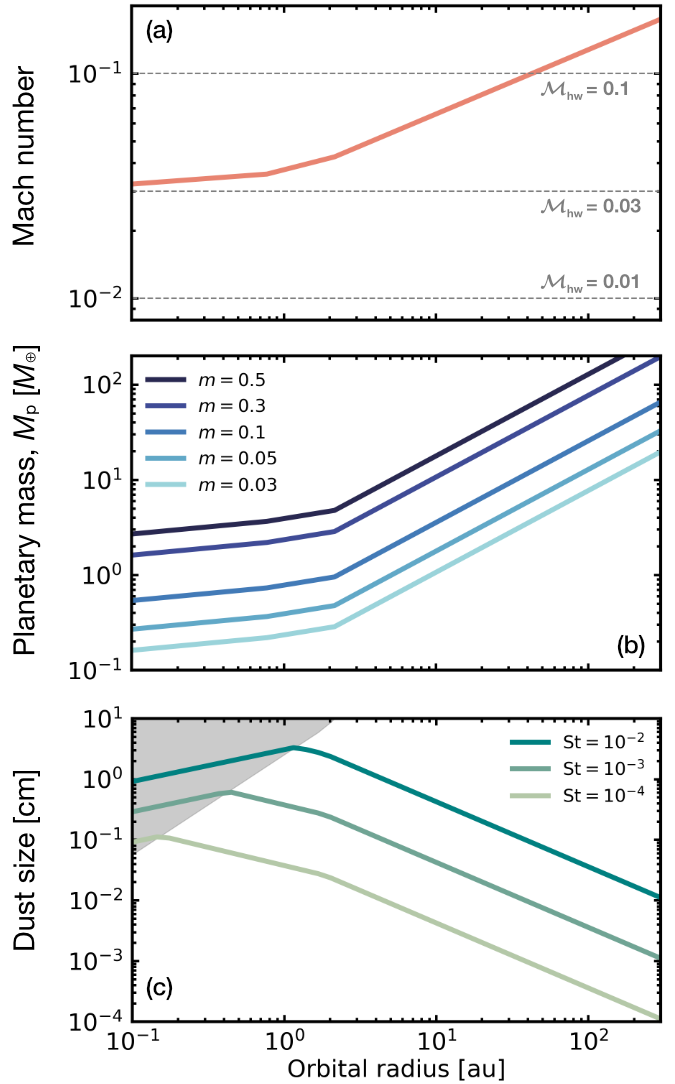}
    \caption{\add{Mach number, planetary mass, and the dust size as a function of the orbital radius. The gray shaded region in \textit{panel c} denotes the Stokes regime.}}
    \label{fig:params}
\end{figure}

\subsection{Disk model}\label{sec:Disk model}
We considered the typical steady accretion disk model with a dimensionless viscous alpha parameter \citep{Shakura:1973}, $\alpha_{\rm acc}$, including viscous heating due to the gas accretion and stellar irradiation heating \citep[e.g.,][]{Ida:2016}. For simplicity, we fixed the stellar mass, the stellar luminosity, the mass accretion rate, and the viscous alpha parameter as $M_\ast=1M_\odot,\,L_\ast=1L_\odot,\,\dot{M}_\ast=10^{-8}M_\odot\text{/yr},$ and $\,\alpha_{\rm acc}=10^{-3}$.

The disk midplane temperature is given by $T_{\rm disk}=\max(T_{\rm vis},\,T_{\rm irr})$, where $T_{\rm vis}$ and $T_{\rm irr}$ are temperatures determined by viscous heating and stellar irradiation \citep{garaud2007effect,oka2011evolution,Ida:2016},
\begin{align}
    T_{\rm vis}\simeq&200\,\Biggl(\frac{M_{\ast}}{1\,M_{\odot}}\Biggr)^{3/10}\Biggl(\frac{\alpha_{\rm acc}}{10^{-3}}\Biggr)^{-1/5}\Biggl(\frac{\dot{M}_{\ast}}{10^{-8}\,M_{\odot}/\text{yr}}\Biggr)^{2/5}\Biggl(\frac{r}{1\,\text{au}}\Biggr)^{-9/10}\,{\rm K},\label{eq:T vis}\\
    T_{\rm irr}\simeq&150\,\Biggl(\frac{L_{\ast}}{1\,L_{\odot}}\Biggr)^{2/7}\Biggl(\frac{M_{\ast}}{1\,M_{\odot}}\Biggr)^{-1/7}\Biggl(\frac{r}{1\,\text{au}}\Biggr)^{-3/7}\,{\rm K}.\label{eq:T irr}
\end{align}

The disk gas scale height is given by:
\begin{align}
    H=\frac{c_{\rm s}}{\Omega}=\sqrt{\frac{k_{\rm B}T_{\rm disk}}{\mu m_{\rm p}}}\frac{1}{\Omega},\label{eq:H gas}
\end{align}
where $k_{\rm B}$ is the Boltzmann constant, $\mu=2.34$ is the mean molecular weight, and $m_{\rm p}$ is the proton mass. Thus, the aspect ratio of the disk is given by
\begin{align}
    h\equiv\frac{H}{r}=\max(h_{\rm g,vis},\,h_{\rm g,irr}),\label{eq:aspect ratio}
\end{align}
where 
\begin{align}
    h_{\rm g,vis}\simeq&0.027\,\Biggl(\frac{M_{\ast}}{M_{\odot}}\Biggr)^{-7/20}\Biggl(\frac{\alpha_{\rm acc}}{10^{-3}}\Biggr)^{-1/10}\Biggl(\frac{\dot{M}_{\ast}}{10^{-8}\,M_{\odot}/\text{yr}}\Biggr)^{1/5}\Biggl(\frac{r}{1\,\text{au}}\Biggr)^{1/20},\label{eq:h vis}\\
    h_{\rm g,irr}\simeq&0.024\,\Biggl(\frac{L_{\ast}}{L_{\odot}}\Biggr)^{1/7}\Biggl(\frac{M_{\ast}}{M_{\odot}}\Biggr)^{-4/7}\Biggl(\frac{r}{1\,\text{au}}\Biggr)^{2/7}.\label{eq:h irr}
\end{align} 
The gas surface density is given by:
\begin{align}
    \Sigma_{\rm g}&=\frac{\dot{M}_\ast}{3\pi\alpha_{\rm acc}H^2\Omega_{\rm K}}=\min(\Sigma_{\rm g,vis},\Sigma_{\rm g,irr}),\\
    \label{eq:Sigma g}
\end{align}
where
\footnotesize
\begin{align}
&\Sigma_{\rm g,vis}\simeq2.1\times10^3\,\text{g/cm}^2\,\Bigg(\frac{0.027}{h}\Bigg)^{-2}\Biggl(\frac{M_{\ast}}{M_{\odot}}\Bigg)^{1/5}\Bigg(\frac{\alpha_{\rm acc}}{10^{-3}}\Bigg)^{-4/5}\Bigg(\frac{\dot{M}{\ast}}{10^{-8}\,M{\odot}/\text{yr}}\Bigg)^{3/5}\Bigg(\frac{r}{1\,\text{au}}\Biggr)^{-3/5},\\
&\Sigma_{\rm g,irr}\simeq2.7\times10^3\,\text{g/cm}^2\,\Biggl(\frac{L_{\ast}}{L_{\odot}}\Biggr)^{-2/7}\Biggl(\frac{M_{\ast}}{M_{\odot}}\Bigg)^{9/14}\Bigg(\frac{\alpha_{\rm acc}}{10^{-3}}\Bigg)^{-1}\Bigg(\frac{\dot{M}{\ast}}{10^{-8}\,M{\odot}/\text{yr}}\Bigg)\Biggl(\frac{r}{1\,\text{au}}\Biggr)^{-15/14}.
\end{align}
\normalsize

\subsection{Orbital radius of the planet}\label{sec:Orbital radius}
When we convert a dimensionless quantity into a dimensionless one, we need to determine the orbital distance of the planet, $r_{\rm p}$. To do this, we specify $r_{\rm p}$ based on the value of the Mach number of the headwind, $\mathcal{M}_{\rm hw}$.

From \Equref{eq:aspect ratio}, the Mach number of the headwind is given by:
\begin{align}
    \mathcal{M}_{\rm hw}&=-\frac{h}{2}\frac{\mathrm{d}\ln p}{\mathrm{d}\ln r}
    \simeq\max\Bigg(0.034\,\biggl(\frac{r_{\rm p}}{1\,\text{au}}\biggr)^{1/20},0.033\biggl(\frac{r_{\rm p}}{1\,\text{au}}\biggr)^{2/7}\Bigg),\label{eq:Mhw value}
\end{align}
where we set $\mathrm{d}\ln p/\mathrm{d}\ln r=-2.55$ for the viscous heating regime and $\mathrm{d}\ln p/\mathrm{d}\ln r=-2.78$ for the irradiation heating regime, respectively \citep{Ida:2016}. From \Equref{eq:Mhw value}, we can set $r_{\rm p}\simeq1$ au when $\mathcal{M}_{\rm hw}=0.03$ ($r_{\rm p}\simeq50$ au when $\mathcal{M}_{\rm hw}=0.1$) as a reference value of the planet location \add{(\Figref{fig:params}a)}. %In Sect. \ref{sec:Discussion}, we show the results of 1 au and 50 au cases.

\subsection{Planet mass}\label{sec:Planet mass}
In the disk model given in Sect. \ref{sec:Disk model}, the planet mass is calculated by:
\begin{align}
    M_{\rm p}&=mM_\ast h^3,\\
    &\simeq\max\Bigg(6.6\,m\biggl(\frac{r_{\rm p}}{1\,\text{au}}\biggr)^{3/20},\,4.6\,m\biggl(\frac{r_{\rm p}}{1\,\text{au}}\biggr)^{6/7}\Bigg)\,M_\oplus\label{eq:Mpl}\\
    &\simeq
    \begin{cases}
    0.66\,\bigg(\frac{m}{0.1}\bigg)\,M_\oplus\quad(r_{\rm p}=1\text{ au}),\\
    13\,\bigg(\frac{m}{0.1}\bigg)\,M_\oplus\quad(r_{\rm p}=50\text{ au}).
    \end{cases}
\end{align}
The pebble isolation mass is given by \citep{Bitsch:2018}:
\begin{align}
    M_{\rm iso}&=25\,M_\oplus\,\Bigg(\frac{h}{0.05}\Bigg)^3\Bigg[0.34\Bigg(\frac{3}{\log\alpha_{\rm diff}}\Bigg)^4+0.66\Bigg].\label{eq:Miso}
\end{align}
\add{Figure \ref{fig:params}b shows $M_{\rm p}$ for different $m$ as a function of the orbital radius.}

\subsection{Dust size}\label{sec:Dust size}
From \Equref{eq:Stokes number}, the physical size of dust, $s$, is calculated by:
\begin{align}
s&=\min\Bigg(\frac{\Sigma_{\rm g}{\rm St}}{\sqrt{2\pi}\rho_\bullet},\bigg(\frac{9\mu m_{\rm H}H{\rm St}}{4\rho_\bullet\sigma_{\rm mol}}\bigg)^{1/2}\Bigg).\label{eq:dust size s}
\end{align}
In \Equref{eq:dust size s}, we used the following equations for the stopping time of dust:
\begin{empheq}
    [left={t_{\rm stop}=\empheqlbrace}]{alignat=2}
    &\frac{\sqrt{2\pi}\rho_\bullet s}{\Sigma_{\rm g}\Omega}\quad (s\leq9\lambda/4\text{: Epstein regime}),\\
    &\frac{4\rho_\bullet\sigma_{\rm mol}s^2}{9\mu m_{\rm p}H\Omega}\quad (s>9\lambda/4\text{: Stokes regime}),\label{eq:t stop}
\end{empheq}
where $\rho_\bullet=2\,\text{g/cm}^3$ is the internal density of dust and $\sigma_{\rm mol}=2\times10^{-15}\,\text{cm}^2$ is the molecular collision cross section.

In our disk model, the gas drag regime switches from the Stokes to the Epstein regime when ${\rm St}\leq6.6\times10^{-3}$ at $r_{\rm p}=1$ au (${\rm St}\leq2.2\times10^3$ at 50 au). In the Epstein regime, we have
\footnotesize\add{
\begin{empheq}  
    [left={s\simeq\empheqlbrace}]{alignat=2}
    &3.7\text{ mm}\,\Bigg(\frac{\rm St}{10^{-3}}\Bigg)\Bigg(\frac{\Sigma_{\rm g}}{2.1\times10^3\,\text{g/cm}^2}\Bigg)\Bigg(\frac{\rho_\bullet}{2\,\text{g/cm}^3}\Bigg)^{-1}\quad (r_{\rm p}=1\text{ au}),\\
    &0.076\text{ mm}\,\Bigg(\frac{\rm St}{10^{-3}}\Bigg)\Bigg(\frac{\Sigma_{\rm g}}{41\,\text{g/cm}^2}\Bigg)\Bigg(\frac{\rho_\bullet}{2\,\text{g/cm}^3}\Bigg)^{-1}\quad (r_{\rm p}=50\text{ au}).
\end{empheq}}
\normalsize
\add{Figure \ref{fig:params}c shows the dust size for different St as a function of the orbital radius.}

\subsection{Time}\label{sec:Time}
\rev{The orbital period is given by:
\begin{align}
    T_{\rm orb}\simeq\Bigg(\frac{t}{2\pi}\Bigg)\Bigg(\frac{r_{\rm p}}{1\,\text{au}}\Bigg)^{3/2}\Bigg(\frac{M_{\ast}}{M_\odot}\Bigg)^{-1/2}\,\text{yr},\label{eq:time}
\end{align}
where $t$ is the dimensionless time in our simulations.}
%T_{\rm orb}=\frac{2\pi}{\Omega}t\simeq\bigg(\frac{r_{\rm p}}{1\,\text{au}}\bigg)^{3/2}\,t\,\text{yr},\label{eq:time}

\subsection{Length scale}\label{sec:Length scale}
For each dust surface density simulation, we used a 1D simulation domain with the constant spatial intervals, $\Delta x=0.01\,H$ (Sect. \ref{sec:Dust surface density calculation}). For a given orbital distance of the planet, $r_{\rm p}$, we convert the units of the grid cell from $H$ to au based on the following methods. We defined the radial distance of the $i$-th grid from the central star in au units as $r_i$, where $i=0$ corresponds to the planet location and $i=1,2,\dots$ ($i=-1,-2,\dots$) corresponds to the position outside (inside) the planetary orbit. Assuming that the disk aspect ratio is constant, we computed $r_i$ by the following equation: %Here we define the location of the grid cell as $x_i$, where $i=0$ corresponds to the planet location and $i=1,2,\dots$ ($i=-1,-2,\dots$) corresponds to the position outside (inside) the planetary orbit. %Although the disk gas scale height is a function of the orbital radius, $H(r)$ (\Equref{eq:aspect ratio}), assuming that the gas scale height is constant in the small spatial range, we compute $r_i$ by the following equation: 
\begin{align}
    r_0=&r_{\rm p},\\
    r_i\simeq&
    \begin{cases}
    r_{i-1}\left(1+h(r_{\rm p})\times\Delta x\right)\quad(i=1,2,\dots),\\
    r_{i+1}\left(1-h(r_{\rm p})\times\Delta x\right)\quad(i=-1,-2,\dots).\\
    \end{cases}
\end{align}

%============================================
\def\thesection{B}
\setcounter{equation}{0}
\def\theequation{B.\arabic{equation}}
\setcounter{figure}{0}
\def\thefigure{B.\arabic{figure}}

\begin{figure}
    \centering
    \includegraphics[width=1\linewidth]{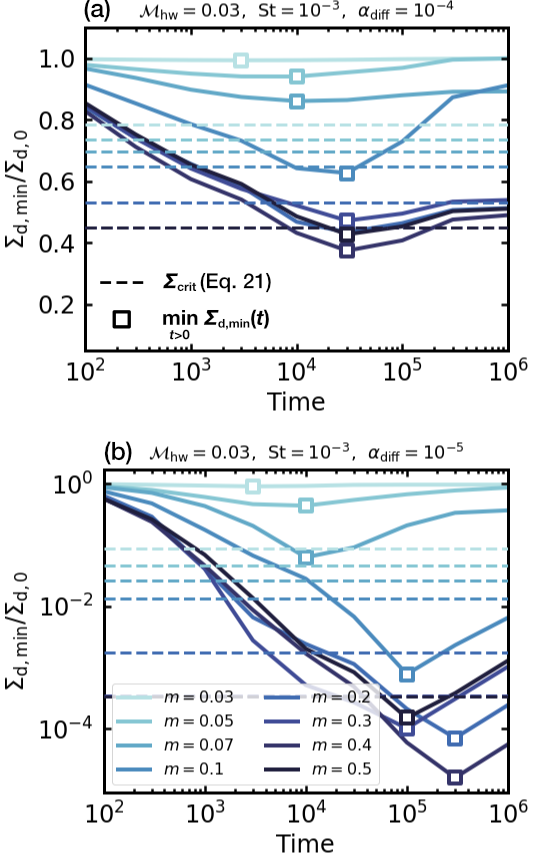}
    \caption{Time evolution of the minimum dust surface density for different planetary masses. We fixed the Stokes number and the Mach number ${\rm St}=10^{-3}$ and $\mathcal{M}_{\rm hw}=0.03$. We set $\alpha_{\rm diff}=10^{-4}$ in \textit{panel a} and $\alpha_{\rm diff}=10^{-5}$ in \textit{panel b}. The square symbols denote the global minimum of $\Sigma_{\rm d,min}(t)$. The horizontal dashed lines mark $\Sigma_{\rm crit}$ (Eq. \ref{eq:Sigma crit}).}
    \label{fig:Smin_time}
\end{figure}

\begin{figure}
    \centering
    \includegraphics[width=1\linewidth]{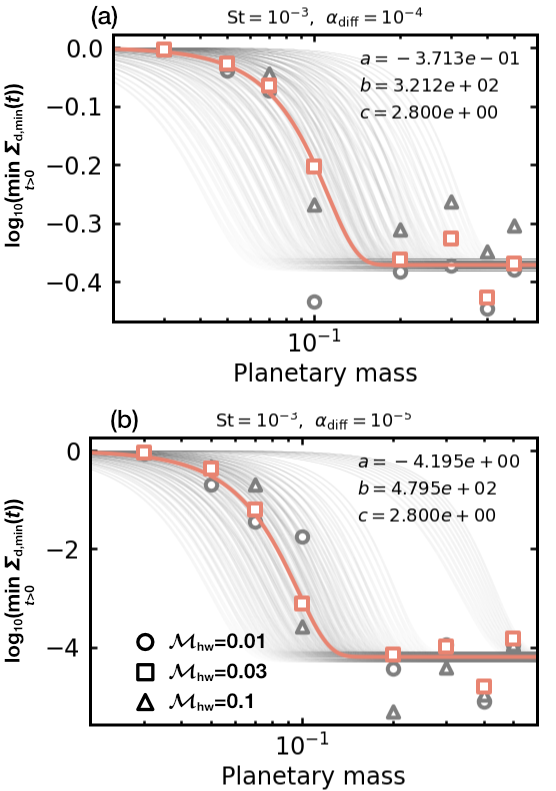}
    \caption{Global minimum of the dust surface density as a function of the planetary mass. We fixed the Stokes number ${\rm St}=10^{-3}$. We set $\alpha_{\rm diff}=10^{-4}$ in \textit{panel a} and $\alpha_{\rm diff}=10^{-5}$ in \textit{panel b}. Different symbols correspond to different Mach numbers. The \rev{red} solid line is the fitting formula for the numerical results of $\mathcal{M}_{\rm hw}=0.03$ (Eq. \ref{eq:sigma min fit appendix}). \rev{The gray thin lines show the uncertainties of \Equref{eq:sigma min fit appendix}.}}
    \label{fig:Smin_fit}
\end{figure}

\begin{figure}
    \centering
    \includegraphics[width=1\linewidth]{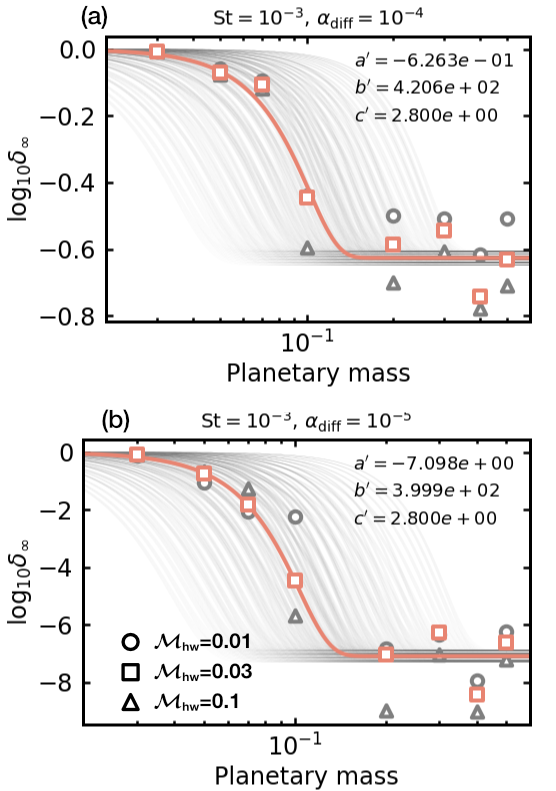}
    \caption{Steady-state dust gap depth as a function of the planetary mass. We fixed the Stokes number ${\rm St}=10^{-3}$. We set $\alpha_{\rm diff}=10^{-4}$ in \textit{panel a} and $\alpha_{\rm diff}=10^{-5}$ in \textit{panel b}. Different symbols correspond to different Mach numbers. The \rev{red} solid line is the fitting formula for the numerical results of $\mathcal{M}_{\rm hw}=0.03$ (Eq. \ref{eq:delta inf fit appendix}). \rev{The gray thin lines show the uncertainties of \Equref{eq:delta inf fit appendix}.}}
    \label{fig:delta_inf}
\end{figure}

\begin{figure}
    \centering
    \includegraphics[width=1\linewidth]{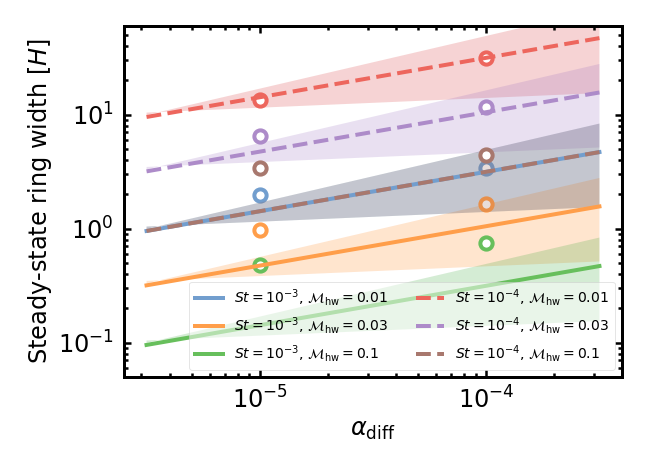}
    \caption{Steady-state dust ring width as a function of the turbulent parameter. The numerical results for different planetary masses were averaged and plotted with the circle symbols. The solid \rev{and dashed} \addsec{lines are} given by \Equref{eq:W ring fit}. \rev{The shaded regions show the uncertainties.}}
    \label{fig:ring_width_inf}
\end{figure}

\begin{figure}
    \centering
    \includegraphics[width=1\linewidth]{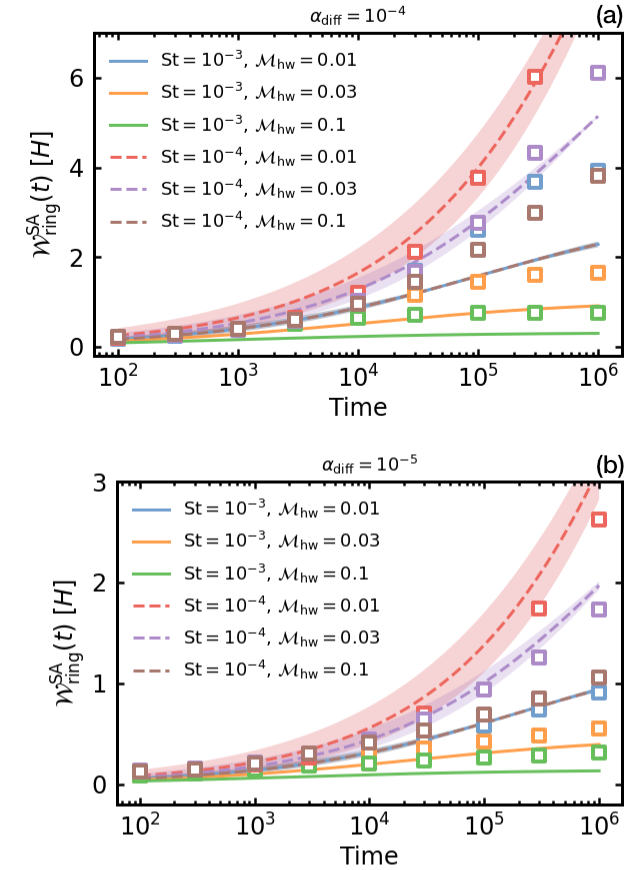}
    \caption{Time-dependent dust ring width as a function of time. The numerical results for different planetary masses were averaged and plotted with the square symbols. The solid \rev{and dashed} \addsec{lines are} given by \Equref{eq: W ring SA appendix}. \rev{The shaded regions show the uncertainties.}}
    \label{fig:ringwidth_time_fit}
\end{figure}

%---------------------
\section{Fitting formulae for $\Sigma_{\rm min}^{\rm fit}$, $\delta_\infty^{\rm fit}$, and $\mathcal{W}_{\rm ring,\infty}^{\rm fit}$}\label{sec:Appendix Fitting formulae}
We introduce the fitting formulae for the global minimum of the time-dependent dust surface density, $\Sigma_{\rm min}^{\rm fit}$, the steady-state dust gap depth, $\delta_\infty^{\rm fit}$, and the steady-state dust ring width, $\mathcal{W}_{\rm ring,\infty}^{\rm fit}$. For the fitting processes, we used the numerical results of ${\rm St}\leq10^{-3}$. 

Figure \ref{fig:Smin_time} shows the time evolution of the minimum dust surface density for different planetary masses, $\Sigma_{\rm d,min}(t)$, \add{obtained from our numerical simulations}. The minimum dust surface density has the complex dependence on time. We marked the global minimum of the time-dependent dust surface density, $\displaystyle\min_{t>0}\Sigma_{\rm d,min}(t)$, with the square symbol in \Figref{fig:Smin_time}. We \add{considered} that when $\displaystyle\min_{t>0}\Sigma_{\rm d,min}(t)<
\Sigma_{\rm crit}$ the dust gap expands with time (Sect. \ref{sec:Dust gap width}). %This condition is satisfied when $m\geq0.07$ and $\alpha_{\rm diff}=10^{-5}$ (\Figref{fig:Smin_time}b). 

Figure \ref{fig:Smin_fit} shows $\displaystyle\min_{t>0}\Sigma_{\rm d,min}(t)$ as a function of the planetary mass. We found that the dependence of $\displaystyle\min_{t>0}\Sigma_{\rm d,min}(t)$ on the Mach number is weak. Thus, for the fitting process we used the numerical results of $\mathcal{M}_{\rm hw}=0.03$ as the representative value.

We assumed that the numerical result can be fitted by the following sigmoid curve:\rev{
\begin{align}
    &\log_{10}\Sigma_{\rm min}^{\rm fit}=a\times{\rm erf}(|b|m^c).\label{eq:fitting eq1}
\end{align}
Using the \texttt{scipy.optimize.curve\_fit} library \add{of python}, we constrained the fitting coefficients in \Equref{eq:fitting eq1}: $a,\,b,$ and $c$. We obtained
\begin{align}
    \begin{cases}
        a=-0.3713\pm9.512\times10^{-3},\\
        b=321.2\pm455.3,\\
        c=2.8\pm0.5927,
    \end{cases}\label{eq:Smin fit 1}
\end{align}
when $\alpha_{\rm diff}=10^{-4}$ and
\begin{align}
    \begin{cases}
        a=-4.195\pm9.930\times10^{-2},\\
        b=479.5\pm469.5,\\
        c=2.8\pm0.4009,
    \end{cases}\label{eq:Smin fit 2}
\end{align}
when $\alpha_{\rm diff}=10^{-5}$. From Eqs. (\ref{eq:Smin fit 1}) and (\ref{eq:Smin fit 2}), we obtained
\begin{align}
    \log_{10}\Sigma_{\rm min}^{\rm fit}=-0.37\,\bigg(\frac{\alpha_{\rm diff}}{10^{-4}}\bigg)^{-1.1}\times\mathrm{erf}\Bigg(3.2\times10^2\,\bigg(\frac{\alpha_{\rm diff}}{10^{-4}}\bigg)^{-0.17}m^{2.8}\Bigg),\label{eq:sigma min fit appendix}
\end{align}
where we used the best fit values of the fitting coefficients.} Equation (\ref{eq:fitting eq1}) gives $\Sigma_{\rm min}^{\rm fit}>1$ when $\alpha_{\rm diff}\gtrsim10^{-4}$, which is \add{unphysical}. Thus, we set the upper limit and obtained the semi-analytic formula for the global minimum of the dust surface density:\add{
\begin{align}
    \Sigma_{\rm min}^{\rm global}=\min(1,\,\Sigma_{\rm min}^{\rm fit}).\label{eq:Smin fit}
\end{align}}
We plotted \Equref{eq:Smin fit} in \Figref{fig:Smin_fit} with the solid line. 

Figure \ref{fig:delta_inf} shows the steady-state dust gap depth as a function of the planetary mass. Similar to the fitting process of $\Sigma_{\rm min}^{\rm fit}$, we used the numerical results of $\mathcal{M}_{\rm hw}=0.03$ as the representative value for the fitting and assumed that the numerical result can be fitted by:\rev{
\begin{align}
    \log_{10}\delta_{\infty}^{\rm fit}=a^{\prime}\times{\rm erf}(|b^{\prime}|m^{c^{\prime}}).\label{eq:fitting eq3}
\end{align}
Using the \texttt{scipy.optimize.curve\_fit} library, we obtained the fitting coefficients $a^{\prime},b^{\prime},$ and $c^{\prime}$ as follows:
\begin{align}
    \begin{cases}
        a^{\prime}=-0.6263\pm2.123\times10^{-3},\\
        b^{\prime}=420.6\pm639.6,\\
        c^{\prime}=2.8\pm0.6286,
    \end{cases}\label{eq:delta inf fit 1}
\end{align}
when $\alpha_{\rm diff}=10^{-4}$ and
\begin{align}
    \begin{cases}
        a^{\prime}=-7.098\pm0.2133,\\
        b^{\prime}=399.9\pm558.9,\\
        c^{\prime}=2.8\pm0.5792,
    \end{cases}\label{eq:delta inf fit 2}
\end{align}
when $\alpha_{\rm diff}=10^{-5}$. From Eqs. (\ref{eq:delta inf fit 1}) and (\ref{eq:delta inf fit 2}), we obtained
\begin{align}
        \log_{10}\delta_{\infty}^{\rm fit}=-0.63\,\bigg(\frac{\alpha_{\rm diff}}{10^{-4}}\bigg)^{-1.1}\times\mathrm{erf}\Bigg(4.2\times10^2\,\bigg(\frac{\alpha_{\rm diff}}{10^{-4}}\bigg)^{0.022}\,m^{2.8}\Bigg).\label{eq:delta inf fit appendix}
\end{align}
Equation} (\ref{eq:fitting eq3}) gives $\delta_\infty^{\rm fit}>1$ when $\alpha_{\rm diff}\gtrsim10^{-4}$. To avoid unphysical solutions, we set the upper limit:
\begin{align}
    \delta_\infty=\min(\delta_0,\delta_\infty^{\rm fit}).\label{eq:delta inf fit}
\end{align}
We plotted \Equref{eq:delta inf fit} in \Figref{fig:delta_inf} with the solid line.

Figure \ref{fig:ring_width_inf} shows the steady-state dust ring width as a function of $\alpha_{\rm diff}$. As shown in \Figref{fig:ringwidth}, the dust ring width is weakly dependent on the planetary mass. Thus, we averaged the numerical results for different planetary masses and plotted them in \Figref{fig:ring_width_inf} with the circle symbols. We assumed that the steady-state dust ring width is proportional to the characteristic length in which the drift timescale coincides with the diffusion timescale, $\mathcal{L}_{\rm eq}$ (Eq. \ref{eq:L eq}): $\mathcal{W}_{\rm ring,\infty}^{\rm fit}=C\times\mathcal{L}_{\rm eq}$. With the least-squares method, we \add{derived} the coefficient:\rev{
\begin{align}
    C=
    \begin{cases}
        0.63\pm0.35\quad(\alpha_{\rm diff}=10^{-4}),\\
        2.8\pm0.54\quad(\alpha_{\rm diff}=10^{-5}),
    \end{cases}\label{eq:delta inf fit 2}
\end{align}
and} then obtained
\begin{align}
    \mathcal{W}_{\rm ring,\infty}^{\rm fit}=0.63\,\Bigg(\frac{\alpha_{\rm diff}}{10^{-4}}\Bigg)^{-0.65}\times\mathcal{L}_{\rm eq}.\label{eq:W ring fit}
\end{align}
We plotted \Equref{eq:W ring fit} in \Figref{fig:ring_width_inf} with the solid lines.

Figure \ref{fig:ringwidth_time_fit} shows the dust ring width as a function of time. Since the dust ring width is weakly dependent on the planetary mass, same as in \Figref{fig:ring_width_inf}, we averaged the numerical results for different planetary masses at each time and then plotted them in \Figref{fig:ringwidth_time_fit} with the squares symbols. We assumed that the time-dependent dust ring width can be fitted by the following sigmoid curve:
\begin{align}
    \mathcal{W}_{\rm ring}^{\rm SA}(t)=\mathcal{W}_{\rm ring,\infty}^{\rm fit}\Bigg(1-\frac{1}{1+(t/\tau_{\rm ring})^q}\Bigg).\label{eq: W ring SA appendix}
\end{align}
With the least-squares method, we determined the power of \addsec{$(t/\tau_{\rm ring})^q$: \rev{$q=0.42\pm0.045$}}. We plotted \Equref{eq: W ring SA appendix} in \Figref{fig:ringwidth_time_fit} with the solid lines.

%--------------------------------
\def\thesection{C}
\setcounter{equation}{0}
\def\theequation{C.\arabic{equation}}
\setcounter{figure}{0}
\def\thefigure{C.\arabic{figure}}

\section{Additional figures}

\begin{figure*}[htbp]
    \centering
    \includegraphics[width=\linewidth]{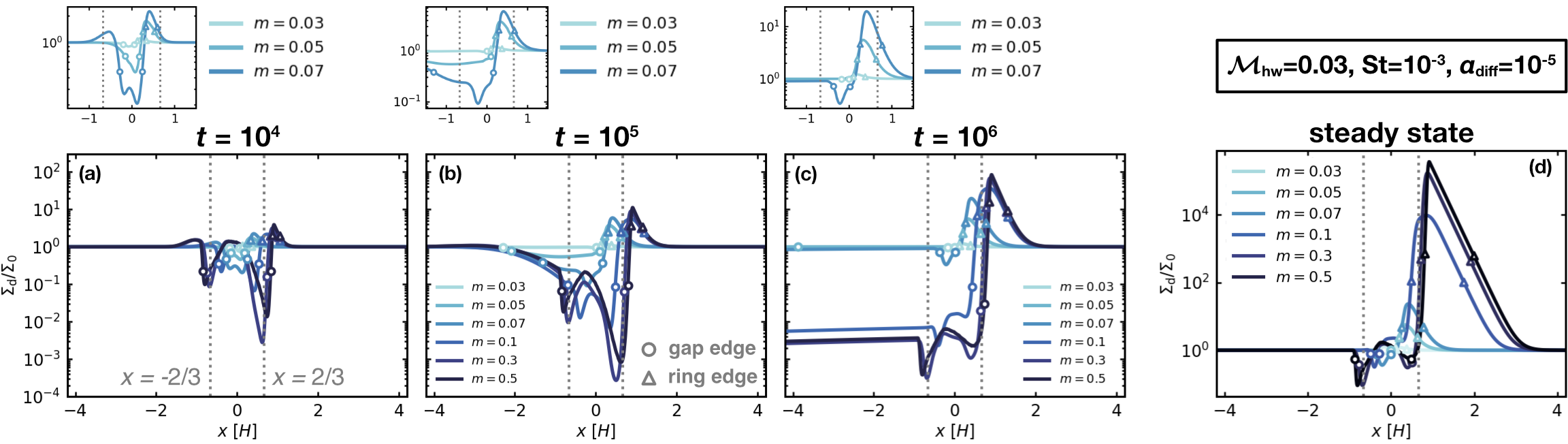}
    \caption{Dependence of $\Sigma_{\rm d}(t)$ on the planetary mass. \add{We set} $\mathcal{M}_{\rm hw}=0.03,\,{\rm St}=10^{-3}$, and $\alpha_{\rm diff}=10^{-5}$. The vertical dotted lines correspond to $|x|=4/3$ (the $x$-coordinate of the edge of the outflow region for $m\gtrsim0.3$; \Equref{eq:width of outflow region}). The figures on the upper left corners of the panels a--c are the zoom-in views for $m=0.03,\,0.05,$ and $0.07$.}
    \label{fig:planet mass dependence appendix}
\end{figure*}

\begin{figure*}[htbp]
    \centering
    \includegraphics[width=\linewidth]{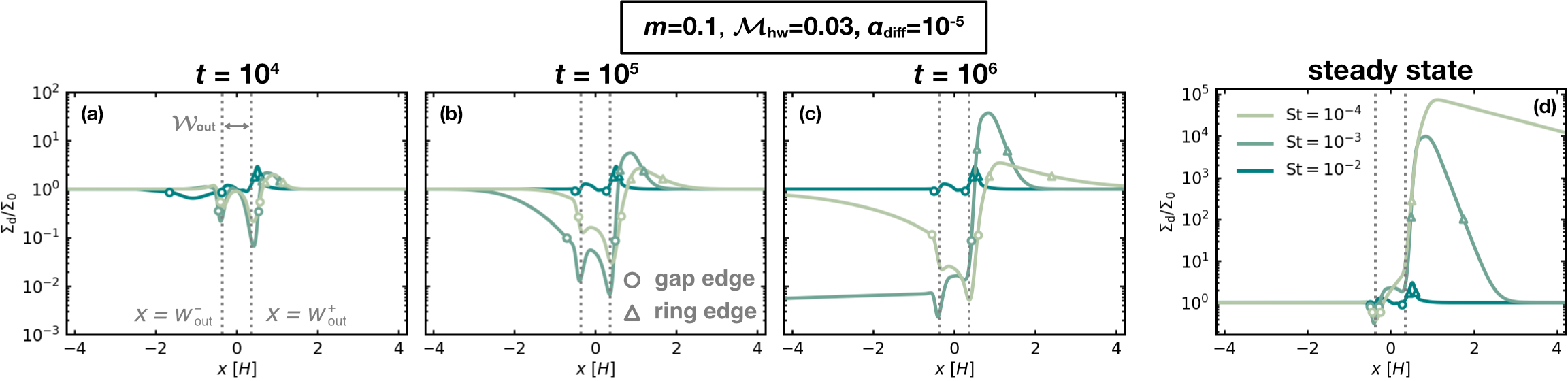}    
    \caption{Dependence of $\Sigma_{\rm d}(t)$ on the Stokes number. \add{We set} $m=0.1,\,\mathcal{M}_{\rm hw}=0.03$, and $\alpha_{\rm diff}=10^{-5}$. The vertical dotted lines correspond to $x=w_{\rm out}^\pm$.}
    \label{fig:St dependence appendix}
\end{figure*}

\begin{figure*}[htbp]
    \centering
    \includegraphics[width=\linewidth]{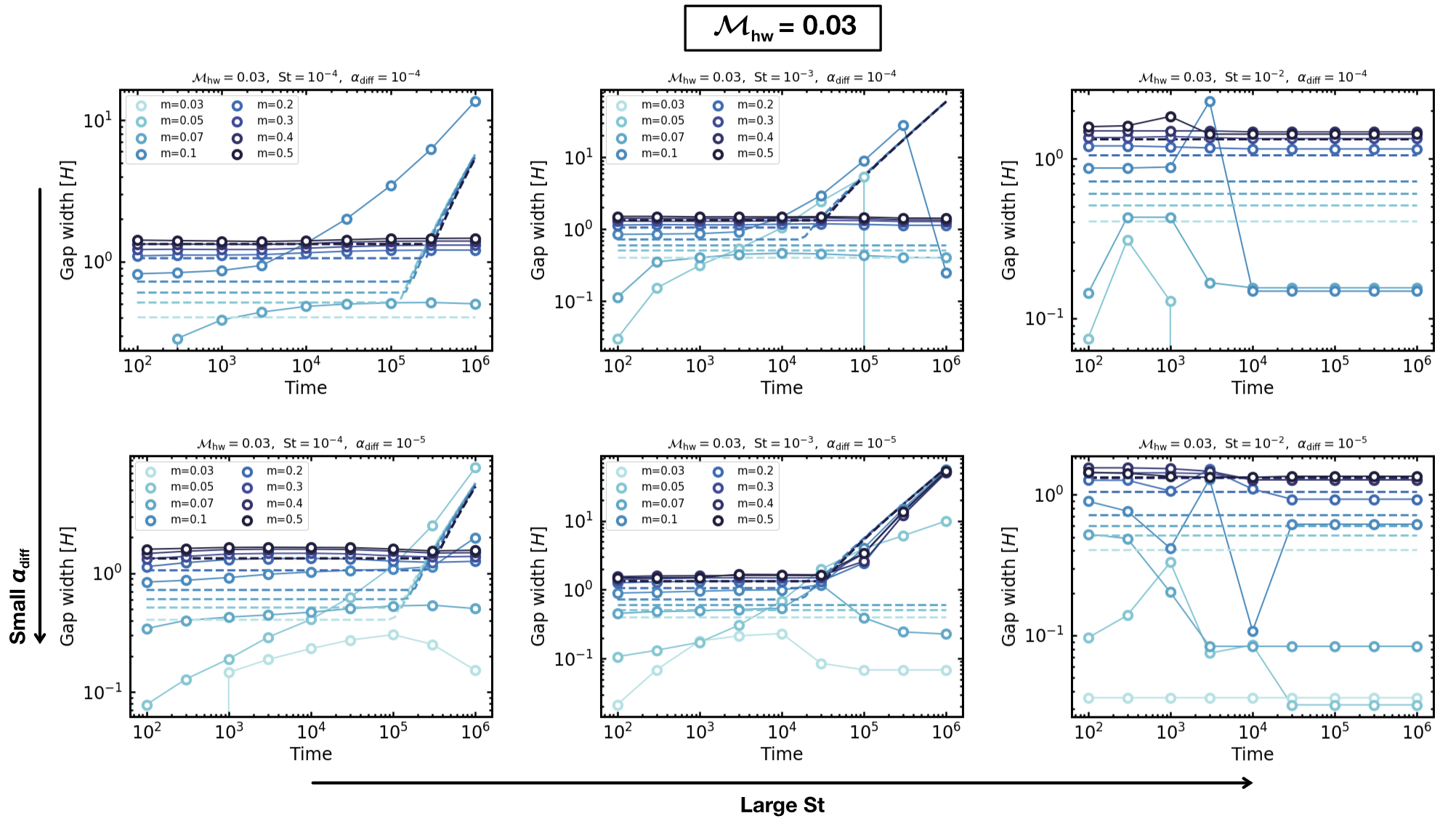}    
    \caption{Time evolution of the dust gap width for different planetary masses. We fixed the Mach number of the headwind as $\mathcal{M}_{\rm hw}=0.03$. We set $\alpha_{\rm diff}=10^{-4}$ in the top row and $\alpha_{\rm diff}=10^{-5}$ in the bottom row. We set ${\rm St}=10^{-4}$ (\textit{left column}), ${\rm St}=10^{-3}$ (\textit{middle column}), and ${\rm St}=10^{-2}$ (\textit{right column}). \addsec{The solid lines with the circle symbols and the dashed lines are the numerically-calculated and the semi-analytic dust gap widths, respectively} (Eq. \ref{eq:W gap SA}).}
    \label{fig:gapwidth_time_appendix}
\end{figure*}

\begin{figure*}[htbp]
    \centering
    \includegraphics[width=\linewidth]{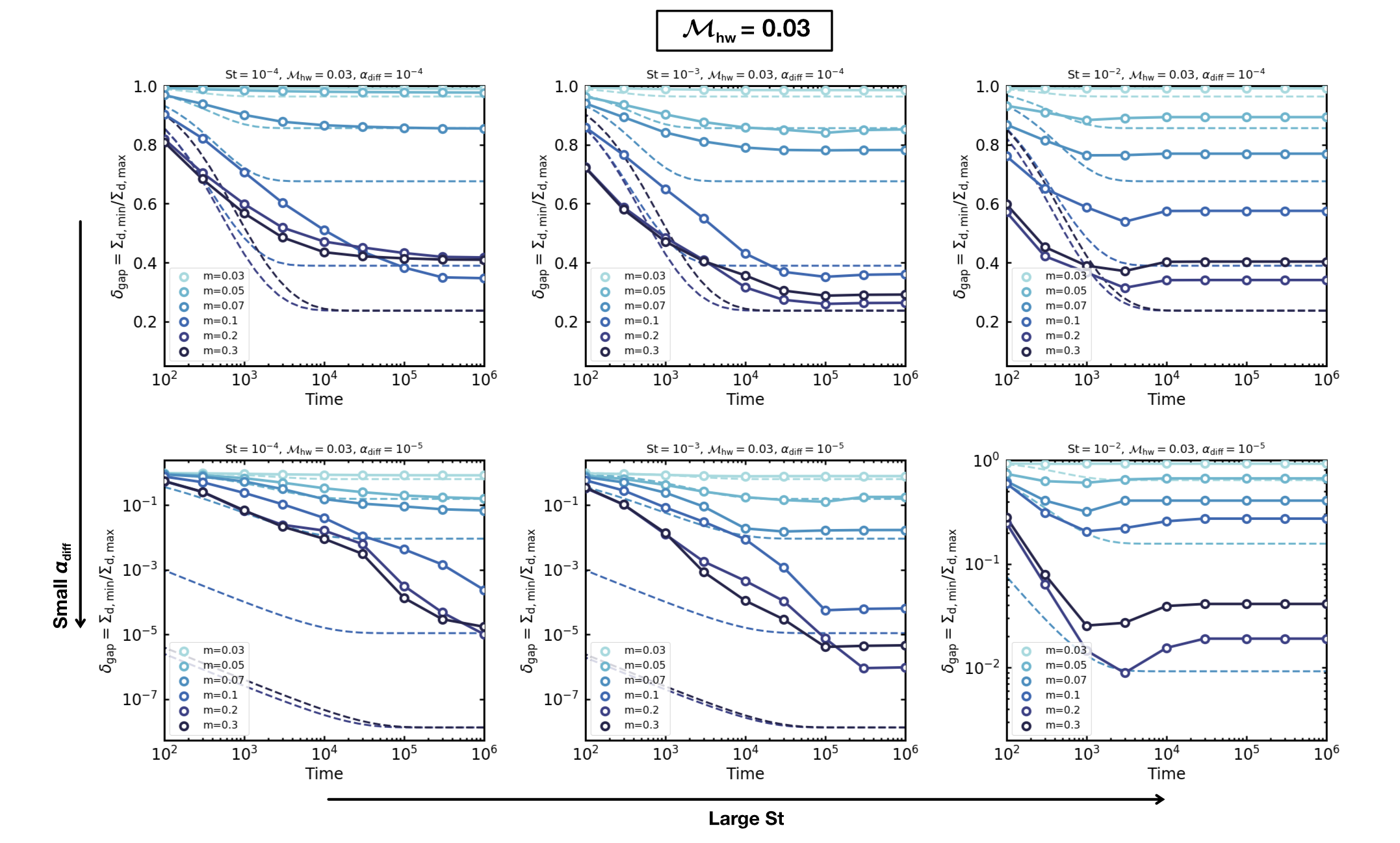}    
    \caption{Time evolution of the dust gap depth for different planetary masses.  We fixed the Mach number of the headwind as $\mathcal{M}_{\rm hw}=0.03$. We set $\alpha_{\rm diff}=10^{-4}$ in the top row and $\alpha_{\rm diff}=10^{-5}$ in the bottom row, respectively. We set ${\rm St}=10^{-4}$ (\textit{left column}), ${\rm St}=10^{-3}$ (\textit{middle column}), and ${\rm St}=10^{-2}$ (\textit{right column}). \addsec{The solid lines with the circle symbols and the dashed lines are the numerically-calculated and the semi-analytic dust gap depths, respectively} (Eq. \ref{eq:delta SA}; Sect. \ref{sec:Semi-analytic models of dust rings and gaps}). We note that the semi-analytic dust gap depth for $m\geq0.1$ is out of the range of the plot when ${\rm St}=10^{-2}$ and $\alpha_{\rm diff}=10^{-5}$, which deviates significantly from the numerical result.}
    \label{fig:gapdepth_time_appendix}
\end{figure*}

\begin{figure*}[htbp]
    \centering
    \includegraphics[width=\linewidth]{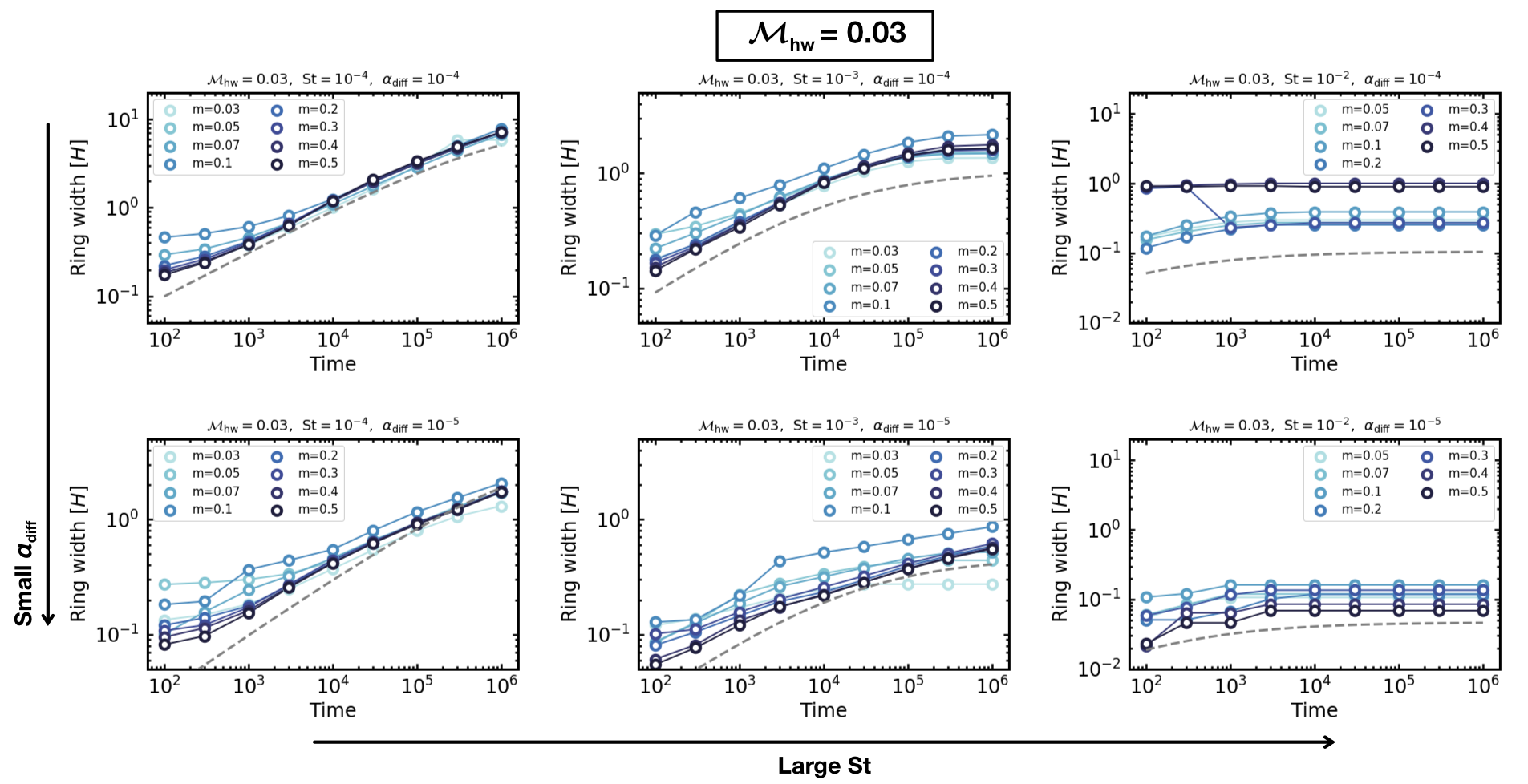}    
    \caption{Time evolution of the dust ring width for different planetary masses. We fixed the Mach number of the headwind as $\mathcal{M}_{\rm hw}=0.03$. We set $\alpha_{\rm diff}=10^{-4}$ in the top row and $\alpha_{\rm diff}=10^{-5}$ in the bottom row, respectively. We set ${\rm St}=10^{-4}$ (\textit{left column}), ${\rm St}=10^{-3}$ (\textit{middle column}), and ${\rm St}=10^{-2}$ (\textit{right column}).  The solid lines with the circle symbols and the dashed lines are the numerically-calculated and the semi-analytic dust ring widths, respectively (Eq. \ref{eq: W ring SA}; Sect. \ref{sec:Semi-analytic models of dust rings and gaps}), respectively.}
    \label{fig:ringwidth_appendix}
\end{figure*}

\end{appendix}

\end{document}